\documentclass[aps,prc,twocolumn,superscriptaddress,showpacs,floatfix]{revtex4}
\usepackage{xspace}
\usepackage{longtable}
\usepackage{graphicx}
\usepackage{afterpage}
\usepackage{placeins}
\newcommand{\GeV}{\ensuremath{\mbox{GeV}}\xspace}

\newcommand{\GeVc}{\ensuremath{\mbox{GeV}/c}\xspace}
\newcommand{\MeVc}{\ensuremath{\mbox{MeV}/c}\xspace}

\newcommand{\mm}{\ensuremath{\mbox{mm}}\xspace}

\newcommand{\mrad}{\ensuremath{\mbox{mrad}}\xspace}
\newcommand{\rad}{\ensuremath{\mbox{rad}}\xspace}

\newcommand{\ps}{\ensuremath{\mbox{ps}}\xspace}

\newcommand{\pip}{\ensuremath{\pi^+}\xspace}
\newcommand{\pim}{\ensuremath{\pi^-}\xspace}

\newcommand{\bfpip}{\ensuremath{\mathbf {\pi^+}}\xspace}
\newcommand{\bfpim}{\ensuremath{\mathbf {\pi^-}}\xspace}

\def\be{\begin{equation}}
\def\ee{\end{equation}}
\def\bea{\begin{eqnarray}}
\def\eea{\end{eqnarray}}

\newcommand{\bfGeVc}{\ensuremath{\mathbf {\mbox{\bf GeV}/c}}\xspace}

\begin{document}
\title{{\Large EUROPEAN ORGANIZATION FOR NUCLEAR RESEARCH} \\
\vskip 2cm
\hspace*{10cm}
{\rm CERN-PH-EP/2009-024} \\
\hspace*{10cm}
{\rm 15 September 2009} \\

\vskip 3cm

{\large Forward production of  charged pions 
with incident protons on  nuclear targets at the 
CERN PS} \\
\vskip 3cm
{\rm  Measurements of the  double-differential $\pi^{\pm}$ production
  cross-section in the range of momentum $0.5~\GeVc \leq p \le 8.0~\GeVc$ 
  and angle $0.025~\rad \leq \theta  \le 0.25~\rad$
  in collisions of protons on
  beryllium, carbon, nitrogen, oxygen, aluminium, copper,
  tin, tantalum and lead are presented. 
  The data were taken with the large acceptance HARP detector in the T9 beam
  line of the CERN PS.
  %The incident pion had a momentum in the range from
  %3~\GeVc to  12.9~\GeVc and impinged on target with a thickness of
  %5\% of a nuclear interaction length.  
  %
  Incident particles were identified by an elaborate system of beam
  detectors.
  Thin targets of 5\% of a nuclear interaction length were used. 
  The tracking and identification of the
  produced particles were performed using the forward system of the 
  HARP experiment.
  Results are obtained for the double-differential cross-sections 
  $
  %{{d^2 \sigma^{\pi}}}/{{dpd\Omega }}
  %%{{d^2 \sigma}}/{{dpd\Omega }}
  {{\mathrm{d}^2 \sigma}}/{{\mathrm{d}p\mathrm{d}\Omega }}
  $
  mainly at four incident proton beam momenta (3~\GeVc, 5~\GeVc, 8~\GeVc and 
  12~\GeVc). 
  Measurements are compared with the  GEANT4 and MARS Monte Carlo
  generators.
  A global parametrization is provided as an approximation of all
  the collected datasets which can serve as a tool for quick yields estimates.} 

\vskip 5cm
\centerline{\bf{(submitted to Physical Review C)}}
\clearpage
}

\clearpage

\author{M.~Apollonio} 
\altaffiliation{Now at Imperial College, University of London, UK.}
\affiliation{Universit\`{a} degli Studi e Sezione INFN, Trieste, Italy}
\author{A.~Artamonov}   % and ITEP, Moscow
\altaffiliation{ITEP, Moscow, Russian Federation.}
\affiliation{ CERN, Geneva, Switzerland}
\author{A. Bagulya} 
\affiliation{P. N. Lebedev Institute of Physics (FIAN), Russian Academy of
Sciences, Moscow, Russia}
\author{G.~Barr}
\affiliation{Nuclear and Astrophysics Laboratory, University of Oxford, UK} 
\author{A.~Blondel}
\affiliation{Section de Physique, Universit\'{e} de Gen\`{e}ve, Switzerland} 
\author{F.~Bobisut} 
\affiliation{Sezione INFN$^{(a)}$ and Universit\'a degli Studi$^{(b)}$, 
Padova, Italy}
\author{M.~Bogomilov}
\affiliation{ Faculty of Physics, St. Kliment Ohridski University, Sofia,
  Bulgaria}
 \author{M.~Bonesini}
\thanks{Corresponding author (M.~Bonesini).~E-mail: 
maurizio.bonesini@mib.infn.it}
\affiliation{Sezione INFN Milano Bicocca, Milano, Italy} 
\author{C.~Booth} 
\affiliation{ Dept. of Physics, University of Sheffield, UK}
\author{S.~Borghi}  % new footnote or CERN
\altaffiliation{Now at CERN}
\affiliation{Section de Physique, Universit\'{e} de Gen\`{e}ve, Switzerland}
\author{S.~Bunyatov}
\affiliation{Joint Institute for Nuclear Research, JINR Dubna, Russia} 
\author{J.~Burguet--Castell}
\affiliation{Instituto de F\'{i}sica Corpuscular, IFIC, CSIC and Universidad de Valencia, Spain}
\author{M.G.~Catanesi}
\affiliation{Sezione INFN, Bari, Italy} 
\author{A.~Cervera--Villanueva}
\affiliation{Instituto de F\'{i}sica Corpuscular, IFIC, CSIC and Universidad de Valencia, Spain}
\author{P.~Chimenti}  % also supported Geneva
\affiliation{Universit\`{a} degli Studi e Sezione INFN, Trieste, Italy}
\author{L.~Coney} 
\altaffiliation{MiniBooNE Collaboration.}
\affiliation{Columbia University, New York, USA}
\author{E.~Di~Capua}
\affiliation{Universit\`{a} degli Studi e Sezione INFN, Ferrara, Italy} 
\author{U.~Dore}
\affiliation{ Universit\`{a} ``La Sapienza'' e Sezione INFN Roma I, Roma,
  Italy}
\author{J.~Dumarchez}
\affiliation{ LPNHE, Universit\'{e}s de Paris VI et VII, Paris, France}
\author{R.~Edgecock}
\affiliation{Rutherford Appleton Laboratory, Chilton, Didcot, UK} 
\author{M.~Ellis}          %, OK
\altaffiliation{Now at FNAL, Batavia, Illinois, USA.} %%%R.~Engel$^{h}$,
\affiliation{Rutherford Appleton Laboratory, Chilton, Didcot, UK}
\author{F.~Ferri}           % ?
\affiliation{Sezione INFN Milano Bicocca, Milano, Italy}
\author{U.~Gastaldi}
\affiliation{Laboratori Nazionali di Legnaro dell' INFN, Legnaro, Italy}
\author{S.~Giani} 
\affiliation{ CERN, Geneva, Switzerland}
\author{G.~Giannini} 
\affiliation{Universit\`{a} degli Studi e Sezione INFN, Trieste, Italy}
\author{D.~Gibin}
\affiliation{Sezione INFN$^{(a)}$ and Universit\'a degli Studi$^{(b)}$, 
Padova, Italy}
\author{S.~Gilardoni}       %$^3$, %supported by DOCT
\affiliation{ CERN, Geneva, Switzerland} 
\author{P.~Gorbunov}  %,                new footnote
\altaffiliation{ITEP, Moscow, Russian Federation.}
\affiliation{ CERN, Geneva, Switzerland}
\author{C.~G\"{o}\ss ling}
\affiliation{ Institut f\"{u}r Physik, Universit\"{a}t Dortmund, Germany}
\author{J.J.~G\'{o}mez--Cadenas} % also supported Geneva
\affiliation{Instituto de F\'{i}sica Corpuscular, IFIC, CSIC and Universidad de Valencia, Spain}
\author{A.~Grant}  
\affiliation{ CERN, Geneva, Switzerland}
\author{J.S.~Graulich}
\altaffiliation{Now at Section de Physique, Universit\'{e} de Gen\`{e}ve, Switzerland.}
\affiliation{Institut de Physique Nucl\'{e}aire, UCL, Louvain-la-Neuve,
  Belgium} 
\author{G.~Gr\'{e}goire}
\affiliation{Institut de Physique Nucl\'{e}aire, UCL, Louvain-la-Neuve,
  Belgium} 
\author{V.~Grichine}  % also supported Geneva
\affiliation{P. N. Lebedev Institute of Physics (FIAN), Russian Academy of
Sciences, Moscow, Russia}
\author{A.~Grossheim} %$^1$, %supported by DOCT
\altaffiliation{Now at TRIUMF, Vancouver, Canada.}
\affiliation{ CERN, Geneva, Switzerland} 
\author{A.~Guglielmi$^{(a)}$}
\affiliation{Sezione INFN$^{(a)}$ and Universit\'a degli Studi$^{(b)}$, 
Padova, Italy}
\author{L.~Howlett}
\affiliation{ Dept. of Physics, University of Sheffield, UK}
\author{A.~Ivanchenko}
\altaffiliation{ On leave from Novosibirsk University,  Russia.}
\affiliation{ CERN, Geneva, Switzerland}
\author{V.~Ivanchenko}  %,                new footnote
\altaffiliation{On leave  from Ecoanalitica, Moscow State University,
Moscow, Russia}
\affiliation{ CERN, Geneva, Switzerland}
\author{A.~Kayis-Topaksu}
\altaffiliation{Now at \c{C}ukurova University, Adana, Turkey.}
\affiliation{ CERN, Geneva, Switzerland}
\author{M.~Kirsanov}
\affiliation{Institute for Nuclear Research, Moscow, Russia}
\author{D.~Kolev} 
\affiliation{ Faculty of Physics, St. Kliment Ohridski University, Sofia,
  Bulgaria}
\author{A.~Krasnoperov} 
\affiliation{Joint Institute for Nuclear Research, JINR Dubna, Russia}
\author{J. Mart\'{i}n--Albo}
\affiliation{Instituto de F\'{i}sica Corpuscular, IFIC, CSIC and Universidad de Valencia, Spain}
\author{C.~Meurer}
\affiliation{Institut f\"{u}r Physik, Forschungszentrum Karlsruhe, Germany}
\noaffiliation{}
\author{M.~Mezzetto$^{(a)}$}
\affiliation{Sezione INFN$^{(a)}$ and Universit\'a degli Studi$^{(b)}$, 
Padova, Italy}
\author{G.~B.~Mills}
\altaffiliation{MiniBooNE Collaboration.}
\affiliation{Los Alamos National Laboratory, Los Alamos, USA}
\author{M.C.~Morone}
\altaffiliation{Now at University of Rome Tor Vergata, Italy.}   
\affiliation{Section de Physique, Universit\'{e} de Gen\`{e}ve, Switzerland}
\author{P.~Novella} 
\affiliation{Instituto de F\'{i}sica Corpuscular, IFIC, CSIC and Universidad de Valencia, Spain}
\author{D.~Orestano}
\affiliation{Sezione INFN$^{(c)}$ and Universit\'a$^{(d)}$  Roma Tre, 
Roma, Italy}
\author{V.~Palladino}
\affiliation{Universit\`{a} ``Federico II'' e Sezione INFN, Napoli, Italy}
\author{J.~Panman}
\affiliation{ CERN, Geneva, Switzerland}
 \author{I.~Papadopoulos}  
\affiliation{ CERN, Geneva, Switzerland}
\author{F.~Pastore} 
\affiliation{Sezione INFN$^{(c)}$ and Universit\'a$^{(d)}$  Roma Tre, 
Roma, Italy}
\author{S.~Piperov}
\affiliation{ Institute for Nuclear Research and Nuclear Energy,
Academy of Sciences, Sofia, Bulgaria}
\author{N.~Polukhina}
\affiliation{P. N. Lebedev Institute of Physics (FIAN), Russian Academy of
Sciences, Moscow, Russia}
\author{B.~Popov} 
\altaffiliation{Also supported by LPNHE, Paris, France.}
\affiliation{Joint Institute for Nuclear Research, JINR Dubna, Russia}
\author{G.~Prior}   %supported by DOCT 
\altaffiliation{Now at CERN}
\affiliation{Section de Physique, Universit\'{e} de Gen\`{e}ve, Switzerland}
\author{E.~Radicioni}
\affiliation{Sezione INFN, Bari, Italy}
\author{D.~Schmitz}
\altaffiliation{MiniBooNE Collaboration.}
\affiliation{Columbia University, New York, USA}
\author{R.~Schroeter}
\affiliation{Section de Physique, Universit\'{e} de Gen\`{e}ve, Switzerland}
\author{V.~Serdiouk}
\affiliation{Joint Institute for Nuclear Research, JINR Dubna, Russia}
\author{G~Skoro}
\affiliation{ Dept. of Physics, University of Sheffield, UK}
\author{M.~Sorel}
\affiliation{Instituto de F\'{i}sica Corpuscular, IFIC, CSIC and Universidad de Valencia, Spain}
\author{E.~Tcherniaev}
\affiliation{ CERN, Geneva, Switzerland}
 \author{P.~Temnikov}
\affiliation{ Institute for Nuclear Research and Nuclear Energy,
Academy of Sciences, Sofia, Bulgaria}
\author{V.~Tereschenko}  
\affiliation{Joint Institute for Nuclear Research, JINR Dubna, Russia}
\author{A.~Tonazzo}
\affiliation{Sezione INFN$^{(c)}$ and Universit\'a$^{(d)}$  Roma Tre, 
Roma, Italy}
\author{L.~Tortora$^{(c)}$}
\affiliation{Sezione INFN$^{(c)}$ and Universit\'a$^{(d)}$  Roma Tre, 
Roma, Italy}
\author{R.~Tsenov}
\affiliation{ Faculty of Physics, St. Kliment Ohridski University, Sofia,
  Bulgaria}
\author{I.~Tsukerman}   % and ITEP, Moscow
\altaffiliation{ITEP, Moscow, Russian Federation.}
\affiliation{ CERN, Geneva, Switzerland}
\author{G.~Vidal--Sitjes}  % now at ..
\altaffiliation{Now at Imperial College, University of London, UK.}
\affiliation{Universit\`{a} degli Studi e Sezione INFN, Ferrara, Italy}
\author{C.~Wiebusch}    % now at ..
\altaffiliation{Now at III Phys. Inst. B, RWTH Aachen, Germany.}
\affiliation{ CERN, Geneva, Switzerland}
\author{P.~Zucchelli} %on leave of absence from INFN-Ferrara
\altaffiliation{Now at SpinX Technologies, Geneva, Switzerland;\\
on leave  from INFN, Sezione di Ferrara, Italy.}
\affiliation{ CERN, Geneva, Switzerland}
\newpage

\vskip 2cm
\collaboration{\bf HARP Collaboration}
\noaffiliation
\date{\today}

\maketitle
\clearpage
%\tableofcontents
%\clearpage
%\listoffigures
%\clearpage
%\listoftables
%\clearpage

\section{Introduction}
Final HARP measurements of the double-differential cross-section, 
$
%{{d^2 \sigma}}/{{dpd\Omega }}
{{\mathrm{d}^2 \sigma}}/{{\mathrm{d}p\mathrm{d}\Omega }}
$,
%of positive and negative pion production for 
for $\pi^{\pm}$ forward production 
(in the range of momentum $0.5~\GeVc \leq p \le 8.0~\GeVc$ 
and angle $0.025~\rad \leq \theta  \le 0.25~\rad$) 
by incident protons of 3~\GeVc, 5~\GeVc, 8~\GeVc, 8.9~\GeVc (Be only),
12~\GeVc  and  12.9~\GeVc (Al only) momentum impinging
on thin solid beryllium, carbon, aluminium, copper, tin, tantalum  
or lead targets of 5\% 
nuclear interaction length are presented. 
Our results on the forward production of $\pip$ in p--Al interactions 
at 12.9 GeV/c \cite{ref:alPaper} 
and p--Be interactions at 8.9 GeV/c \cite{ref:bePaper}, 
useful for the understanding of the the 
K2K \cite{K2K-2} and MiniBoone, SciBooNE neutrino beams, 
have been reanalyzed now with
a consistent binning.
In addition, results at 12~\GeVc on thin carbon targets and 
on cryogenic N$_2$ and O$_2$ targets, 
relevant for a precise
calculation of atmospheric neutrino fluxes and improved modeling of 
extensive air showers, previously reported in
references \cite{ref:carbonfw} and \cite{ref:cnofw}, are 
included for completeness. 

The HARP experiment~\cite{ref:harp-prop} at the CERN PS
was designed to measure hadron yields from a large range
of solid and cryogenic nuclear targets and for incident particle momenta from
 1.5~GeV/c to 15~GeV/c.
This corresponds to a proton momentum region of great interest for
neutrino beams and far from coverage by earlier dedicated hadroproduction
experiments~\cite{ref:na56,ref:atherton}.
The main motivations were to measure pion yields 
to allow substantially improved calculations of
the atmospheric neutrino
flux~\cite{ref:atm_nu_flux} to be made,
to measure particle yields as input for the flux
calculation of accelerator neutrino experiments~\cite{ref:physrep},
to help to design the targetry for a future neutrino factory~\cite{ref:nf}
%,such as K2K~\cite{ref:k2k,ref:k2kfinal},
%MiniBooNE~\cite{ref:miniboone} and SciBooNE~\cite{ref:sciboone}
and to improve the reliability of extensive air shower simulations,
by reducing the uncertainties of hadronic interaction models
in the low energy range~\cite{ref:christine_phd}. 

Covering an extended range of solid targets in
the same experiment, it is possible to perform systematic
comparisons of hadron production models with measurements 
at different incoming beam momenta over a large range 
of target atomic number $A$.
Pion production data at low momenta  ($\leq 25$ \GeVc)
are extremely scarce and HARP
is the first experiment to provide a large data set, taken with many 
different targets, full particle identification and large detector 
acceptance.
These data, together with the ones published in reference 
\cite{ref:pionForward}, are also of great interest for corrections of secondary
interactions in detector studies in particle physics experiments.

Existing data are mainly at fixed production angles with Be targets
\cite{ref:expBe} and are affected usually 
by large uncertainties. Only two experiments
\cite{ref:Allaby,ref:Eichten} at 19.2 \GeVc and 24 \GeVc were 
performed with an extended set of nuclear targets. 
The E910 experiment at BNL has recently published data at 12.3 \GeVc and
17.5 \GeVc with protons on Be, Cu and Au targets \cite{ref:E910}.

Data were taken in the T9 beam of the CERN PS.

\section{Experimental apparatus}
\label{subsec:harp_det}

\begin{figure*}[htb]
\centering
\includegraphics[width=0.8\textwidth]{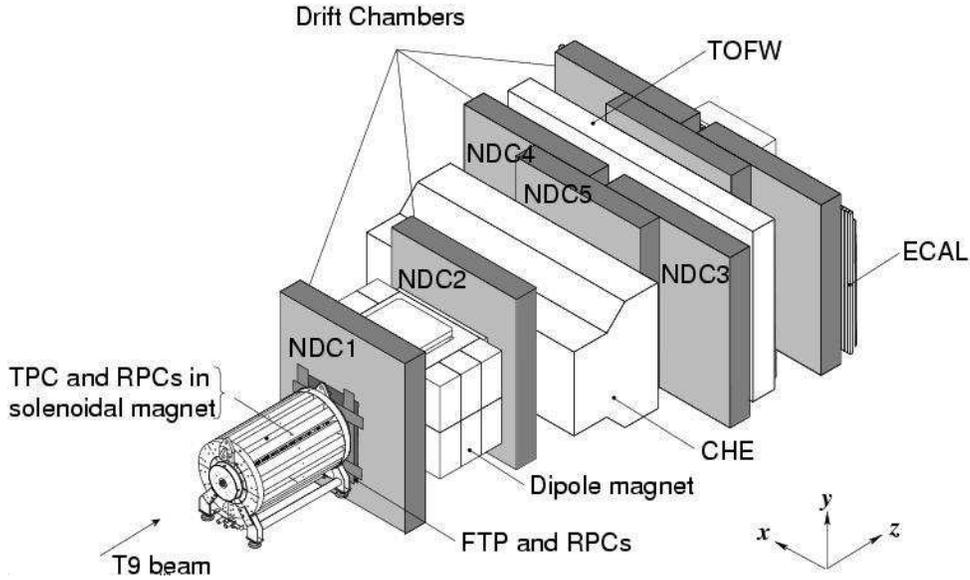} 
\caption{
\label{fig:harp} 
Schematic layout of the HARP detector.
The convention for the coordinate system is shown in the lower-right
corner.
The three most downstream (unlabelled) drift chamber modules are only partly
equipped with electronics and are not used for tracking.
The detector covers a total length of 13.5 m along the beam axis and has 
a maximum width of 6.5 m perpendicular to the beam.
}
\end{figure*}

 The HARP experiment
 makes use of a large-acceptance spectrometer consisting of a
 forward and a large-angle detection system.
 The HARP detector is shown in Fig.~\ref{fig:harp} and is fully described
 in Ref.~\cite{ref:harpTech}.
 The forward spectrometer -- 
 based on five modules of large area drift chambers
 (NDC1-5)~\cite{ref:NOMAD_NIM_DC} and a dipole magnet
 complemented by a set of detectors for particle identification (PID): 
 a time-of-flight wall (TOFW)~\cite{ref:tofPaper}, a large Cherenkov detector (CHE) 
 and an electromagnetic calorimeter (ECAL) --
 covers polar angles up to 250~mrad, 
 which is well matched to the angular range of interest for the
 measurement of hadron production to calculate the properties of
 conventional neutrino beams.
 As the most downstream detection element visible in the figure, 
 the 1.4 m wide beam muon identifier (BMI) is used to measure the muon contamination in the beam.

The discrimination power of TOFW time-of-flight below 3 GeV/c and the Cherenkov
dectector above 3 GeV/c are combined to provide powerful separation
of forward pions and protons. The calorimeter is 
%presently 
used only 
for separating pions and electrons when characterizing the response
of the other detectors. The resulting pion identification efficiency 
is bigger than 96\% (98\%) for pion momenta larger than 0.5 (3.0) 
GeV/c, with purity around $99.5 \%$. 

 The large-angle spectrometer -- based on a Time Projection Chamber (TPC) 
 and Resistive Plate Chambers (RPCs)
 located inside a solenoidal magnet --
 has a large acceptance in the momentum
 and angular range for the pions relevant to the production of the
 muons in a neutrino factory. 
 %It covers a large majority of the pions accepted in the focusing
 %system of a typical design.
 %The neutrino beam of a neutrino factory originates from
 %the decay of muons which are in turn the decay products of pions.
 For the analysis described here  only  the forward spectrometer and
 the beam instrumentation are used.

%\subsection{Beam, target and trigger detectors}
%\label{sec:beamtrigger}

The HARP experiment took data in 2001
and 2002.
The momentum definition of the T9 beam 
%used by the HARP experiment 
is known with a precision of the order of 1\%~\cite{ref:t9}. 

The target is placed inside the inner field cage of the TPC,
in an assembly that can be moved in and out of the solenoid magnet.
%such that,
%in addition to particles produced in the forward direction, 
%backward-going tracks can be measured.
The solid targets used for the measurements reported here
have a cylindrical shape with a nominal diameter of about 30~\mm.
Their thickness along the beam is equivalent to about 5\% 
of an interaction length $\lambda_\mathrm{I}$~\footnote{A cryogenic target
setup was instead used for N$_2$ and O$_2$ data taking. The target
thickness was equivalent to about 5.5\% (7.5\%) $\lambda_\mathrm{I}$ for
N$_2$ (O$_2$)}.

A set of four multi-wire
proportional chambers (MWPCs) measures the position and direction of
the incoming beam particles with an accuracy of $\approx$1~\mm in
position and $\approx$0.2~\mrad in angle per projection.
A beam time-of-flight system (BTOF)
measures the time difference of particles over a $21.4$~m path-length. 
It is made of two
identical scintillation hodoscopes, TOFA and TOFB (originally built
for the NA52 experiment~\cite{ref:NA52}),
which, together with a small target-defining trigger counter,
also used for the trigger, provide particle
identification at low energies. This provides separation of pions, kaons
and protons up to 5~\GeVc and determines the initial time at the
interaction vertex ($t_0$). 
The timing resolution of the combined BTOF system is about 70~\ps.
A system of two N$_2$-filled Cherenkov detectors (BCA and BCB) is
used to tag electrons at low energies and pions at higher energies. 
The electron and pion tagging efficiency is found to be close to
100\%.
A set of trigger detectors completes the beam instrumentation.

The positive beam used in this analysis contains mainly positrons, pions
and protons, with a small admixture of kaons and deuterons and heavier
ions. The proton fraction in the incoming beam goes from
$\sim 35 \%$ at 3 \GeVc to $\sim 92 \%$ at 12 \GeVc. 
The selection of beam protons is done requiring a pulse height consistent
with the absence of signal in both beam Cherenkov detectors (BCA and BCB). 
At the lowest beam energy, 3~\GeVc, the BTOF is used to reject pions
which at that momentum do not produce Cherenkov light.
At 5 GeV/c, the $\pi/p$ separation is obtained by the BTOF system and
only one of the Cherenkov (usually BCB), while the other is used to
tag $e^{\pm}$.
Ions are rejected by the BTOF system at all momenta.
In the 12.9~\GeVc beam the two Cherenkov counters were operated with
different pressures to make it possible to measure the kaons separately
from the protons and pions.  
After tagging the beam contamination from all particle species is negligible.
%This measurement ensures that the kaon content is negligible at all beam
%settings.  
Only events with a single reconstructed beam track in the four MWPCs,
good timing measurements in BTOF and no
signal in the beam halo counters are accepted. 

A downstream trigger in the forward scintillator trigger plane 
was required to record the event,
accepting only tracks with a trajectory outside the central hole
(60~mm diameter) which allows beam particles to pass. 
%The FTP is a double plane of scintillation counters
%covering the full aperture of the spectrometer magnet except 
%a 60~mm central hole for allowing
%non-interacting beam particles to pass.  
%The efficiency is measured using tracks recognized by the pattern
%recognition in a sample of events taken with a
%beam-trigger only and with a trigger based on signals in the Cherenkov
%detector. 
Accepting only tracks with a trajectory outside the central hole, the
efficiency  is measured to be $>$99.8\%. Particle identification is done
later, at the analysis stage, via the downstream PID detectors. 

The length of the accelerator spill is 400~ms with a typical intensity
of 15~000 beam particles per spill, and less for the lower momentum
settings. 
The average number of events recorded by the data acquisition ranges
from 300 to 350 per spill.

The absolute normalization of the cross-section was
performed using `incident-proton' triggers. 
These are triggers where the same selection on the beam particle was
applied but no selection on the interaction was performed.
The rate of this trigger was down-scaled by a factor 64.
These unbiased events are used to determine $N_\mathrm{pot}$ in the 
cross-section formula (1), see later.
\section{Data Analysis}
%%% \subsection{Event and particle selection}

%%A detailed description of the experimental techniques 
%%used for data analysis in the HARP forward spectrometer 
%%can be found in Ref.~\cite{ref:alPaper,ref:pidPaper}.
The analysis procedure is similar to the one reported in
Refs.~\cite{ref:carbonfw}, \cite{ref:cnofw},
\cite{ref:bePaper} and
\cite{ref:pionForward} where with respect to our first paper on 
forward pion production 
in p--Al interactions~\cite{ref:alPaper}, a number of improvements to
the analysis techniques and detector simulation had been made.   
Tracks are reconstructed in the drift chambers downstream of the
magnet.
Only tracks with at least one  hit in the TOFW are accepted for the analysis.
Hits are searched for in the Cherenkov detector consistent with these
tracks to complete the particle identification.
Secondary track selection criteria are optimized to ensure the quality
of momentum reconstruction and a clean time-of-flight measurement
while maintaining a high reconstruction efficiency. 
The method to combine the information from these PID detectors is
described in \cite{ref:pidPaper}.
In the kinematic range of the current analysis the pion identification
efficiency is about 98\%, while the background from mis-identified
protons is well below 1\%.
The momentum reconstruction is performed by means of a 
special implementation~\cite{recpack} 
of the Kalman filter~\cite{kalman} which uses
the position of the target as a constraint.
For the final stage of the analysis the same unfolding technique
as in Refs.~\cite{ref:carbonfw}, \cite{ref:cnofw} and
\cite{ref:pionForward} has been applied.

The background induced by
interactions of beam particles in the materials outside the target
is measured  by taking data without a
target in the target holder (``empty target data'').  
These data are  subject to the same  event and track
selection criteria as the standard  data sets. 
To take into account this background the number of particles of the
observed particle type  in the ``empty target data''
are subtracted bin-by-bin (momentum and angular bins) from the number
of particles of the same type. The uncertainty induced by
this method is labeled ``empty target subtraction''. 
The collected event statistics on the different solid targets 
is summarised in Table~\ref{tab:events}. 
\begin{table*}[hbp!] 
\caption{Total number of events and tracks used in the various nuclear 
  5\%~$\lambda_{\mathrm{I}}$ solid target data sets and the number of
  accepted protons on target as calculated from the pre-scaled 
  incident beam triggers. Numbers are
  for incident proton in units of $10^3$ events. } 
\label{tab:events}
{\small
\begin{center}
\begin{tabular}{ll|c|c|c|c|c|c} \hline
%\bf{Data set}          &         &\bf{3 \bfGeVc}&\bf{5 \bfGeVc}&\bf{8 \bfGeVc}&\bf{8.9 \bfGeVc} & \bf{12 \bfGeVc} & \bf{12.9 \bfGeVc}\\ \hline
\bf{Data set (\bfGeVc)}          &         &\bf{3}&\bf{5}&\bf{8}&\bf{8.9}&\bf{12} & \bf{12.9}\\ \hline
    Total DAQ events     & (Be)     & 1113  & 1296   & 1935     & 5868 &  1207 &                \\
                         &  (C)     & 1345  & 2628       & 1846 &    &    1062 &              \\
                         & (Al)     & 1159  & 1789      & 1707    &  &  619   & 4713            \\
                         & (Cu)     & 624   & 2079       & 2089     & &    745   &              \\
                         & (Sn)     & 1637  & 2828       & 2404    & &   1803  &        \\
                         & (Ta)     & 1783  & 2084      & 1965    &  &  866   &       \\   
                         & (Pb)     & 1911  & 2111      & 2266     &  &  487  &        \\ 
\hline  
  Accepted  beam protons &  (Be)    & 99   &  289      &  761   & 2103  &    580  & \\
  with forward interaction & (C)    & 101  &  542      &  709   &       &    470 &  \\
                         &  (Al)    & 86   &  376      &  637    &   &  306     & 2116                 \\
                         &  (Cu)    & 73   & 408      &  741  &  &    363   &               \\
                         &  (Sn)    & 217   & 528      &  818    & &   856     &               \\
                         &  (Ta)    & 281   & 398      &  668    & &   403     &               \\
                         &  (Pb)    & 310   & 387      &  758    & &   221     &                 \\
\hline
  \bf{Final state $\bfpim$ ($\bfpip$) } & (Be)  & 0.08 (0.4)   &   1.2 (2.9)     &  8.0  (15.5) & 26.5 (48.6)   &  11.0 (18.0) &         \\
     {selected with PID}     & (C)   & 0.06 (0.3)   &   1.9 (4.9)     &  6.9  (13.8) &  &  8.2  (13.5) &          \\
                             & (Al)  & 0.05 (0.2)    &   1.3 (3.3)     &  6.4 (12.9)  &  &  5.8  (9.3) &  45.9  (70.9)   \\ 
                             & (Cu)  & 0.03 (0.1)   &   1.3 (3.1)     &  7.1 (12.6) &  &  6.8 (10.4) &             \\
                             & (Sn)  & 0.07 (0.2)   &   1.5 (3.4)     &  7.6  (13.6)&  &  15.2 (23.5) &              \\
                             & (Ta)  & 0.08 (0.2)   &   1.0 (2.2)     &  5.5 (9.7) &  &  7.1 (10.4)  &               \\
                             & (Pb)  & 0.08 (0.2)   &   0.9 (2.0)     &  5.7 (10.3) &  &  3.7 (5.4)  &                \\
\hline
\end{tabular}
\end{center}
}
\end{table*}

\subsection{Cross-section calculation}

The double-differential cross-section is calculated as follows:
\begin{eqnarray}
\frac{d^2 \sigma^{\alpha}}{dp d\Omega}(p_i,\theta_j)   =   
\frac{A}{N_A \rho t} \cdot \frac{1}{N_{\rm pot}} \cdot \frac{1}{\Delta p_i \Delta \Omega_j} \cdot \\ \nonumber
\sum_{p'_i,\theta'_j,\alpha'} \mathcal{M}^{\rm cor}_{p_i\theta_j\alpha p'_i\theta'_j\alpha'} \cdot 
N^{\alpha'}(p'_i,\theta'_j)\hspace{0.1cm}
\end{eqnarray} 
where 
\begin{itemize}
\item $\frac{d^2 \sigma^{\alpha}}{dp d\Omega}(p_i,\theta_j)$ is the
  cross-section in mb/(\GeVc sr) for the produced particle type $\alpha$ (p,
  $\pi^+$ or $\pi^-$) for each true momentum and angle bin ($p_i,\theta_j$)
  covered in this analysis;
\item $N^{\alpha'}(p'_i,\theta'_j)$  is the number of reconstructed 
  particles of
  type $\alpha'$ in bins of reconstructed momentum $p'_i$ and angle
  $\theta_j'$ in the raw data, after subtraction of empty target data 
  (due to beam protons interacting in material other than the nuclear
   target). These particles must satisfy the event, track 
  and PID selection criteria.
\item $\mathcal{M}^{\rm cor}_{p\theta\alpha p'\theta'\alpha'}$ is the
  correction matrix which accounts for the finite efficiency and resolution of
  the detector. It unfolds the true variables $p_i, \theta_j, \alpha$ from
  the reconstructed variables $p'_i, \theta'_j, \alpha'$ and corrects the
  observed particle number to take into account effects such as reconstruction
  efficiency, acceptance, absorption, pion decay, tertiary production, PID
  efficiency and PID misidentification rate.
\item $\frac{A}{N_A \rho t}$, $\frac{1}{N_{\rm pot}}$ and
  $\frac{1}{\Delta p_i \Delta \Omega_j}$ are normalization factors,
  namely:
\subitem $\frac{N_A \rho t}{A}$ is the number of target nuclei per unit area 
\footnote{$A$ - atomic  mass, $N_A$ - Avogadro number, $\rho$ - target
  density and $t$ - target thickness};
\subitem $N_{\rm pot}$ is the number of protons on
  target;
\subitem $\Delta p_i $ and $\Delta \Omega_j $ are the bin sizes in
  momentum and solid angle, respectively 
\footnote{$\Delta p_i = p^{\rm max}_i-p^{\rm min}_i$,\hspace{0.2cm}
  $\Delta \Omega_j = 2 \pi (\cos(\theta^{\rm min}_j)- 
  \cos(\theta^{\rm max}_j))$}.
\end{itemize}
We do not make a correction for the attenuation
of the proton beam in the target, so that strictly speaking the
cross-sections are valid for $\lambda_{\mathrm{I}}=5\%$
targets.

%While the number of particles $N^{\alpha'}(p'_i,\theta'_j)$ is
%relatively easy to count in the raw data, the 
The  calculation of the
correction matrix $M^{\rm cor}_{p_i\theta_j\alpha
  p'_i\theta'_j\alpha'}$ is done with
the unfolding method introduced 
by D'Agostini~\cite{ref:DAgostini}~\footnote{
The  unfolding method tries to put in correspondence the
vector of measured observables (such as particle momentum, polar
angle and particle type) $x_{\rm meas}$ with the vector of true values
$x_{\rm true}$ using a migration matrix: $x_{meas} = {\sl M}_{migr} \times x_{true}$.
The goal of the method is to compute a transformation 
(correction matrix) to obtain the expected
values for $x_{true}$ from the measured ones. The most simple
and obvious solution, based on simple matrix inversion 
${\sl M}^{-1}_{\rm migr}$, 
is usually unstable and is dominated by large variances and strong negative
correlations between neighbouring bins.
In the method of D' Agostini, 
the correction matrix ${\sl M}^{\rm UFO}$ tries to connect
the measurement space (effects) with the space of the true values (causes) 
using an iterative Bayesian approach, based on Monte Carlo simulations to
estimate the probability for a given effect to be produced by a certain 
cause.}. 
This method has been used in the recent HARP 
publications~\cite{ref:carbonfw, ref:cnofw,ref:pionForward} 
and it is also applied in the analysis described here.

The Monte Carlo simulation of the HARP setup is based on 
GEANT4~\cite{ref:geant4}. 
%(GEometry ANd Tracking)~\cite{ref:geant4}. 
The detector
materials are accurately 
%reproduced 
described
in this simulation as well as the
relevant features of the detector response and the digitization
process. All relevant physics processes are considered, including
multiple scattering, energy loss, absorption and
re-interactions. 
The simulation is independent of the beam particle type
because it only generates for each event
%starts the interaction process by generating 
exactly one secondary particle of a specific particle type inside the
target material and propagates it through the 
%\fix{target}
complete detector. 
A small difference (at the few percent level) is observed between the
efficiency calculated for 
events simulated with the single-particle Monte Carlo and with a
simulation using a multi-particle hadron-production model.
A similar difference is seen between the single-particle Monte Carlo and
the efficiencies measured directly from the data.
A momentum-dependent correction factor determined using the efficiency
measured with the  data is applied to take this into account. 
The track reconstruction used in this analysis and the simulation are
identical to the ones used for the $\pi^+$ production in p--Be
collisions~\cite{ref:bePaper}. 
A detailed description of the corrections and their magnitude can be
found there. 

The reconstruction efficiency (inside the geometrical acceptance) is
larger than 95\% above 1.5~\GeVc and drops to 80\% at 0.5~\GeVc. 
The requirement of a match with a TOFW hit has an efficiency between
90\% and 95\% independent of momentum.
The electron veto rejects about 1\% of the pions and protons below
3~\GeVc with a remaining background of less than 0.5\%.
Below Cherenkov threshold ( $\sim 3$ GeV/c) the TOFW separates pions and protons with
negligible background and an efficiency of $\approx$98\% for pions.
Above Cherenkov threshold the efficiency for pions is greater than 99\%
with only 1.5\% of the protons mis-identified as a pion.
The kaon background in the pion spectra is smaller than 1\%.

%GB point A
The absorption and decay of particles is simulated by the Monte Carlo.
The generated single particle can re-interact and produce background
particles by hadronic or electromagnetic processes, thus giving rise to
tracks in the dipole spectrometer.
In such cases also the additional tracks are entered into the
migration matrix thereby taking into account the combined effect of the
generated particle and any secondaries it creates.
The absorption correction is on average 20\%, approximately independent
of momentum.
Uncertainties in the absorption of secondaries in the dipole
spectrometer material are taken into account by
a variation of 10\% of this effect in the simulation. 
The effect of pion decay is treated in the same way as the absorption
and is 20\% at 500~\MeVc and negligible at 3~\GeVc. 

The uncertainty in the production of background due to tertiary
particles is larger. 
The average correction is $\approx$10\% and up to 20\% at
1~\GeVc. 
The correction includes reinteractions in the detector material 
(mainly carbon) as well
as a small component coming from reinteractions in the target.
The subtraction may be computed by MonteCarlo simulations and as
most of the encountered material is carbon, the check of the inelastic
interactions of low-energy protons or pions in carbon is essential.  
The validity of the generators used in the simulation was checked by the
analysis of our data with incoming protons and charged pions on
aluminium and carbon targets at lower momenta (3~\GeVc and 5~\GeVc).
A 30\% uncertainty on the secondary production was assumed.

The average empty-target subtraction amounts to $\approx$20\%.

%\enlargethispage{-2.5cm}

Owing to the redundancy of the tracking system downstream of the
target the detection efficiency is very robust under the usual
variations of the detector performance during the long data taking
periods. 
Since the momentum is reconstructed without making use of the upstream
drift chamber module (which is more sensitive in its performance to the beam
intensity) the reconstruction efficiency is uniquely determined by the
downstream system.
No variation of the overall efficiency has been observed.
The performance of the TOFW and CHE system have been monitored to be
constant for the data taking periods used in this analysis.
The calibration of the detectors was performed on a day-by-day basis.

\subsection{Error estimation}
\label{errorest}

The total statistical error takes into account the direct error
propagation of the statistical errors in the raw data and the statistical
error which is incurred while obtaining the unfolding matrix from the
data. The latter component increases the direct error by a factor two. 
The procedure is outlined in references~\cite{ref:carbonfw},
\cite{ref:grossheim}.

Different types of sources 
%produce ~\cite{ref:carbonfw},\cite{ref:grossheim}.
induce
systematic errors for the analysis
%of the fixed target data 
described here: 
track yield corrections ($\sim 5 \%$), particle identification ($\sim 0.1 \%$),
momentum and angular reconstruction ($\sim 1 \%$)~\footnote{
The quoted error in parenthesis refers to fractional error 
of the integrated cross-section ($\delta^{\pi}_\mathrm{int} (\%)$)
in the kinematic range covered by the HARP experiment}.
The strategy to calculate these systematic errors and the different
methods used for their evaluation are described in~\cite{ref:carbonfw}.
An additional source of error is due to misidentified secondary kaons, 
which are not considered in the particle identification method used for
this analysis and are subtracted on the
basis of a Monte Carlo simulation, as in~\cite{ref:carbonfw}. 
No explicit correction is made for pions coming from decays of other 
particles created in the target, as they give a very small contribution
according to the selection criteria applied in the analysis.
As a result of these systematic error studies, each error source 
can be represented by a covariance matrix. The
sum of these matrices describes the total systematic error,
as explained in~\cite{ref:carbonfw}. 

\begin{figure*}[tbp]
  \begin{center}
\includegraphics[width=0.60\textwidth]{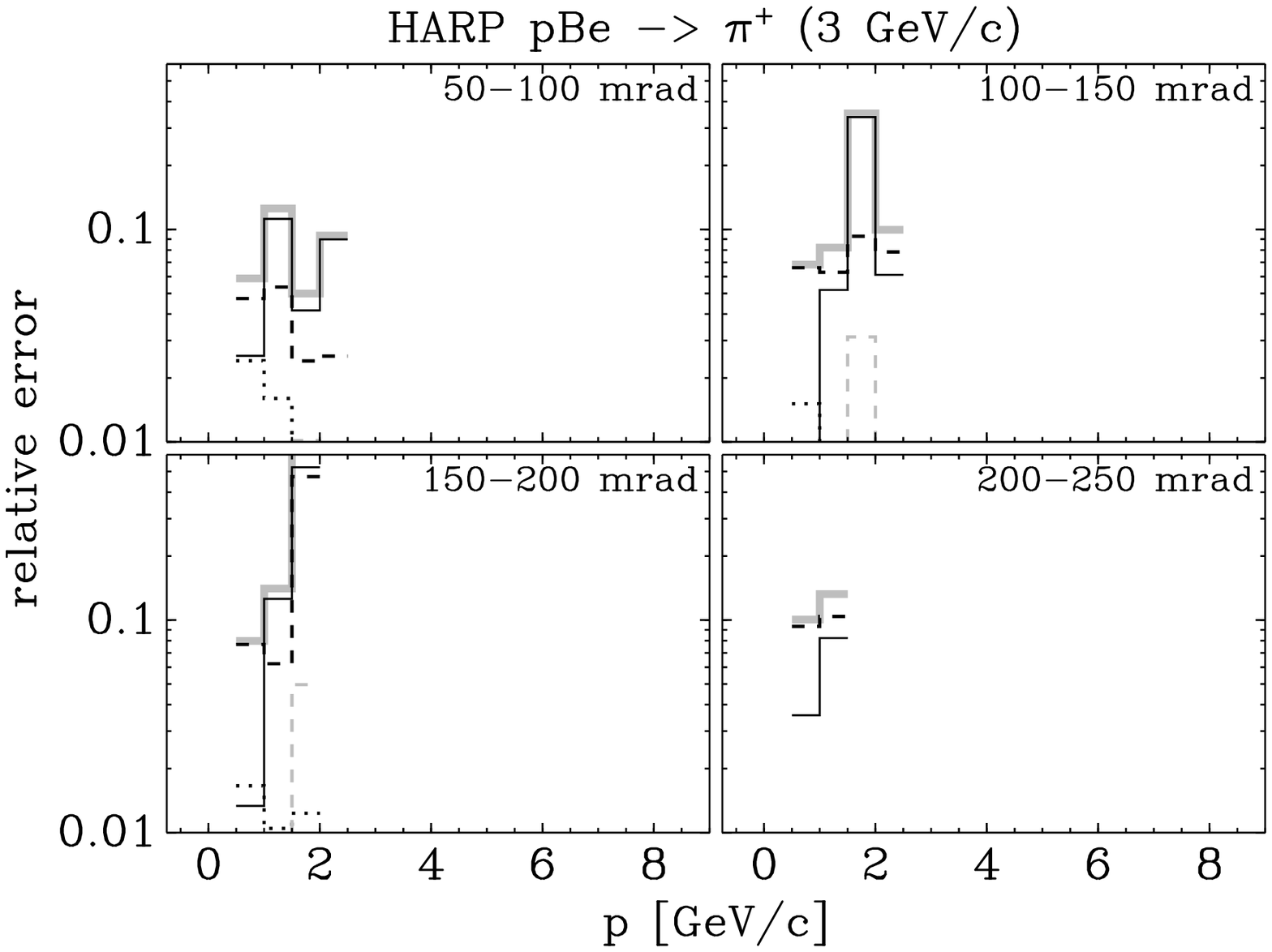}
\includegraphics[width=0.60\textwidth]{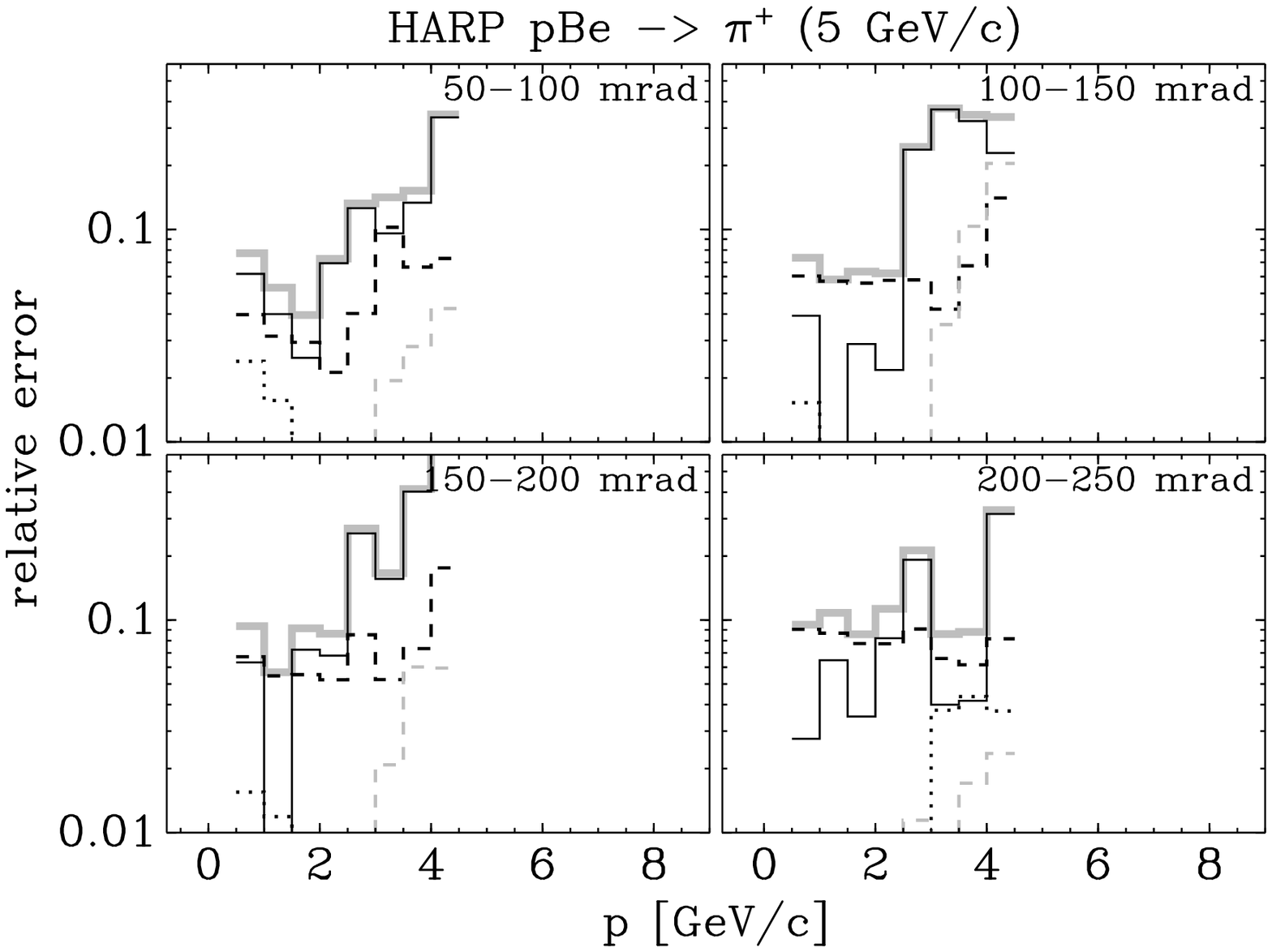}
\includegraphics[width=0.60\textwidth]{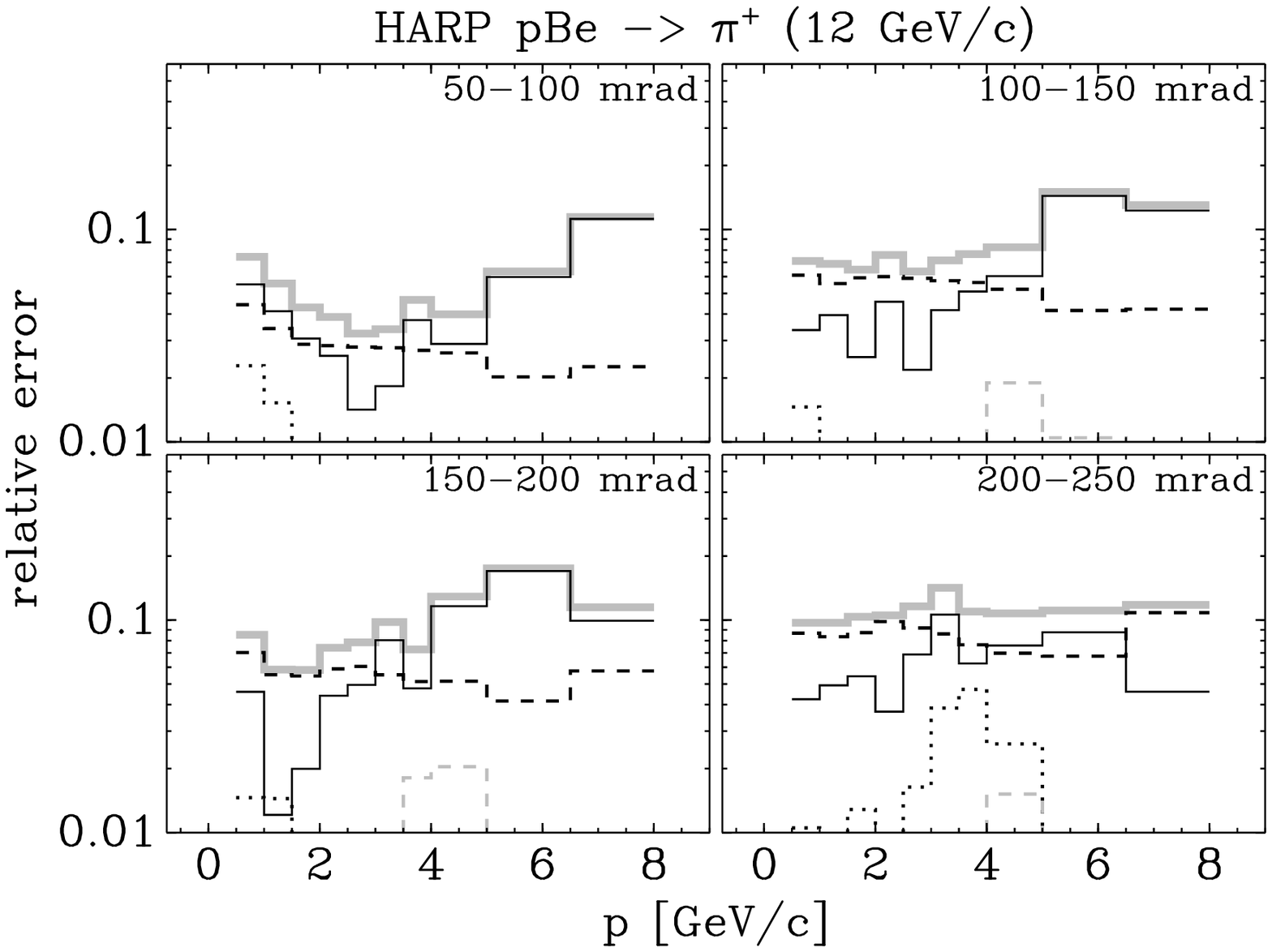}
\end{center}
\caption{Total systematic error (grey solid line) and main components:
black short-dashed line for absorption+tertiares interactions,
black dotted line for track efficiency and target pointing efficiency
(mostly $\leq 1 \%$), black solid line for
momentum scale+resolution and angle scale 
for a typical target (Be) at some typical incident momenta.
}
\label{fig:syst}
\end{figure*}

The experimental uncertainties are shown for a typical light target (Be)
in Figure~\ref{fig:syst}  for $\pi^{+}$ at some  typical 
incident beam momenta
and in Table \ref{tab:syst} for another target (C) at the incident beam momentum 12 GeV/c.
They are very similar for $\pi^{-}$ and at the other beam energies. 
Going from lighter (Be, C) to heavier targets (Ta, Pb)
the corrections for $\pi^{0}$ (conversion) and absorption/tertiaries increase.

\begin{table}
\caption{
\label{tab:syst}
Summary of the systematic uncertainties affecting the computed $\pi^{+}$ double-differential 
cross sections and the integrated cross-section measurements
for p--C interactions at 12 GeV/c. The entries of the table are
weighted bin by bin with the pion production yields.}

\begin{tabular}{| l c  | c |} \hline

{\bf Error Category} & $\delta_{\mathrm{\small diff}}^{\pi}$ (\%) & $\delta_{\mathrm{\small int}}^{\pi}$ (\%) \\ \hline

Track yield corrections: & &  \\
\hspace{0.5cm}{\small Reconstruction efficiency}  & 1.1 & 0.5 \\
\hspace{0.5cm}{\small Pion, proton absorption} &  3.7 & 3.2 \\
\hspace{0.5cm}{\small Tertiary subtraction} &  8.6 & 3.7 \\
\hspace{0.5cm}{\small Empty target subtraction} &  1.2 & 1.2 \\
\hspace{0.5cm}{\small \bf Sub-total} &   {\bf 9.5} & {\bf 5.1} \\ \hline

Particle Identification: & &  \\
\hspace{0.5cm}{\small Electron veto} &  $<$0.1 & $<$0.1 \\
\hspace{0.5cm}{\small Pion, proton ID correction}   & 0.1 & 0.1 \\ 
\hspace{0.5cm}{\small Kaon subtraction} &  $<$0.1 & $<$0.1 \\
\hspace{0.5cm}{\small \bf Sub-total} &   {\bf 0.1} & {\bf 0.1} \\ \hline

Momentum reconstruction: & &  \\
\hspace{0.5cm}{\small Momentum scale} & 2.8 & 0.3 \\
\hspace{0.5cm}{\small Momentum resolution} &  0.8 & 0.3 \\
\hspace{0.5cm}{\small \bf Sub-total} &  {\bf 2.9} & {\bf 0.4} \\ \hline

Angle reconstruction: & & \\
\hspace{0.5cm}{\small Angular scale} & 1.3 & 0.5 \\ \hline

{\bf Total syst.} &  {\bf 10.0 } & {\bf 5.1} \\ \hline 
{\bf Overall normalization:} & {\bf 2.0} & {\bf 2.0} \\ \hline
\end{tabular}
\end{table}
The overall normalization has an uncertainty of $\sim 2 \%$ and is
mainly due to the uncertainty in the efficiency that beam protons counted
in the normalization actually hit the target, with smaller components
from the target density and the beam particle counting procedure. 

On average the total integrated systematic error is around $5-6\%$,
with a differential bin to bin systematic error of the order of
$10-11 \%$, to be compared with a statistical integrated (bin-to-bin
differential) error of $\sim 2-3 \%$ ($\sim 10-13 \%$).
Systematic and statistical errors are roughly of the same order. 

\FloatBarrier
\section{Experimental results}
\label{sec:results}

\begin{figure*}[htb]
\centering
\includegraphics[width=.49\textwidth]{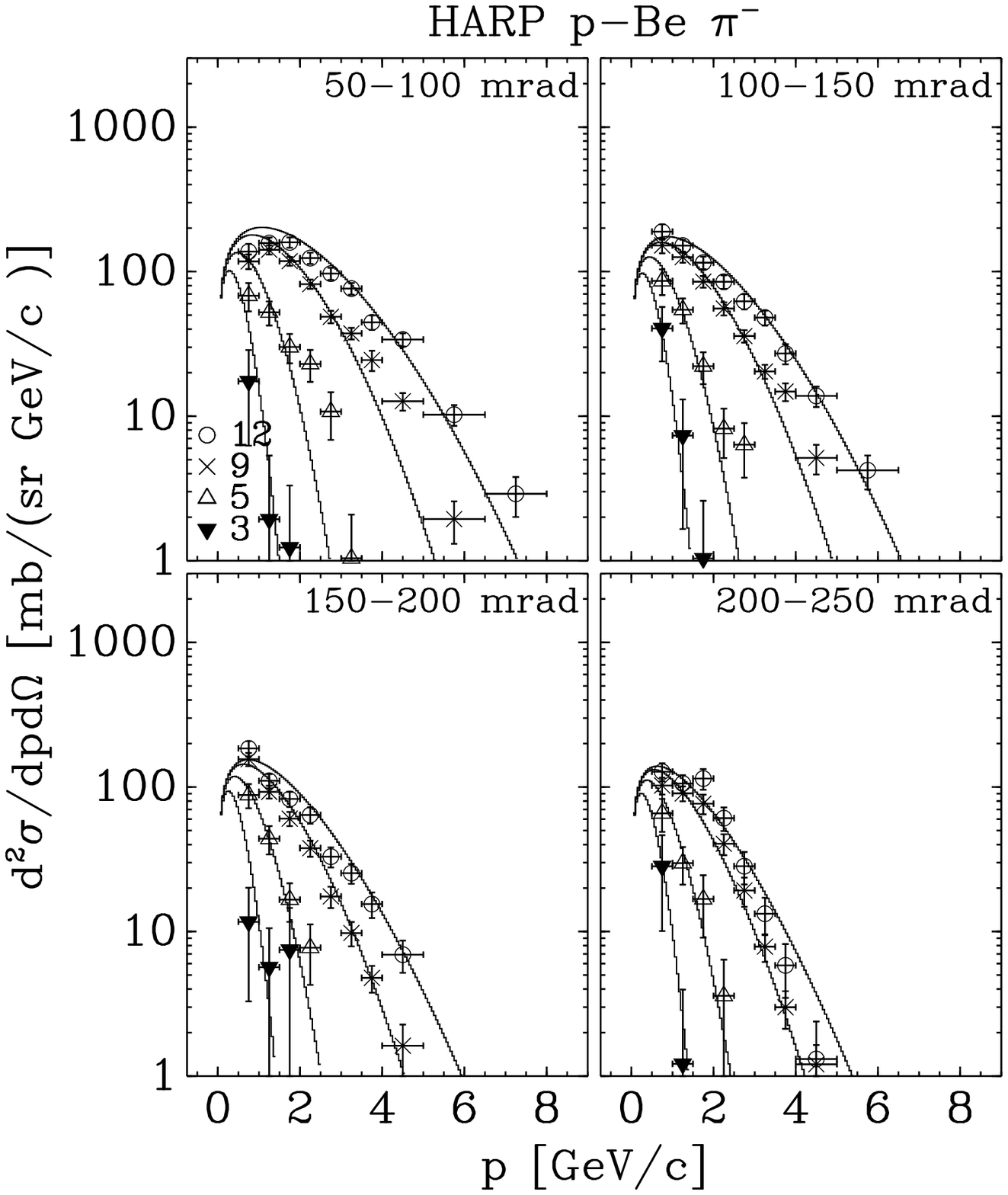}
\includegraphics[width=.49\textwidth]{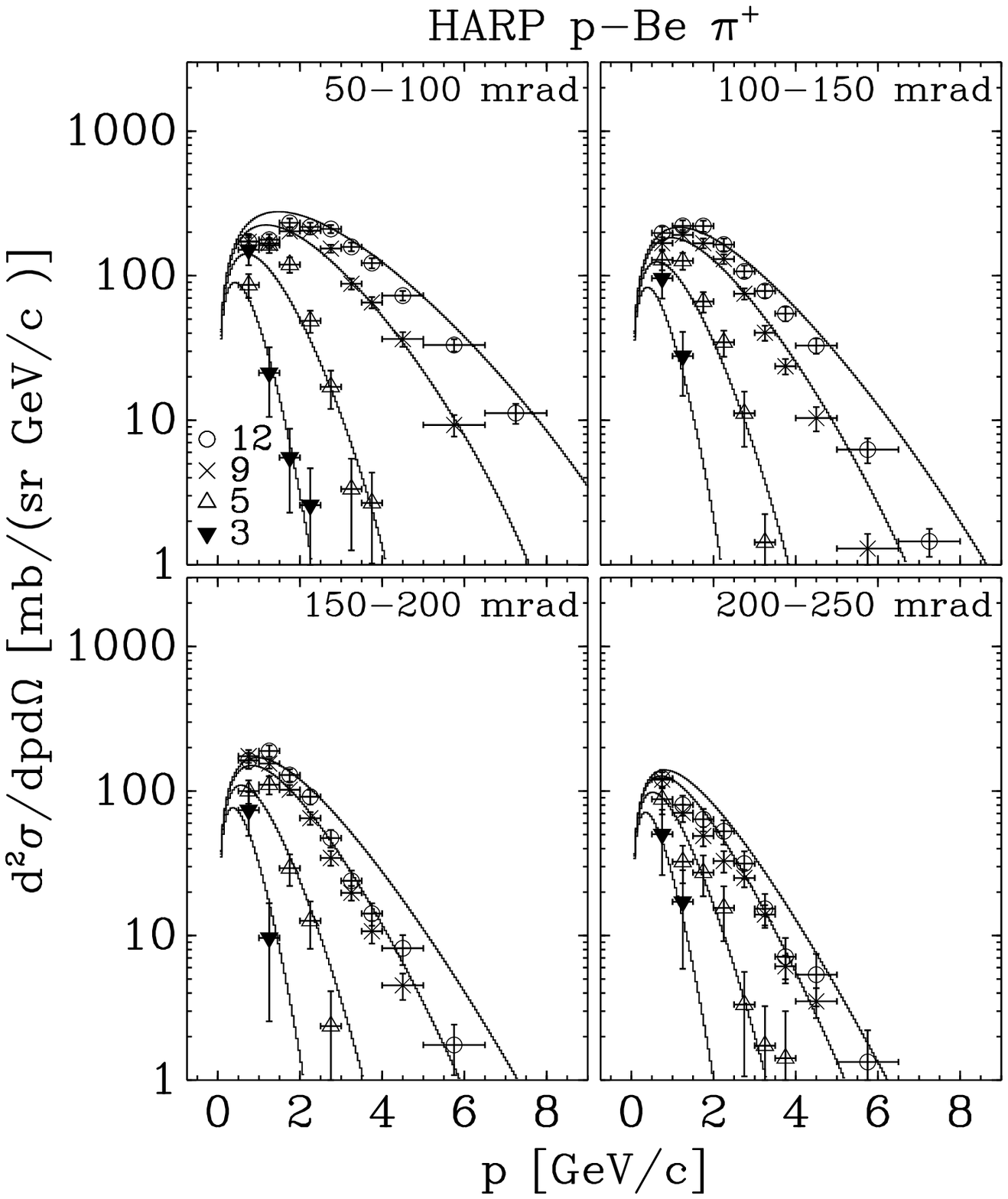}
\caption{p--Be differential cross sections: left panel
$\pi^{-}$ production, right panel $\pi^{+}$ production. 
The curves represent the global parametrization as described in the text.
In the top right corner of each plot the 
covered angular range is shown in mrad.}
\label{fig:Be}
\end{figure*}

\begin{figure*}[htb]
\centering
\includegraphics[width=.49\textwidth]{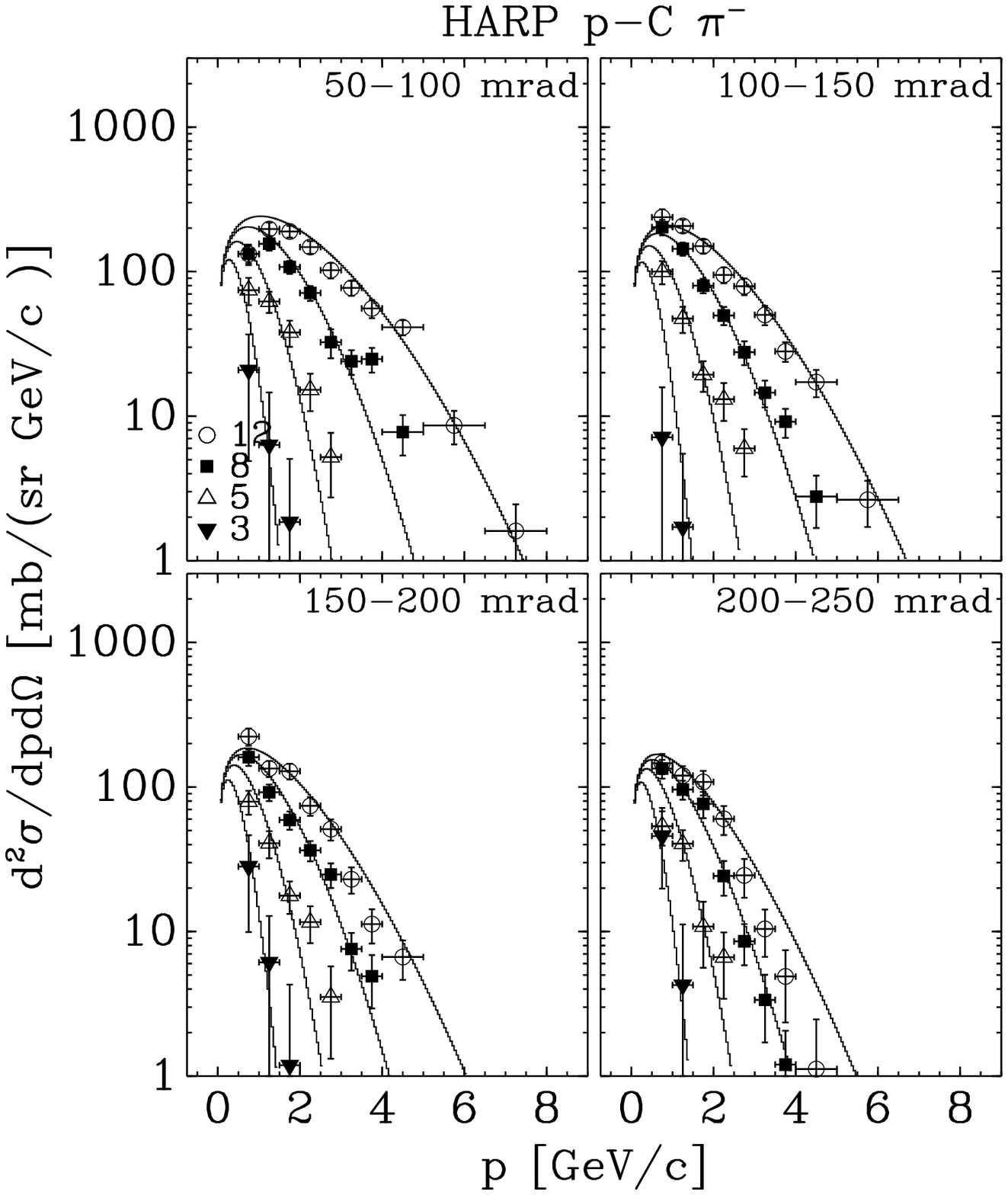}
\includegraphics[width=.49\textwidth]{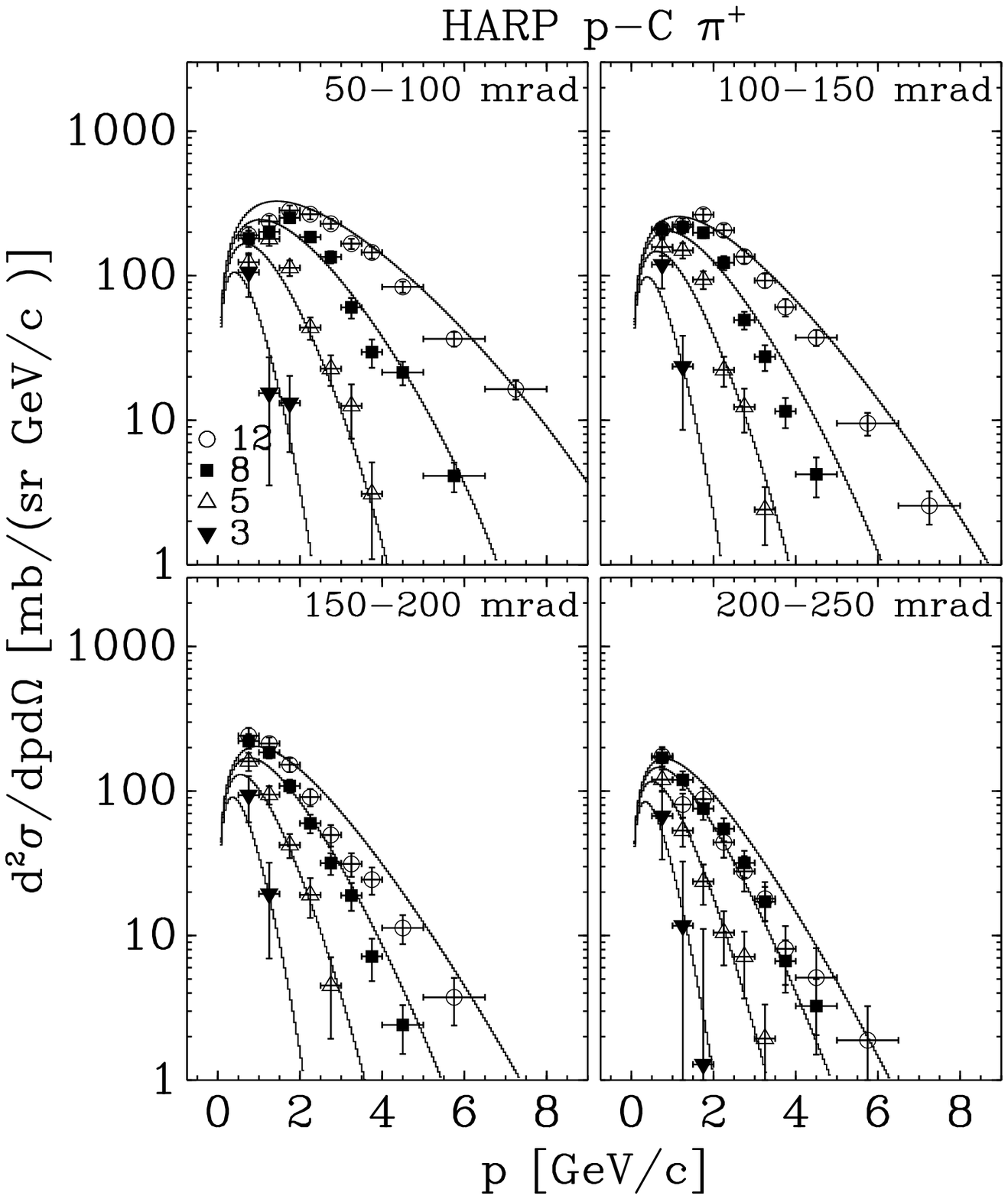}
\caption{p--C differential cross sections: left panel
$\pi^{-}$ production, right panel $\pi^{+}$ production.   
The curves represent the global parametrization as described in the text.
In the top right corner of each plot the 
covered angular range is shown in mrad.}
\label{fig:C}
\end{figure*}

\begin{figure*}[tb]
\centering
\includegraphics[width=.49\textwidth]{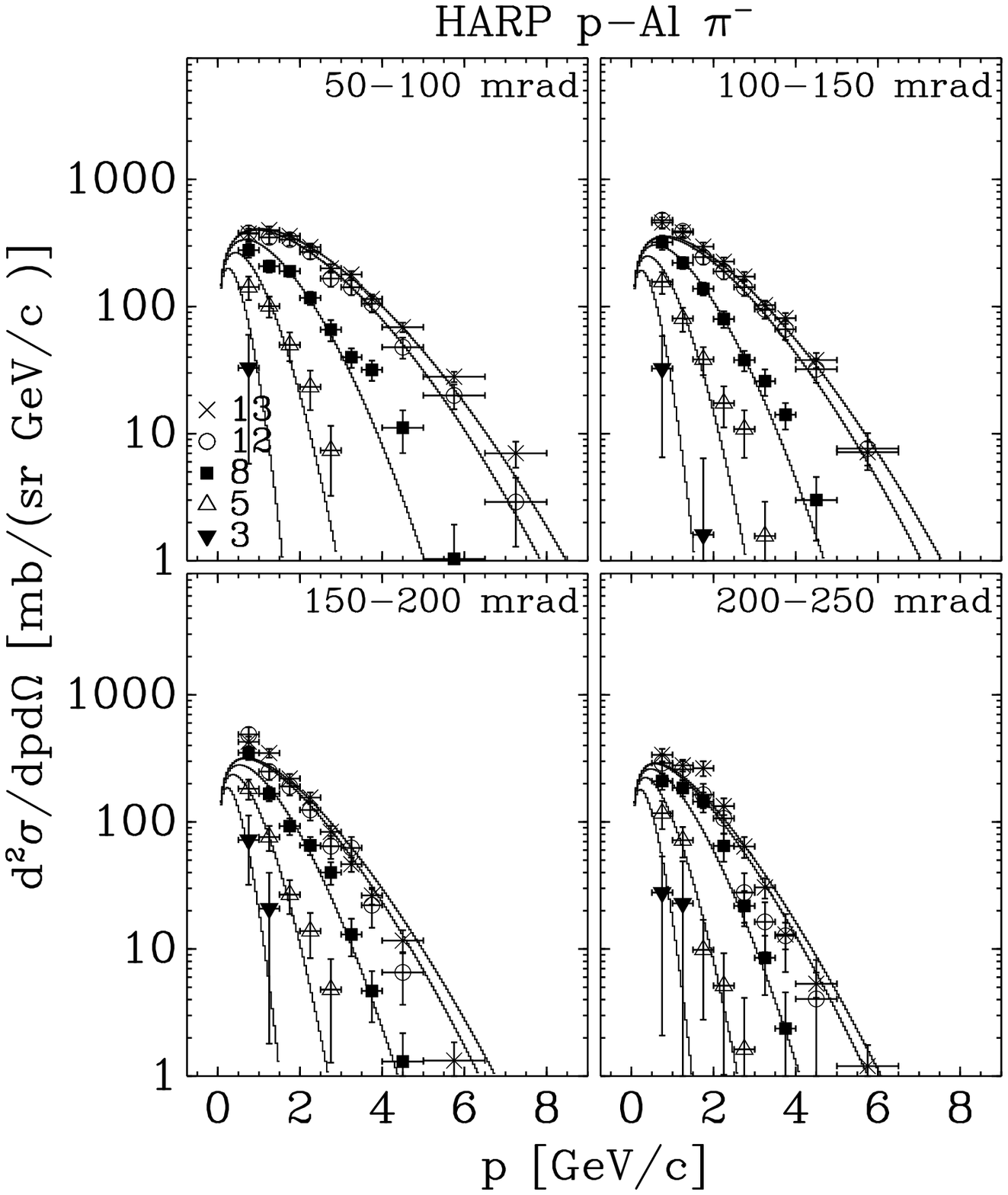}
\includegraphics[width=.49\textwidth]{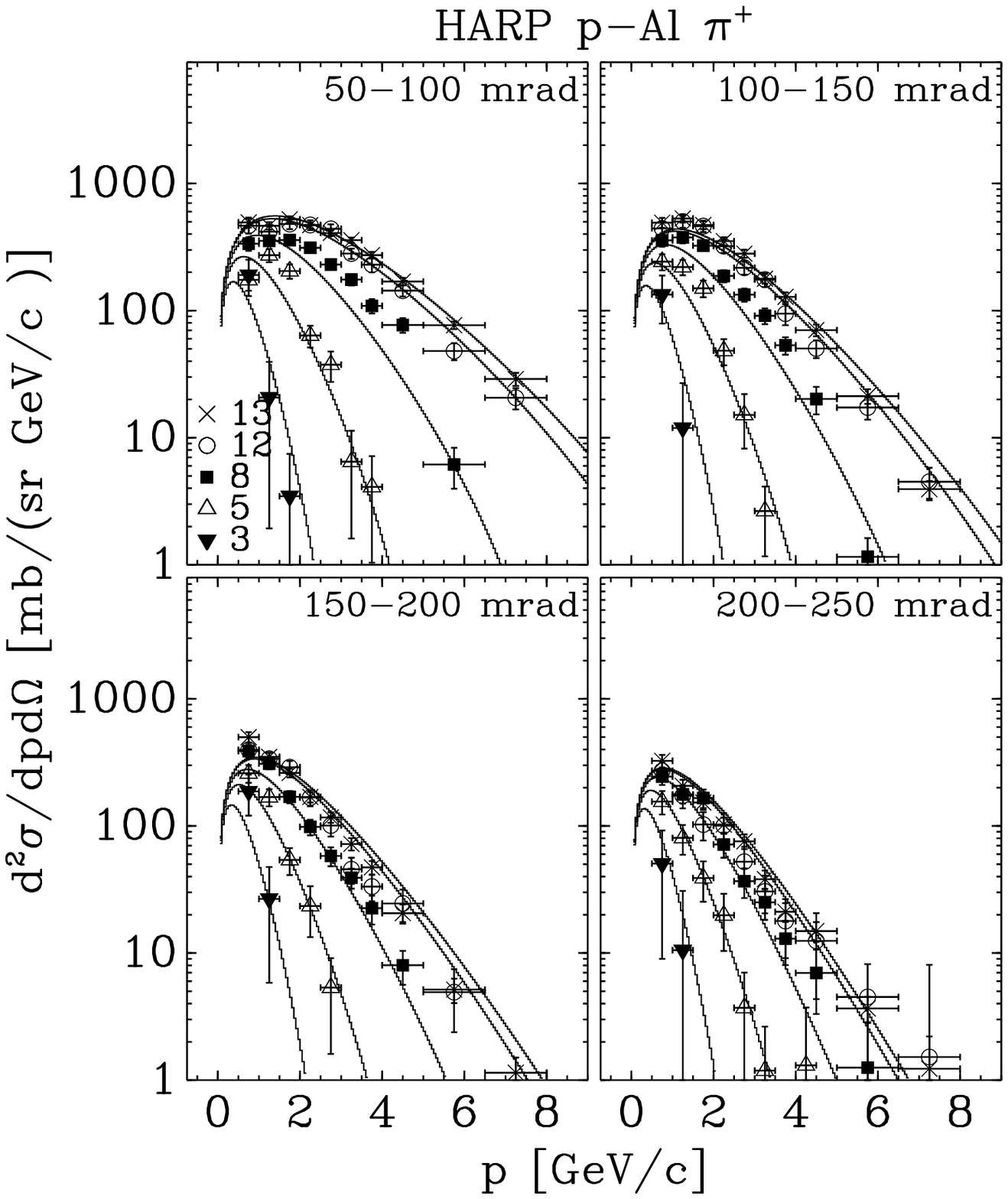}
\caption{p--Al differential cross sections: left panel
$\pi^{-}$ production, right panel $\pi^{+}$ production.   
The curves represent the global parametrization as described in the text.
In the top right corner of each plot the 
covered angular range is shown in mrad.}
\label{fig:Al}
\end{figure*}

\begin{figure*}[tb]
\centering
\includegraphics[width=.49\textwidth]{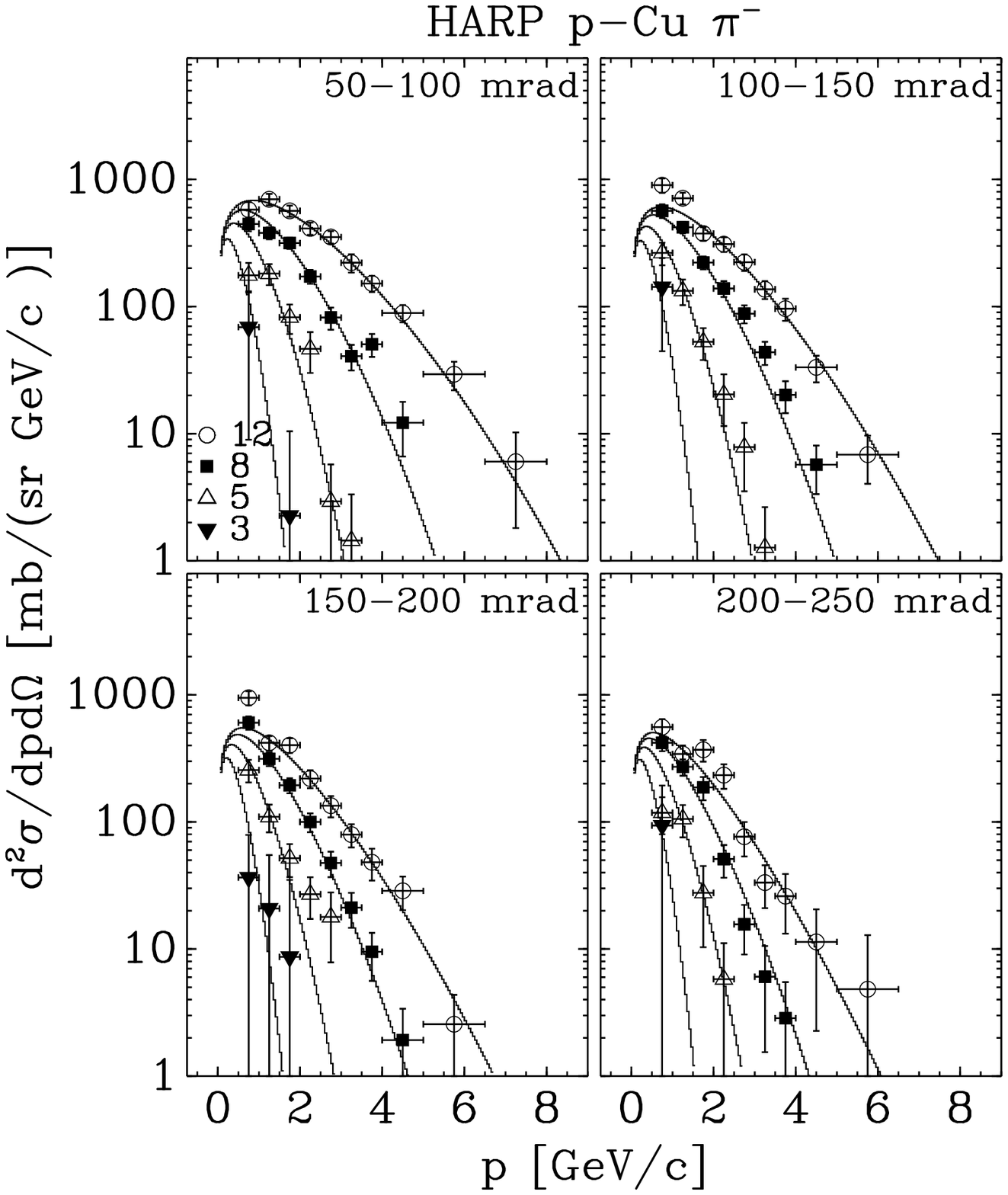}
\includegraphics[width=.49\textwidth]{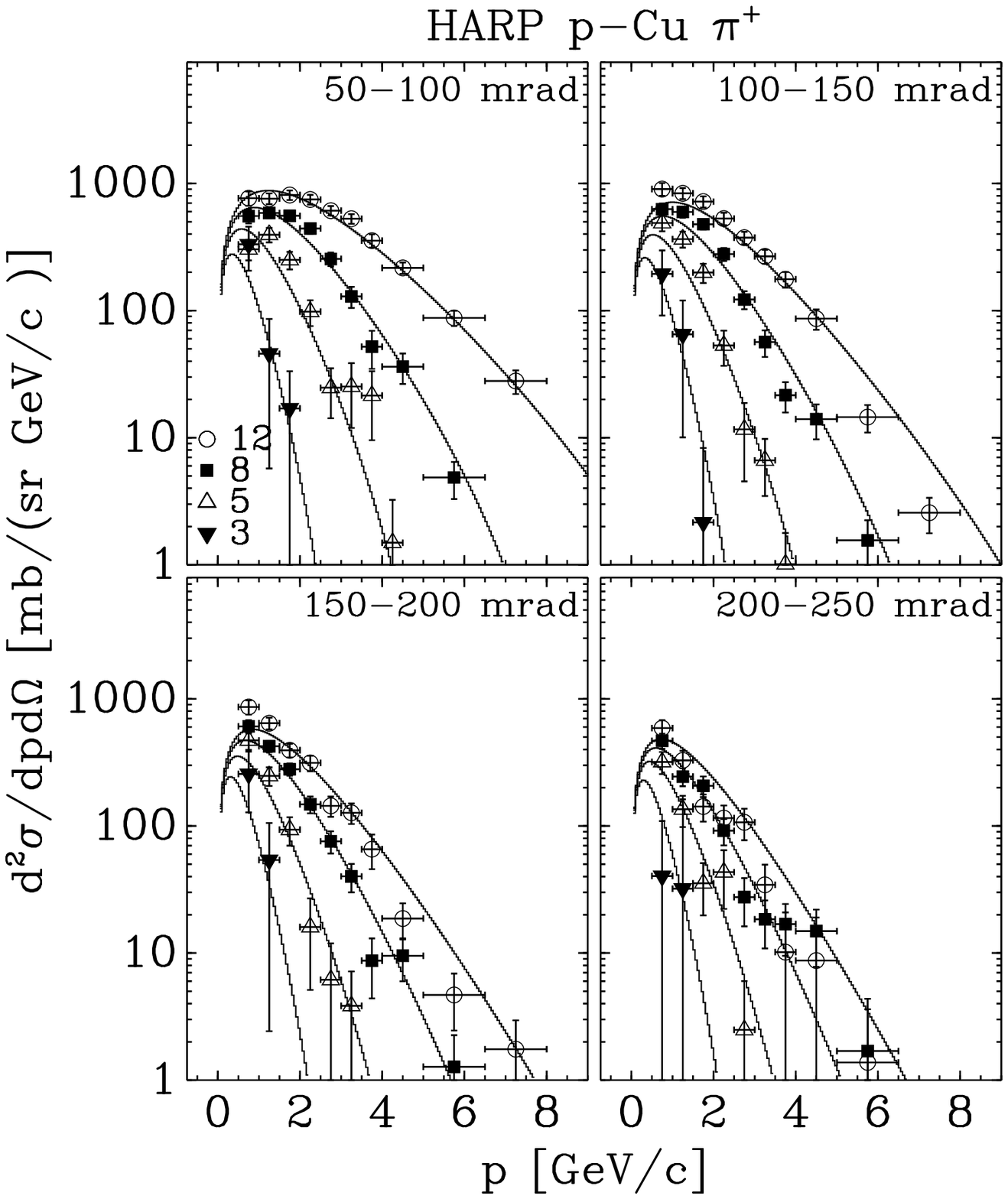}
\caption{p--Cu differential cross sections: left panel
$\pi^{-}$ production, right panel $\pi^{+}$ production.   
The curves represent the global parametrization as described in the text.
In the top right corner of each plot the 
covered angular range is shown in mrad.}
\label{fig:Cu}
\end{figure*}

\begin{figure*}[tb]
\centering
\includegraphics[width=.49\textwidth]{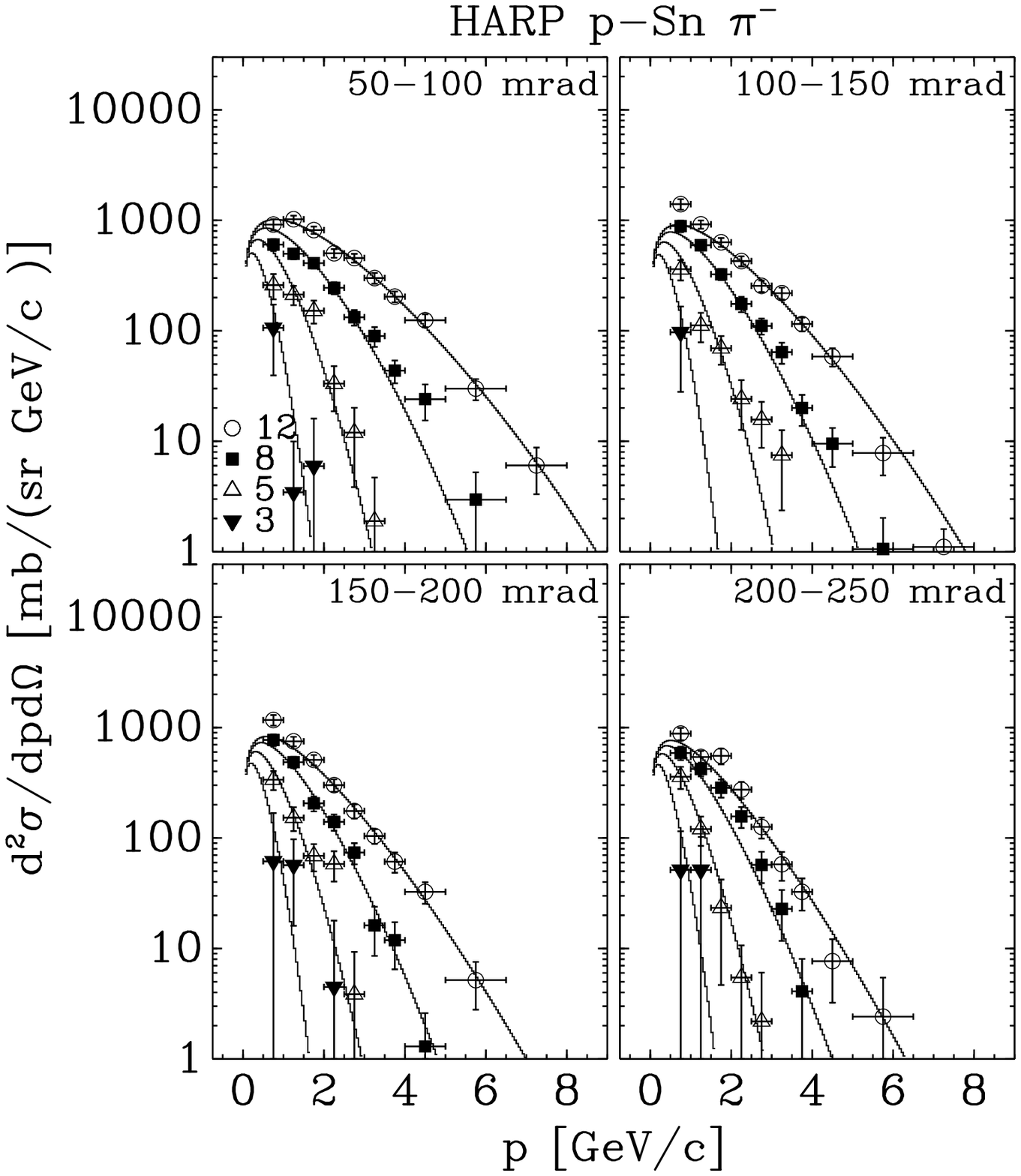}
\includegraphics[width=.49\textwidth]{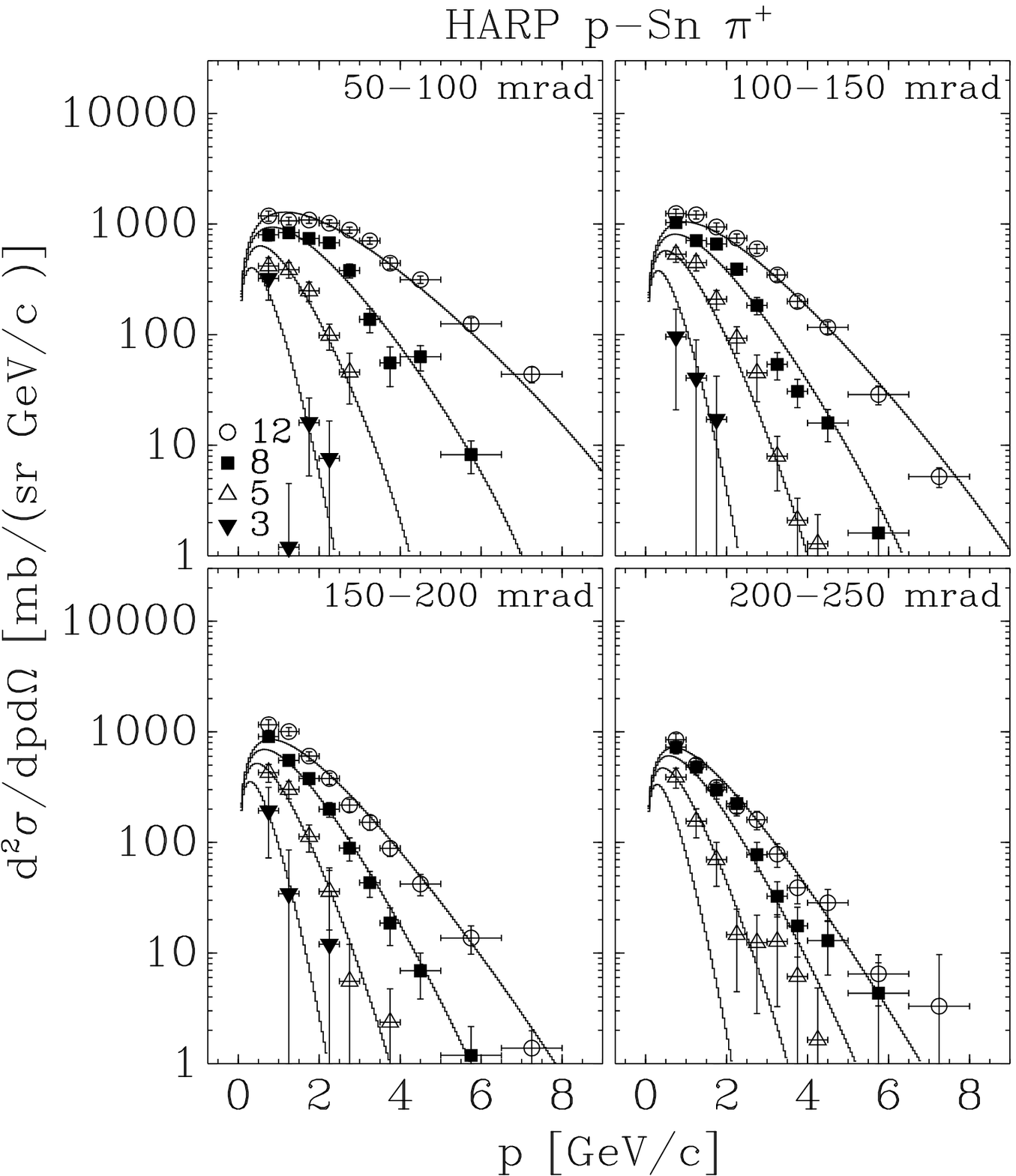}
\caption{p--Sn differential cross sections: left panel
$\pi^{-}$ production, right panel $\pi^{+}$ production.   
The curves represent the global parametrization as described in the text.
In the top right corner of each plot the 
covered angular range is shown in mrad.}
\label{fig:Sn}
\end{figure*}

\begin{figure*}[tb]
\centering
\includegraphics[width=.49\textwidth]{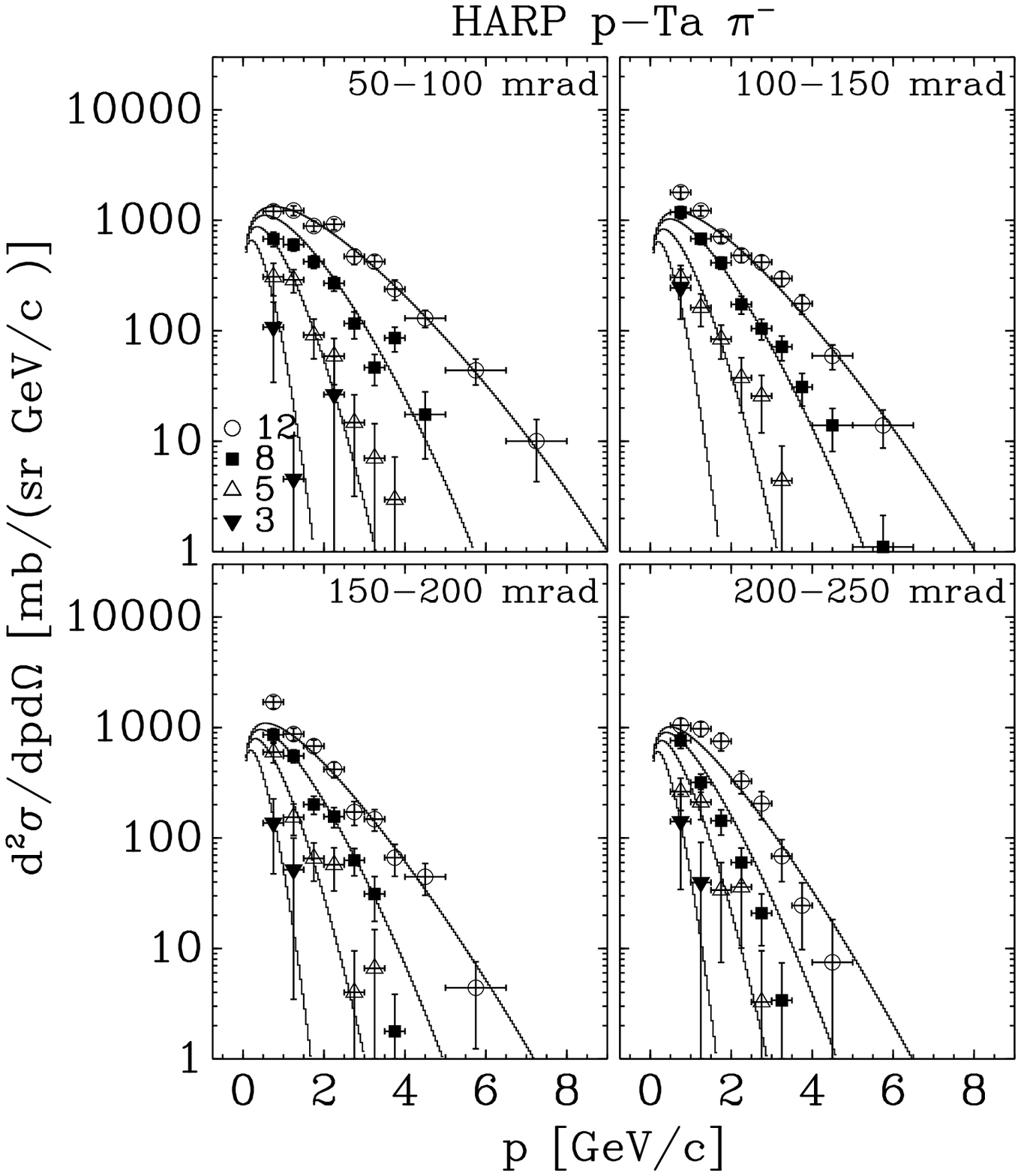}
\includegraphics[width=.49\textwidth]{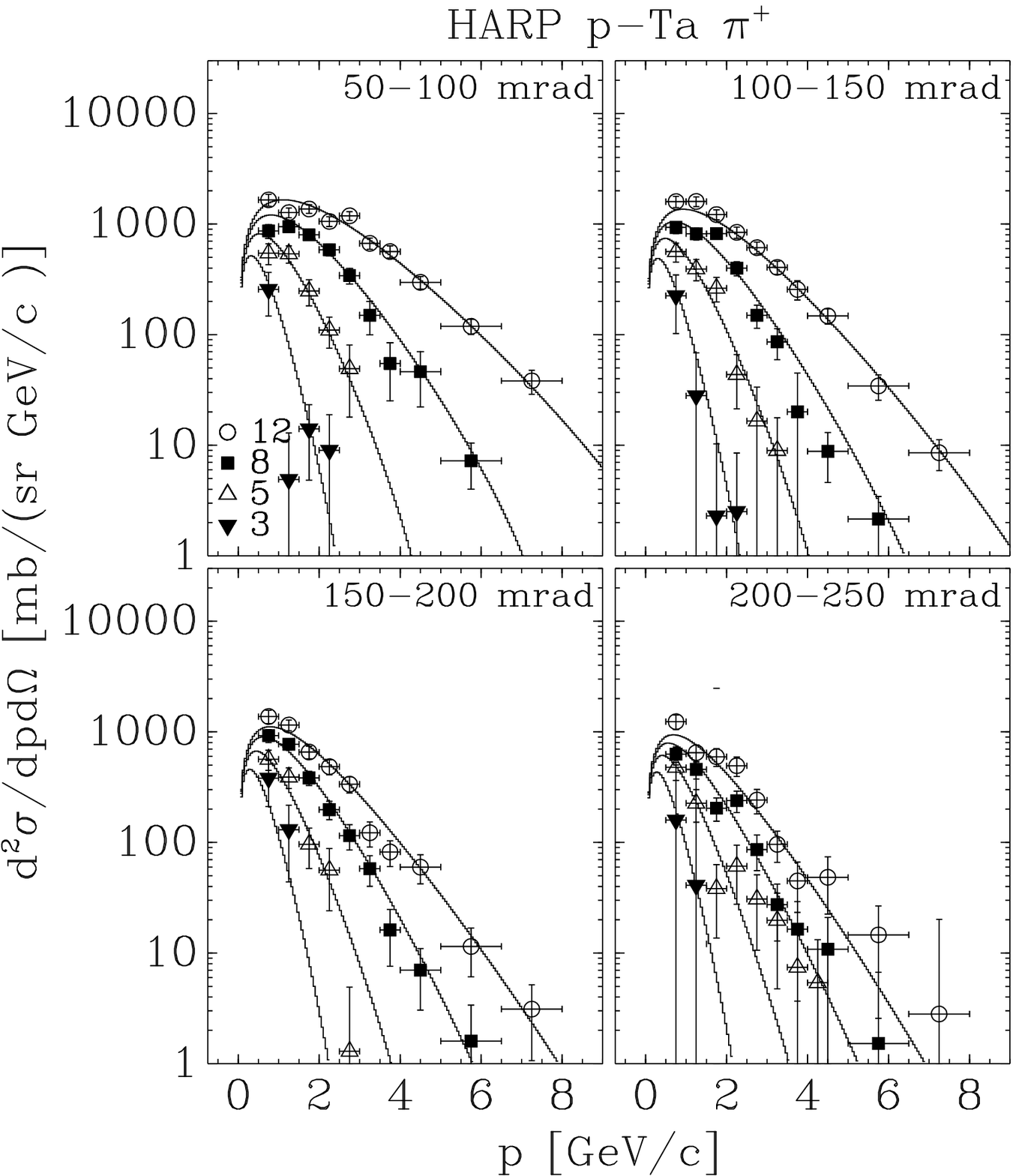}
\caption{p--Ta differential cross sections: left panel
$\pi^{-}$ production, right panel $\pi^{+}$ production.   
The curves represent the global parametrization as described in the text.
In the top right corner of each plot the 
covered angular range is shown in mrad.}
\label{fig:Ta}
\end{figure*}

\begin{figure*}[tb]
\centering
\includegraphics[width=.49\textwidth]{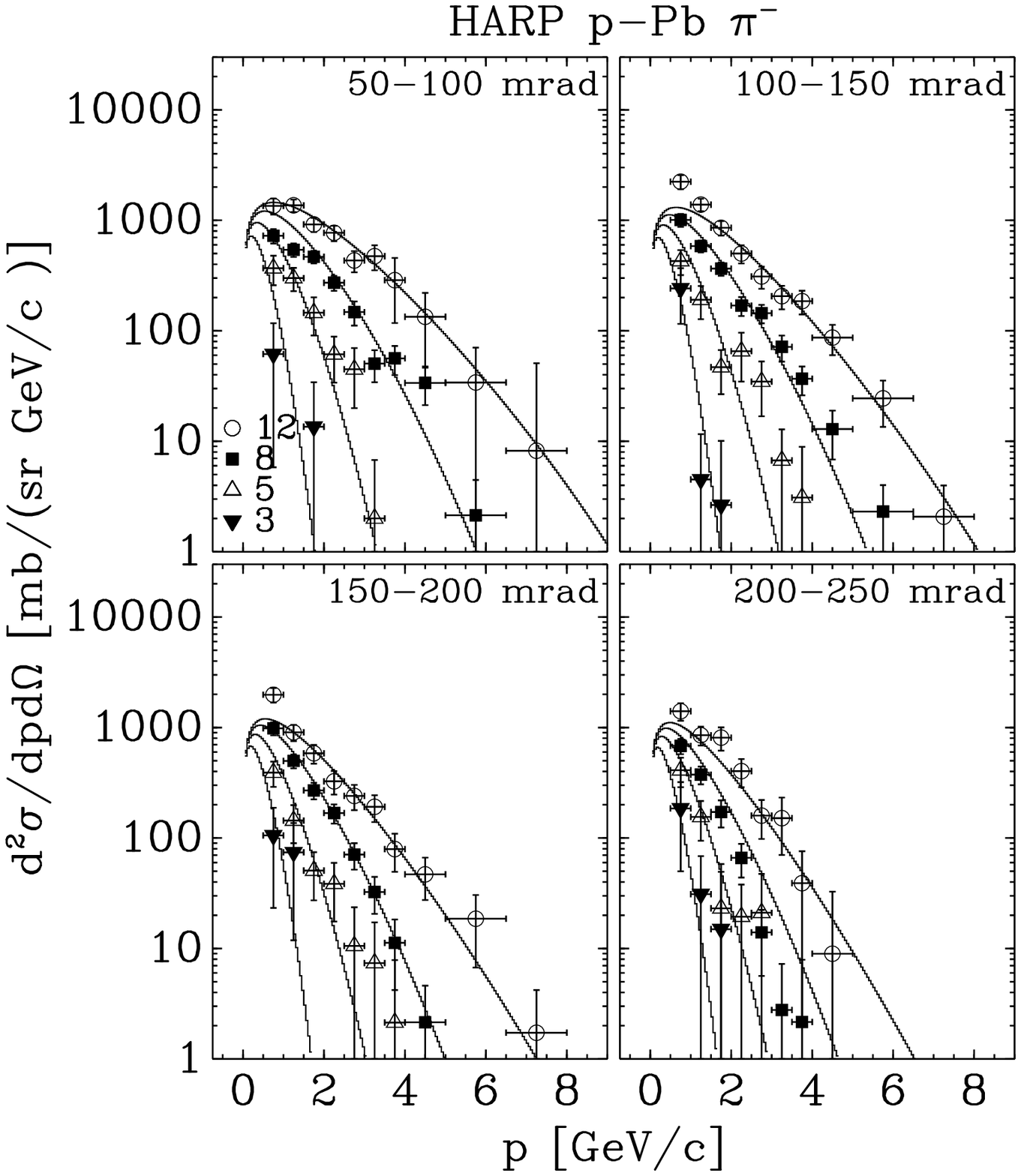}
\includegraphics[width=.49\textwidth]{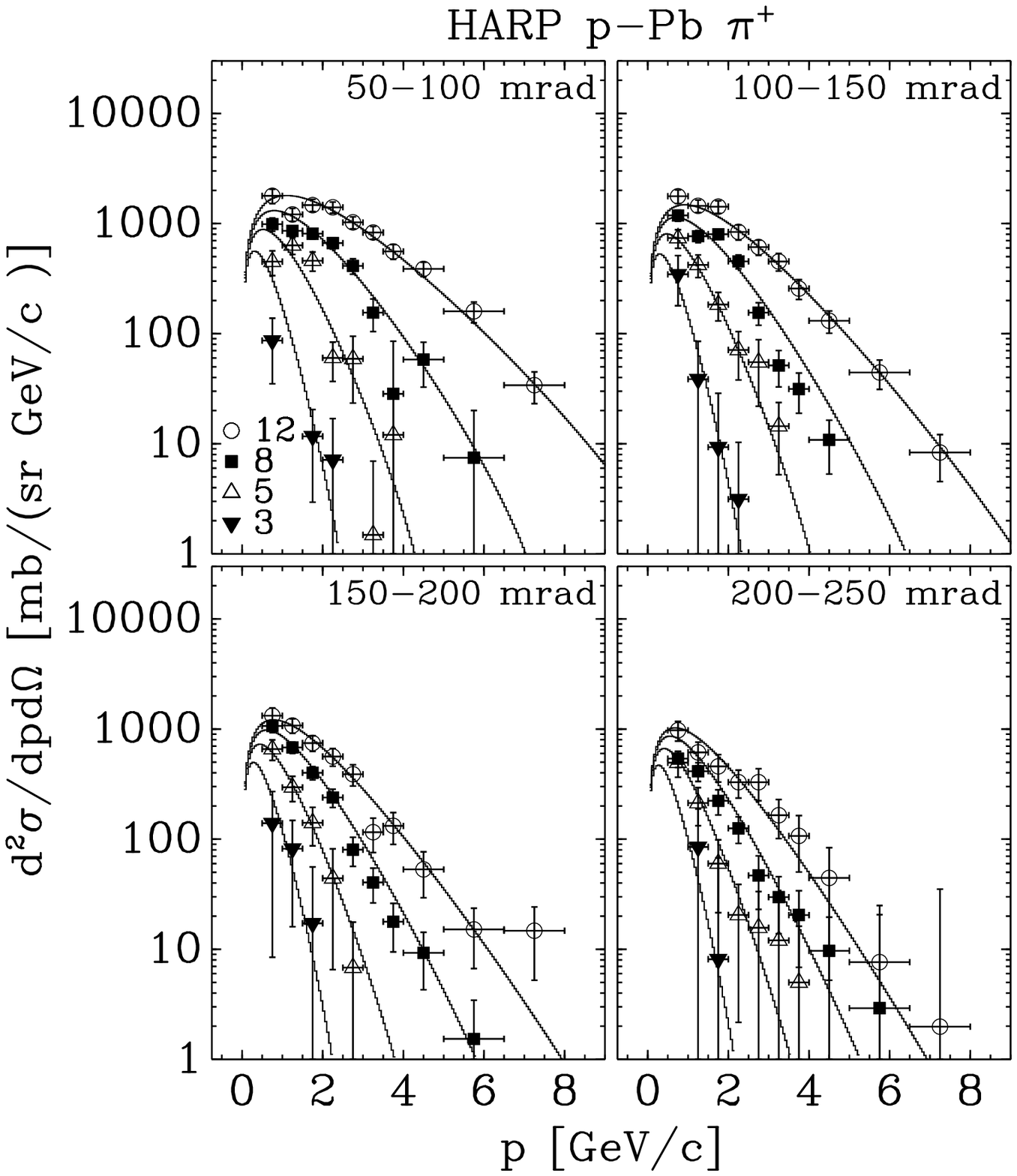}
\caption{p--Pb differential cross sections: left panel
$\pi^{-}$ production, right panel $\pi^{+}$ production.   
The curves represent the global parametrization as described in the text.
In the top right corner of each plot the 
covered angular range is shown in mrad.}
\label{fig:Pb}
\end{figure*}

The measured double-differential cross-sections for the
production of \pip and \pim in the laboratory system as a function of
the momentum and the polar angle for each incident beam momentum are
shown in Figures \ref{fig:Be} to \ref{fig:Pb} for solid
targets from Be to Pb.
The error bars  shown are the
square-roots of the diagonal elements in the covariance matrix,
where statistical and systematic uncertainties are combined
in quadrature.
The correlation of the statistical errors (introduced by the unfolding
procedure) are typically smaller than 20\% for adjacent momentum bins and
even smaller for adjacent angular bins.
The correlations of the systematic errors are larger, typically 80\% for
adjacent bins.
The overall normalization  error ($<2\%$) is not included in the error bars
\footnote{ This makes it possible to calculate e.g.
integrated particle ratios taking these normalization errors 
into account only when applicable, 
i.e. when different beams are compared}.
The results of this analysis are also fully tabulated and reported in
appendix \ref{app:data}. 

An overall fit with a  Sanford-Wang parametrization \cite{ref:SW} 
has been done by using all
solid targets and available beam momenta data ($p \geq 5 \ $ GeV/c), 
see Section~\ref{sec:parametrization} for details, and is shown
as a solid line on all figures.


\begin{figure*}[tbp]
\begin{center}
  \includegraphics[width=0.49\textwidth]{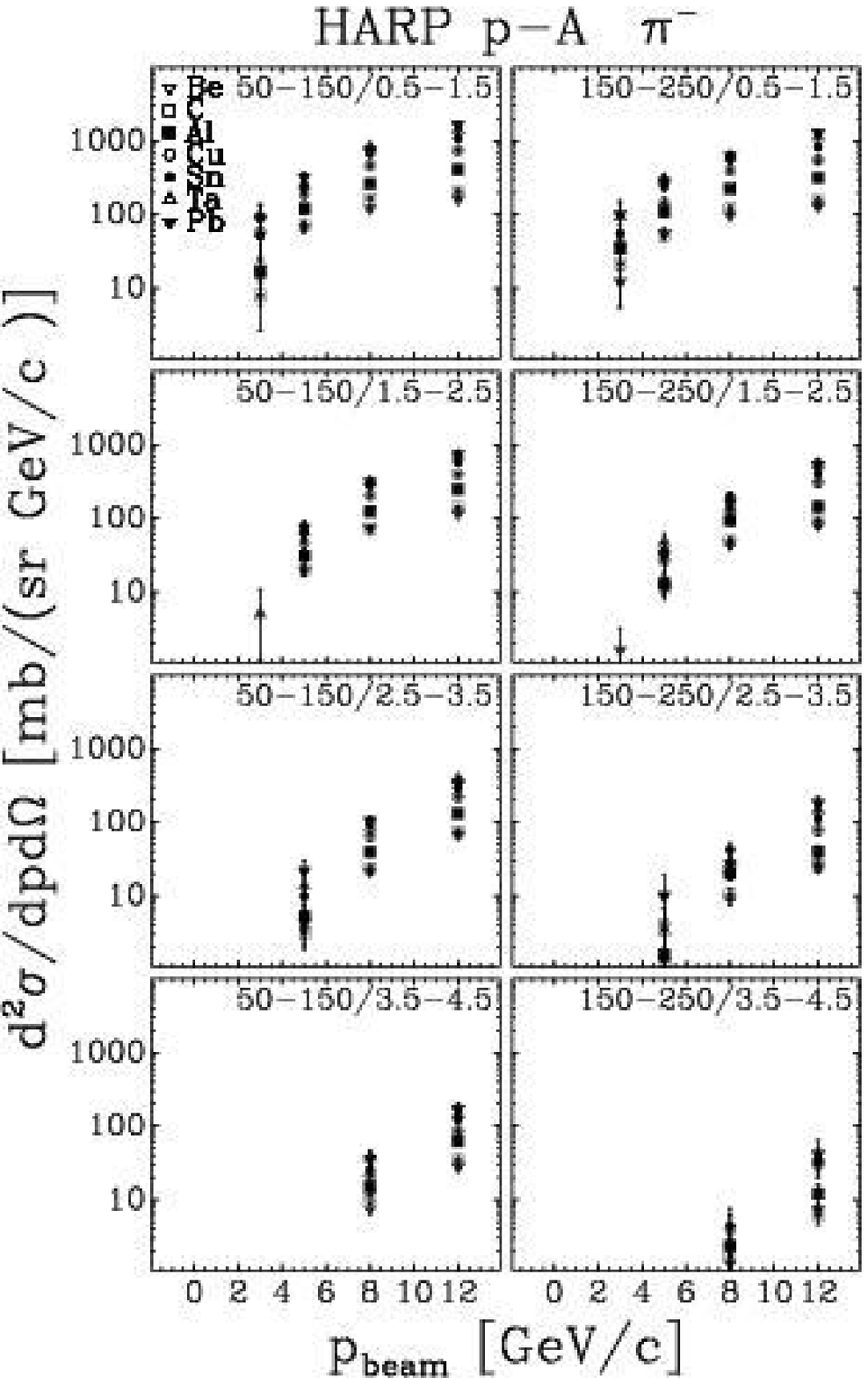}
  \includegraphics[width=0.49\textwidth]{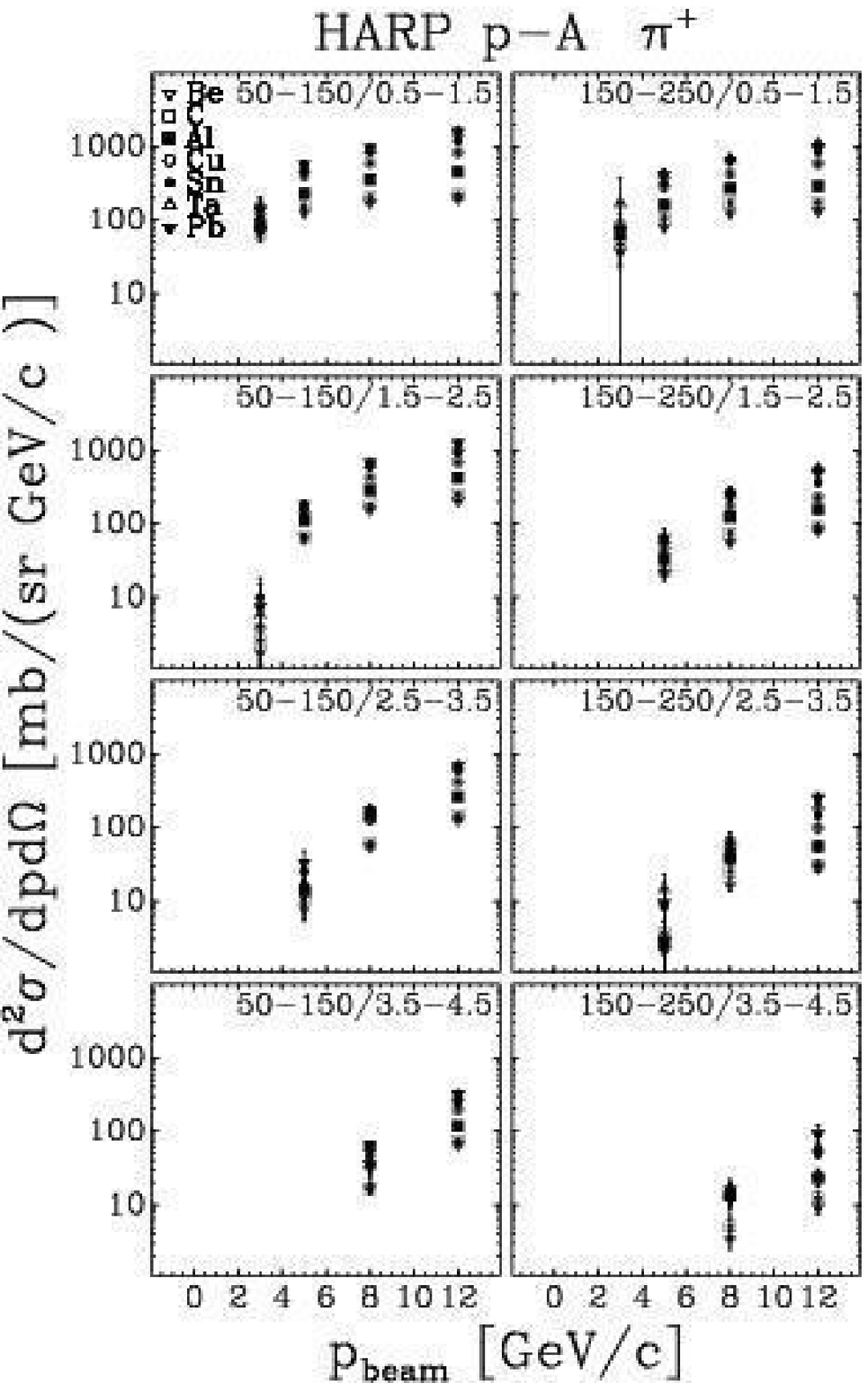}
\end{center}
\caption{
 The dependence on the beam momentum of the \pim (top panel) and \pip (bottom panel)
  production yields
 in p--Be, p--C, p--Al, p--Cu, p--Sn, p--Ta, p--Pb
 interactions averaged over two forward angular regions
 ($0.05~\rad \leq \theta < 0.15~\rad$ and
  $0.15~\rad \leq \theta < 0.25~\rad$)
 and four momentum regions
  ($0.5~\GeVc \leq p < 1.5~\GeV/c$,
   $1.5~\GeVc \leq p < 2.5~\GeV/c$,
   $2.5~\GeVc \leq p < 3.5~\GeV/c$ and
   $3.5~\GeVc \leq p < 4.5~\GeV/c$), for the four different
  incoming beam energies.
Data points for different target nuclei and equal momenta are slightly
 shifted horizontally with respect to each other to increase the visibility.
}
\label{fig:xs-trend}
\end{figure*}

The dependence of the averaged pion yields on the incident beam
momentum is shown in Fig.~\ref{fig:xs-trend}.
The \pim yields, averaged over two angular regions
 ($0.05~\rad \leq \theta < 0.15~\rad$ and 
  $0.15~\rad \leq \theta < 0.25~\rad$)
 and four momentum regions 
  ($0.5~\GeVc \leq p < 1.5~\GeV/c$,
   $1.5~\GeVc \leq p < 2.5~\GeV/c$,
   $2.5~\GeVc \leq p < 3.5~\GeV/c$ and
   $3.5~\GeVc \leq p < 4.5~\GeV/c$),
are
shown in the left panel and the \pip data averaged over the same regions
in the right panel, for four different beam momenta.
Whereas the beam energy dependence of the yields in the
 p--Be, p--C data differs clearly from the dependence in the p--Ta,
p--Pb data one can
observe that the p--Al, p--Cu and p--Sn data display a
smooth transition between them.
The dependence in the p--Be, p--C data is much more flat with a saturation of
the yield between 8~\GeVc and 12~\GeVc
with the p--Al, p--Cu and p--Sn showing an intermediate behaviour.
%% Also the \pip and \pim production yields exhibit a different behaviour.

\begin{figure*}[tbp]
\begin{center}
  \includegraphics[width=0.49\textwidth]{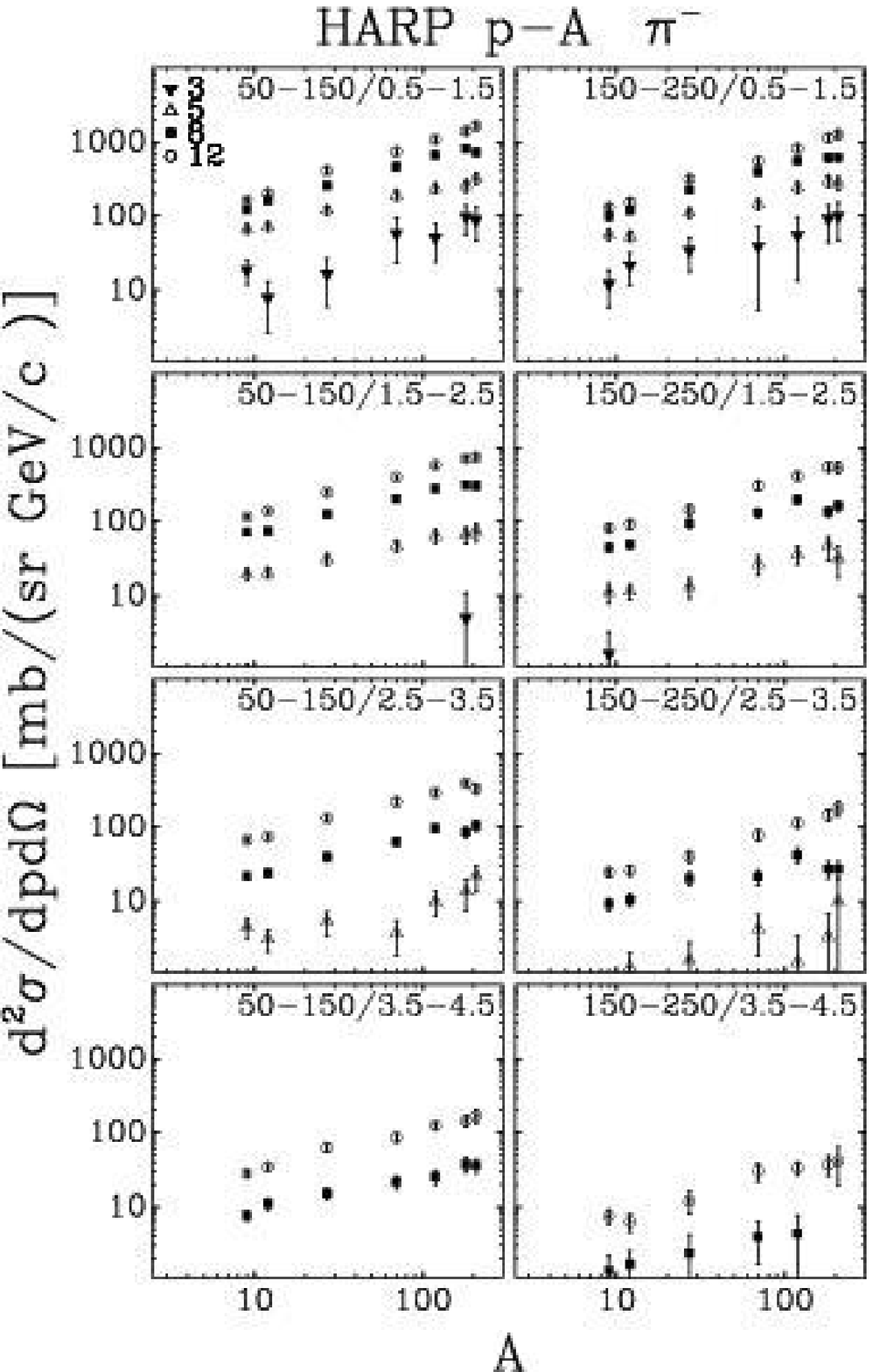}
  \includegraphics[width=0.49\textwidth]{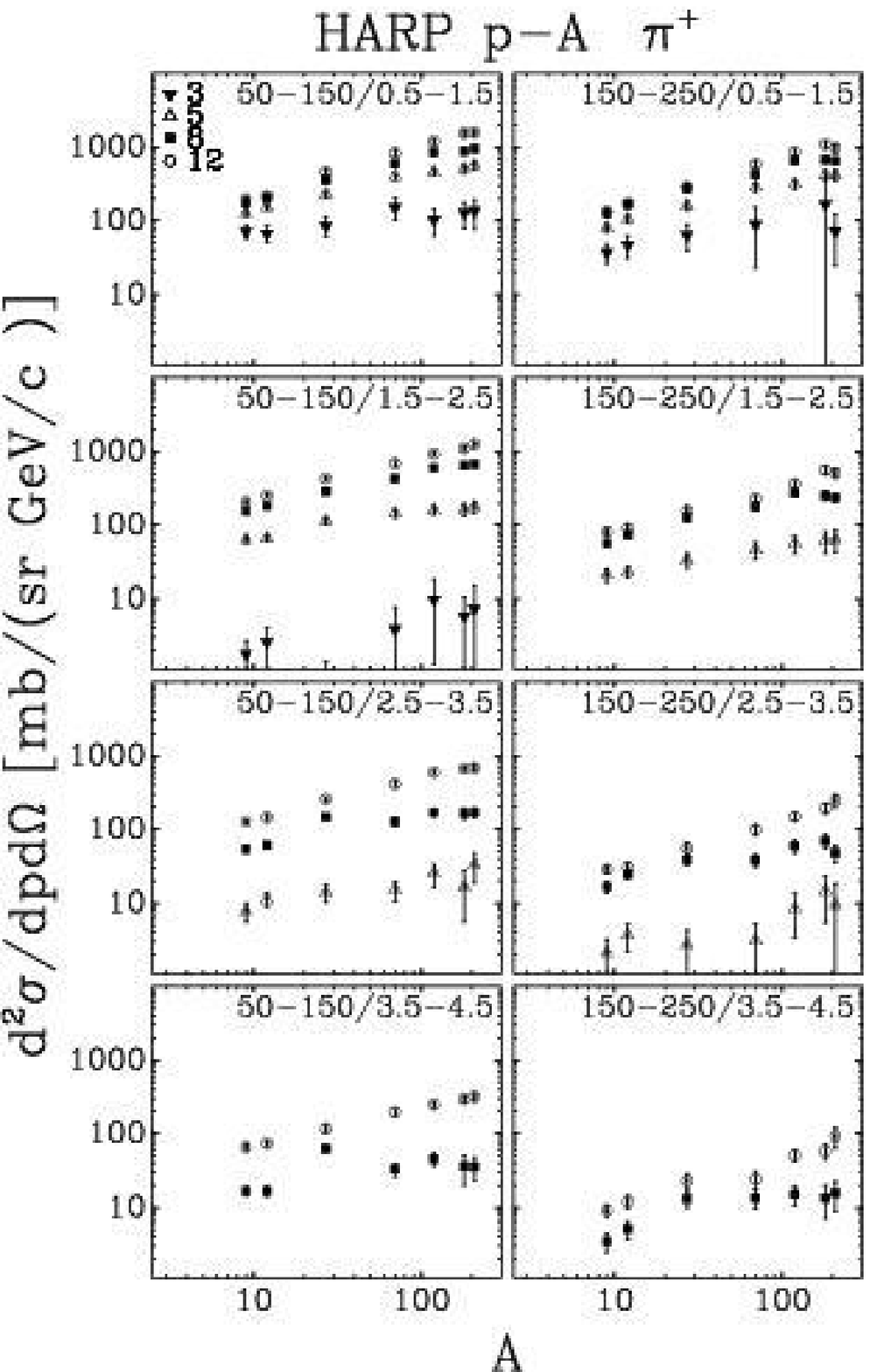}
\end{center}
\caption{
 The dependence on the atomic number $A$ of the pion production yields
 in $p$--Be, $p$--Al, $p$--C, $p$--Cu, $p$--Sn, $p$--Ta, 
  $p$--Pb
 interactions averaged over two forward angular regions
 ($0.05~\rad \leq \theta < 0.15~\rad$ and 
  $0.15~\rad \leq \theta < 0.25~\rad$)
 and four momentum regions 
  ($0.5~\GeVc \leq p < 1.5~\GeV/c$,
   $1.5~\GeVc \leq p < 2.5~\GeV/c$,
   $2.5~\GeVc \leq p < 3.5~\GeV/c$ and
   $3.5~\GeVc \leq p < 4.5~\GeV/c$), for the four different
  incoming beam energies.
 %The results are given in arbitrary units, with a consistent scale
 %between the left and right panel.
}
\label{fig:xs-a-dep}
\end{figure*}

The dependence of the averaged 
pion yields on the atomic number $A$ is instead 
shown in Fig.~\ref{fig:xs-a-dep}.
The \pim yields, averaged over two angular regions
 ($0.05~\rad \leq \theta < 0.15~\rad$ and 
  $0.15~\rad \leq \theta < 0.25~\rad$)
 and four momentum regions 
  ($0.5~\GeVc \leq p < 1.5~\GeV/c$,
   $1.5~\GeVc \leq p < 2.5~\GeV/c$,
   $2.5~\GeVc \leq p < 3.5~\GeV/c$ and
   $3.5~\GeVc \leq p < 4.5~\GeV/c$),
are
shown in the left panel and the \pip data averaged over the same regions
in the right panel, for four different beam momenta.
One observes a smooth behaviour of the averaged yields.
The $A$-dependence is slightly different for \pim and \pip production,
the latter saturating earlier towards higher $A$, especially at lower
beam momenta.

\FloatBarrier

\subsection{Pion production data parametrization}
\label{sec:parametrization}

At low energies, it is common to use the empirical data parametrization 
for pion production in proton-nucleus interactions 
originally developed by Sanford and Wang \cite{ref:SW}.
This parametrization has the functional form: 
%--
\vspace{0.3cm}
\begin{equation}
\label{eq:swformula}
\frac{d^2\sigma (\hbox{pA}\rightarrow \pi^{\pm}X)}{dpd\Omega}=
 c_{1} \exp[B]p^{c_{2}}(1-\frac{p}{p_{\hbox{\footnotesize beam}}}) \ ,
\end{equation} 
%--
where:
%--
\vspace{0.3cm}
\begin{equation}
\label{eq:swformula2}
B=-c_{3}\frac{p^{c_{4}}}{p_{\hbox{\footnotesize beam}}^{c_{5}}}-c_{6}
 \theta (p-c_{7} p_{\hbox{\footnotesize {beam}}} \cos^{c_{8}}\theta ) 
\end{equation} 
%--
$X$ denotes any system of other particles in the final state, 
$p_{\hbox{\footnotesize {beam}}}$ is the proton beam momentum in \GeVc, 
$p$ and $\theta$ are the $\pi^{\pm}$ momentum and angle in units of \GeVc 
and radians, respectively, $d^2\sigma/(dpd\Omega)$ is expressed in 
units of mb/(\GeVc)/sr and 
the parameters $c_1,\ldots ,c_8$ are obtained from fits to the meson 
production data. 

The parameter $c_1$ is an overall normalization factor, 
the four parameters $c_2,c_3,c_4,c_5$ describe the momentum distribution of 
the secondary pions in the forward direction, and the three parameters 
$c_6,c_7,c_8$ describe the corrections to the pion momentum distribution for pion production angles that are different from zero.
The $\pi^{\pm}$ production data on solid targets reported here have been 
fitted simultaneously to the empirical Sanford-Wang formula. 
In the $\chi^2$ minimization, the full error matrix was used. 

To go from a baseline nuclear targets (typically Be) to another nuclear target
$A$, a correction factor 
\begin{equation}
corr = (A/A_{Be})^{\alpha}
\end{equation}
was introduced, with $ \alpha = \alpha_{0} + \alpha_{1} \times x_F +
\alpha_{2} \times x_F^{2}$ where $x_F$ is the Feynman x, see references
\cite{ref:physrep} and \cite{Barton} for details.
A 9-parameter (11-parameter) fit was  done over 24 (25) 
$\pi^{-} (\pi^{+})$ datasets~\footnote{In the fit
procedure only data with $p_{beam} \geq 5$ GeV/c were used, as the 
inclusion of the 3 GeV/c data gave problems in the convergence. The 
$c_{7}$ parameter was
fixed to zero in the $\pi^{-}$ fit, as the results were found insensitive 
to this parameter in an extended range around zero}, 
 corresponding to 1440 (1472) experimental points. 
The goodness-of-fit of the Sanford-Wang parametrization hypothesis for 
the HARP results can be assessed by considering the best-fit $\chi^2$ 
value of $\chi^2_{\hbox{\footnotesize {{min}}}}=13030 \ (8061)$ for 
1431 (1461)  degrees of freedom for the $\pi^{-}$ ($\pi^{+}$) production,
indicating a very poor fit quality. 
In particular, inspection of the HARP inclusive pion production 
double-differential cross-section, and resulting Sanford-Wang parametrization,
points to a description of the ratio $g(\theta)$ of the pion momentum 
distribution at $\theta\neq 0$ with respect to the $\theta = 0$ pion 
momentum distribution that is more complicated than what can be 
accommodated within the Sanford-Wang formula, where this ratio is 
given by $g(\theta)=\exp[-c_6\theta (p-p_c)]$, 
with $p_c\equiv c_{7} p_{\hbox{\footnotesize {beam}}} \cos^{c_{8}}\theta$.

The overall fit may be used as a fast approximation of HARP data valid 
within a factor 2--3 of the quoted experimental errors. The best-fit values of 
the  parameters 
are reported in Table~\ref{tab:swpar_1} together with their errors. 
The fit parameter errors are estimated
by requiring $\Delta\chi^2\equiv \chi^2-\chi^2_{\hbox{\footnotesize {{min}}}} 
= 12.6 \ (10.4) $, corresponding to the 68.27\% confidence level region for eleven (nine) variable parameters. 
Significant correlations among fit parameters are found, as shown by the correlation matrix  given in Tables~\ref{tab:swpar_3} and \ref{tab:swpar_4}.

\begin{table}%[tb]
\caption{\label{tab:swpar_1}
Sanford-Wang parameters and errors obtained by fitting the
$\pi^{+}$  ($\pi^{-}$) datasets. 
The errors refer to the 68.27\% confidence
level for eleven parameters ($\Delta\chi^2=12.6$) for 
$\pi^{+}$ and nine parameters ($\Delta\chi^2=10.4$) for $\pi^{-}$.}
\vspace{5ex}
\begin{tabular}{ c c c} \hline
{\bf Parameter} & $\pi^{+}$      &  $\pi^{-}$ \\ \hline
$c_1$      & $(381.3\pm 40.5)  $ &  $(307.6 \pm 19.4)$ \\
$c_2$      & $(0.88\pm 0.07)$    &  $(0.57 \pm 0.06) $ \\
$c_3$      & $(9.16\pm 0.95)$    &  $(27.36 \pm 1.06) $ \\
$c_4$      & $(1.38\pm 0.09)$    &  $(1.86 \pm 0.04) $ \\
$c_5$      & $(1.66\pm 0.12)$    &  $(2.23 \pm 0.04) $ \\
$c_6$      & $(3.62\pm 0.14)$    &  $(3.04 \pm 0.08) $ \\
$c_7$      & $(0.05\pm 0.04)$    &   -  \\
$c_8$      & $(128.6\pm 61.8)$   &   -  \\
$\alpha_0$ & $(0.69 \pm 0.04)$   &  $(0.72 \pm 0.04)$ \\
$\alpha_1$ & $(-0.91 \pm 0.21)$  &  $(-1.36 \pm 0.20) $ \\
$\alpha_2$ & $(0.34 \pm 0.21)$   & $(2.18 \pm 0.21) $ 
%\hline
\end{tabular}
\end{table}

\begin{table*}%[tb]
\caption{\label{tab:swpar_3}
Correlation coefficients among the Sanford-Wang fit parameters, 
obtained by fitting the data for $\pi^{+}$ production.
 }
\vspace{5ex}
\centerline{
%{\small
%%%\input{swpar_corr_piplus.tex}
%}
%\begin{tabular}{|c|r r r r r r r|}
\begin{tabular}{ c r r r r r r r r r r r}
\hline
{\bf Parameter} & $c_1$  & $c_2$  & $c_3$  & $c_4$  & $c_5$  & $c_6$  & $c_7$ & $c_8$ & $\alpha_0$ & $\alpha_1$ & $\alpha_2$ \\
\hline
$c_1$     &  1.000 &         &        &        &        &        &        &       &   &  & \\
$c_2$     &  0.388 &  1.000  &        &        &        &        &        &       &   &  &  \\
$c_3$     &  0.089 &  -0.349 &  1.000 &        &        &        &        &       &   &  &   \\
$c_4$     & -0.725 & -0.485  &  0.103 &  1.000 &        &        &        &       &   &  &   \\
$c_5$     & -0.580 & -0.651  &  0.585 &  0.840 &  1.000 &        &        &       &   &  &   \\
$c_6$     &  0.309 &  0.358 &   0.069 & -0.262 & -0.106 &  1.000 &        &       &   &  &   \\
$c_7$     & -0.612 & -0.453 &  -0.237 &  0.299 &  0.137 & -0.532 &  1.000 &       &   &  &   \\
$c_8$     & -0.017 &  0.187 &   0.027 &  -0.080&  -0.017 & 0.513 &  -0.319 & 1.000 &   &  &   \\
$\alpha_0$& -0.528 &  0.131 &  -0.030 &  0.364 &  0.289 & -0.008 &  0.013 & -0.025 &  1.000  &  &   \\
$\alpha_1$&  0.451 & -0.095 &  -0.061 & -0.529 & -0.429 &  0.065 & -0.017 & 0.035 & -0.827  & 1.000  &   \\
$\alpha_2$& -0.377 &  0.059 &   0.228 &  0.557 &  0.498 & -0.069  &  0.012 & -0.027 &  0.672  & -0.945       &  1.000    
\end{tabular}
}
\end{table*}

\begin{table*}%[tb]
\caption{\label{tab:swpar_4}
Correlation coefficients among the Sanford-Wang fit parameters, 
obtained  by fitting the data for $\pi^{-}$ production.
}
\vspace{5ex}
\centerline{
%{\small
%%%\input{swpar_corr_piminus.tex}
%\begin{tabular}{|c|r r r r r r r|}
\begin{tabular}{ c r r r r r r r r   r}
\hline
{\bf Parameter} & $c_1$  & $c_2$  & $c_3$  & $c_4$  & $c_5$  & $c_6$   & $\alpha_0$ & $\alpha_1$ & $\alpha_2$ \\
\hline
$c_1$     &  1.000 &        &        &        &        &        &        &       &   \\
$c_2$     & -0.151 &  1.000 &        &        &        &        &        &       &   \\
$c_3$     &  0.040 &  -0.517&  1.000 &        &        &        &        &       &   \\
$c_4$     & -0.544 & -0.263 &  0.047 &  1.000 &        &        &        &       &   \\
$c_5$     & -0.441 & -0.567 &  0.625 &  0.794 &  1.000 &        &        &       &   \\
$c_6$     &  0.134 &  0.181 &  0.365 & -0.269 &  0.027 &  1.000 &        &       &      \\
$\alpha_0$& -0.728 &  0.307 & -0.049 &  0.366 &  0.266 &  0.010 &  1.000 &       &    \\
$\alpha_1$&  0.492 & -0.325 &  0.037 & -0.493 & -0.318 &  0.069 &  -0.809& 1.000  &   \\
$\alpha_2$& -0.342 &  0.285 &  0.040 &  0.492 &  0.330 & -0.073 &  0.636 & -0.950 &  1.000    
\end{tabular}
%}
}
\end{table*}

\begin{figure*}[htb]
\centering
\includegraphics[width=0.49\textwidth]{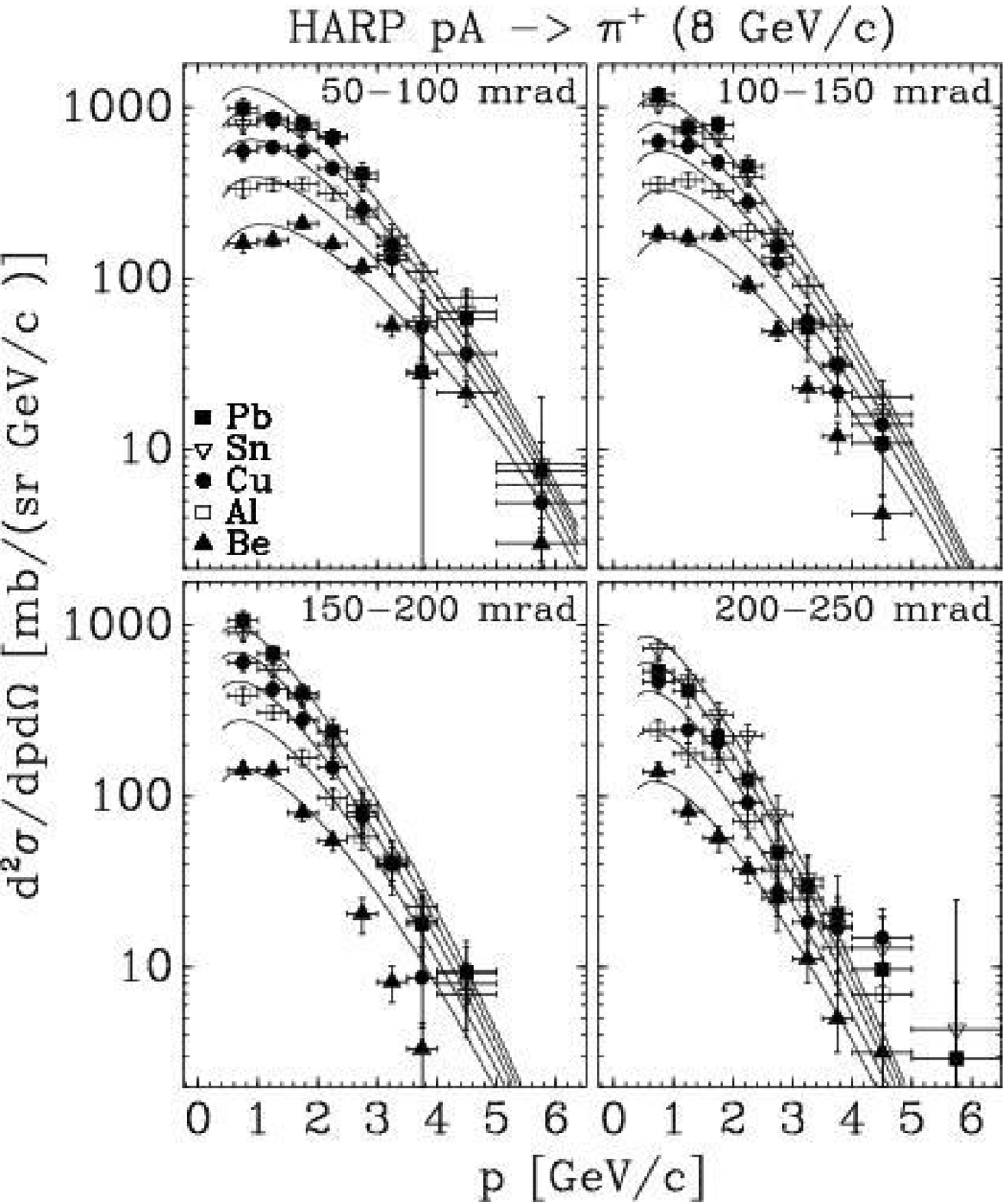}
\includegraphics[width=0.49\textwidth]{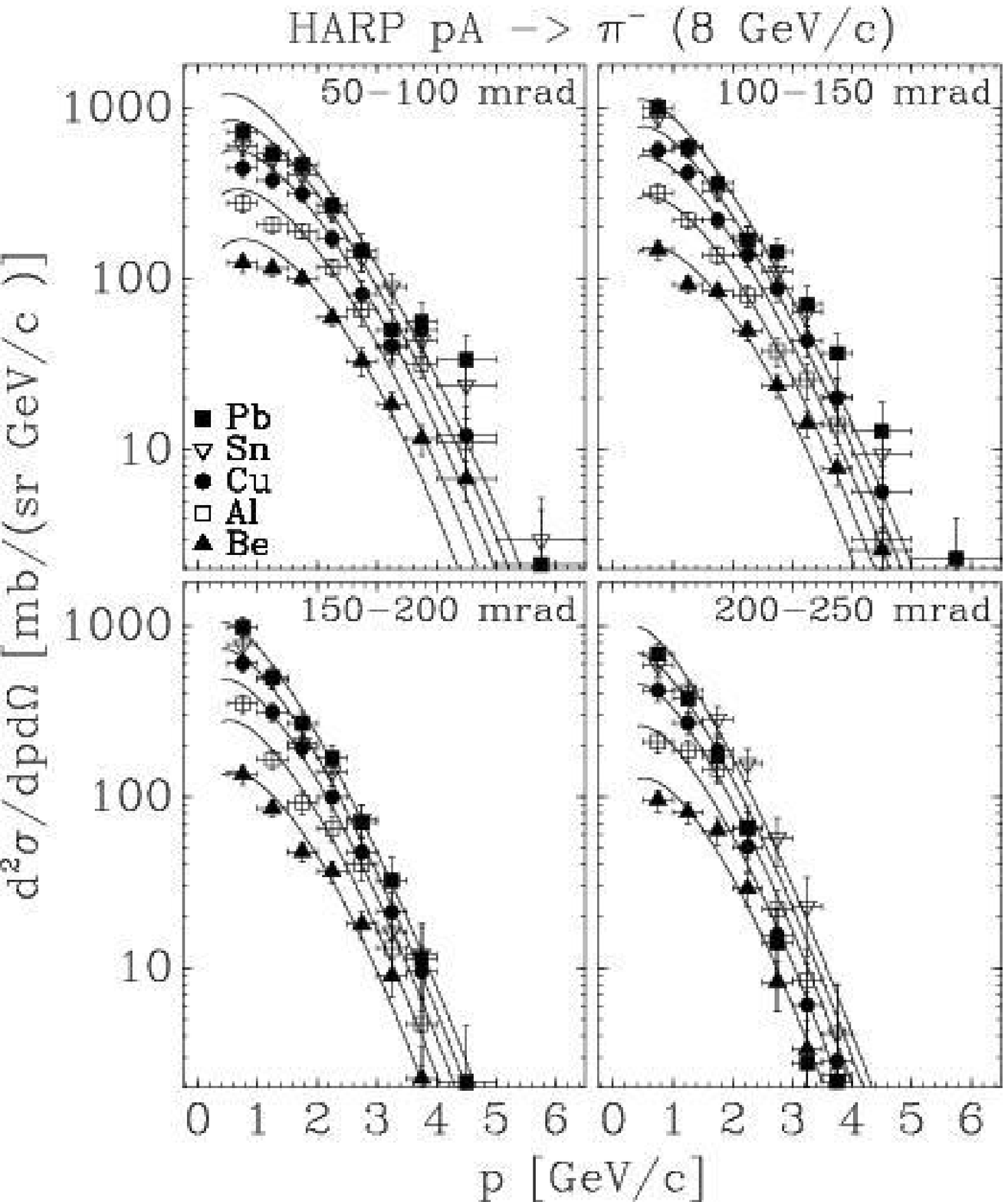}
\includegraphics[width=0.49\textwidth]{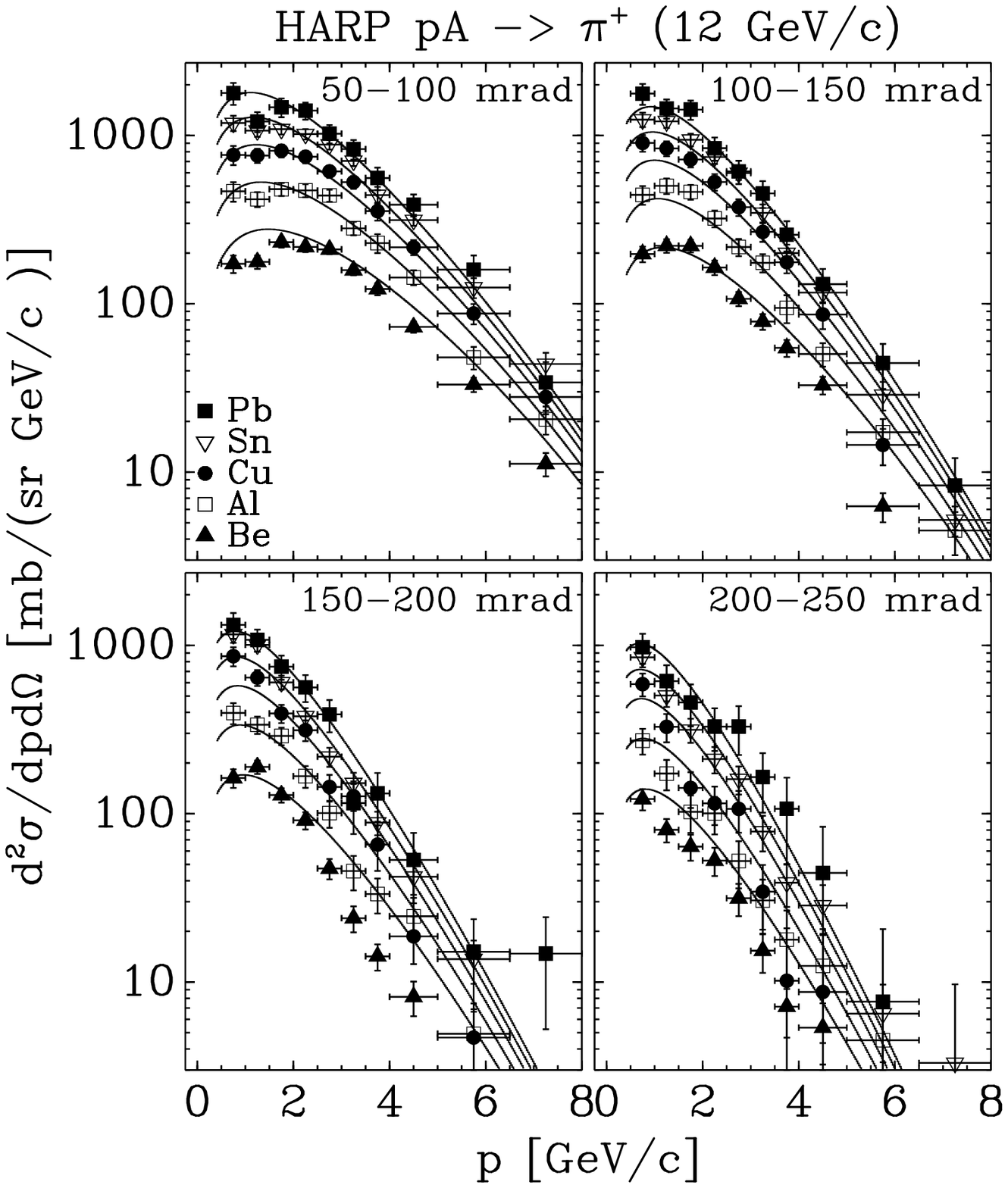}
\includegraphics[width=0.49\textwidth]{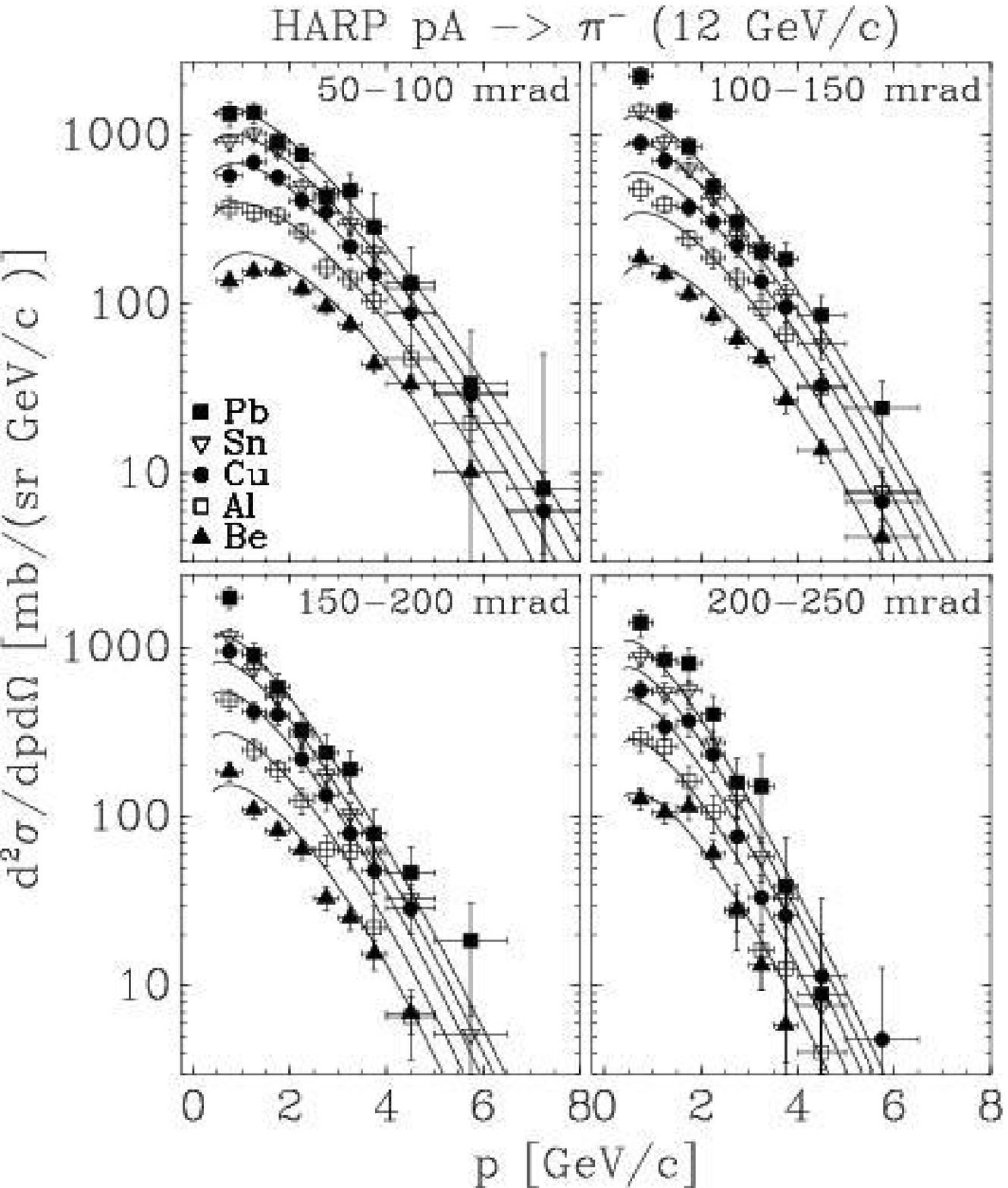}
\caption{Comparison of HARP production data at 8 GeV/c and 12 GeV/c with the Sanford-Wang
global fit.}
\label{fig:SWA}
\end{figure*}

To show the trend of the Sanford-Wang global fit of all HARP datasets,
figure \ref{fig:SWA} reports the comparison, at 8 GeV/c and 12 GeV/c, 
between pion production data and the above parametrization.

For the 8.9 GeV/c MiniBooNE/SciBooNE beamline and the 12.9 GeV/c K2K beamline
two ad-hoc Sanford-Wang parametrizations, using only the relevant HARP
datasets, have been published in references \cite{ref:alPaper} and 
\cite{ref:bePaper}.
Given the poor description of HARP pion production data in terms of the 
original Sanford-Wang parametrization, one extra parameter to better describe
the angular dependence was introduced in the fit reported in reference
\cite{ref:bePaper}.
In the global fit presented here, this extra parameter is not used for
simplicity, as it was found that it did not improve the fit quality in
a significant way.

As a final remark, we stress again that due to the poor fit quality our global 
fit may be just considered as a simple way to summarize an extended set of
data ($\sim 1000$ experimental data points) using a formula with about 
ten parameters. This
may be useful in the intial phase of an experiment design or a Monte Carlo
validation.   
\FloatBarrier
%%\subsection{Comparison with Montecarlo generators and previous experimental  data}
\subsection{Comparison with Monte Carlo generators.}

In the following we will show only some comparisons with  two
widely available Monte Carlo generators: GEANT4 version 7.1~\cite{ref:geant4}
and MARS version 15.07~\cite{ref:mars}, using different
%generators.
models.
%% and previous available data.
The comparison will be shown for a limited set of plots
and only for the Be and Ta targets, as examples of a light and a heavy target.
In both generators, no single model is applicable to all energies and a
transition between low energy models and high energy models, at about
5~\GeVc--10~\GeVc, has to be done.  

At intermediate energies (up to 5~\GeVc--10~\GeVc),
GEANT4 uses two types of intra-nuclear cascade models: the Bertini
model~\cite{ref:bert,ref:bert1} (valid up to $\sim$10~\GeVc) and the Binary
model~\cite{ref:bin} (valid up to $\sim$3~\GeVc). Both models treat the target
nucleus in detail, taking into account density variations and tracking in the
nuclear field.
The Binary model is based on hadron collisions with nucleons, giving
resonances that decay according to their quantum numbers. The Bertini
model is based on the cascade code reported in \cite{ref:bert2}
and hadron collisions are assumed to proceed according to free-space partial
cross sections corrected for nuclear field effects and final state
distributions measured for the incident particle types.

At higher energies, instead, two parton string models,
the quark-gluon string (QGS)  model~\cite{ref:bert,ref:QGSP} and the Fritiof
(FTP) model~\cite{ref:QGSP} are used, in addition to a High Energy
Parametrized model (HEP)
derived from the high energy part of the GHEISHA code used inside
GEANT3~\cite{ref:gheisha}.
The parametrized models of GEANT4 (HEP and LEP) are intended to be fast,
but conserve energy and momentum on average and not event by
event.

A realistic GEANT4 simulation is built by combining models and physics processes
into what is called a ``physics list''. In high energy calorimetry the two
most commonly used are the QGSP physics list, based on the QGS model, the pre-compound
nucleus model and some of the Low Energy Parametrized (LEP) model~\footnote{
Also this model, at low energy, has its root in the GHEISHA code inside
GEANT3.} and the LHEP physics list~\cite{ref:lhep} based on the parametrized
LEP and HEP models.

\begin{table*}
\caption{\label{tab_chi1} Computed $\chi^{2}$ between data and MonteCarlo simulations, assuming a $0 \% (50 \%)$ systematics
on simulation}
\small{
\begin{center}
\begin{tabular}{r|r|r|r|r|r|r|r|r}
& \multicolumn{2}{c|}{3 GeV} & \multicolumn{2}{c|}{5  GeV} & \multicolumn{2}{c|}{8 GeV} & \multicolumn{2}{c}{12 GeV}  \\
& \pip & \pim & \pip & \pim & \pip & \pim & \pip & \pim \\
\hline
ndof & 16 &  16  & 32 & 32 &  36 &  36 & 40 & 40  \\
\hline
model &\multicolumn{8}{c}{\bf Beryllium}\\
\hline
Bertini &   725.0 (39.6)  &   602.7 (35.6)  &  2334.5 (74.0) &  1722.0 (79.9)   &  3372.7 (171.0) &  2461.4 (68.6) &  3501.3 (331.9) &  2390.8 (90.3) \\
Binary  &   690.1 (39.7)  &  1011.1 (41.6)  &  1020.8 (65.6) &  3708.8 (91.8)   &  1411.6 (74.0)  &  3481.4 (77.3)  & \\
LHEP    &   428.4 (36.8)  &   103.7 (14.9)   &   265.4(34.8)  &  1109.7 (71.8)  &   539.5 (29.1)  &   558.2 (31.4)  &   893.2 (34.6) &  1084.4 (33.4) \\
QGSP    &                 &                  &                &                 &   226.6 (35.0)  &   740.3 (49.4)  &   809.7 (31.3) &   972.8 (33.3) \\
FTFP    &                 &                  &                &                 &                 &  1805.7 (65.7)  &   196.2 (16.6) &  1113.6 (36.3)  \\
MARS    &    12.3 (5.1)    &    91.7 (17.2)  &    33.9 (10.1) &    17.0 (9.1)   &    65.4 (9.0)   &   122.3 (16.0)  &    62.6 (7.5)  &    70.0 (7.4)  \\
\hline
model &\multicolumn{8}{c}{\bf Tantalum}\\
\hline
Bertini  &   493.1 (30.6) &   380.6 (25.9) &   517.5 (41.3) &  2144.2 (80.2) &  1186.1 (84.5) &   735.6 (45.1) &  1455.8 (162.4) &   709.7 (68.4) \\
Binary   &   901.3 (35.2) &   784.9 (30.3) &   743.4 (46.3) &  4075.1 (82.4) &  1093.1 (53.0) &  2535.0 (53.8) & \\
LHEP  &   600.7 (28.7)    &    57.5 (7.7)  &   201.0 (28.8) &  1022.6 (61.7) &   371.4 (22.3) &   516.3 (21.7)  &   434.7 (14.8) &   884.9 (25.4) \\
QGSP  &                   &                &                &                &   439.9 (43.2) &   678.1 (36.9)  &   213.3 (12.5) &   636.0 (18.5)\\
FTFP  &                   &                &                &                &                &   365.7 (23.6)  &   135.5 (31.2) &   259.9 (15.7)\\
MARS  &   145.8 (14.3)    &    70.5  (9.1) &   143.2 (32.1) &    94.4 (29.5) &   121.1 (18.1) &    53.3 (6.7)   &   101.2 (7.6)   &    54.0 (5.8)  \\
\end{tabular} 
\end{center}
}
\end{table*}

\begin{table*}
\caption{ \label{tab_norm} Normalization factors data-simulation.} 
\small{
\begin{tabular}{r|rr|rr|rr|rr|rr|rr|rr|rr}
\hline
model        & \multicolumn{2}{c|}{Be 3 GeV} & \multicolumn{2}{c|}{Ta 3 GeV} & \multicolumn{2}{c|}{Be 5  GeV} & \multicolumn{2}{c|}{Ta 5 GeV} 
 & \multicolumn{2}{c|}{Be 8 GeV} & \multicolumn{2}{c|}{Ta 8 GeV} & \multicolumn{2}{c|}{Be 12 GeV} & \multicolumn{2}{c|}{Ta 12 GeV} \\
& \pip & \pim & \pip & \pim & \pip & \pim & \pip & \pim 
& \pip & \pim & \pip & \pim & \pip & \pim & \pip & \pim \\
\hline
Bertini&    0.35 &    1.02 &    0.45 &    0.53 &    0.70 &    1.12 &    0.29 &    0.35 &    1.22 &    1.54 &    0.84 &    1.08 &    1.75 &    1.81 &    1.27 &    1.50 \\
Binary  &    0.36 &    0.75 &    0.28 &    0.34 &    0.73 &    0.88 &    0.16 &    0.23 &    0.99 &    1.05 &    0.50 &    0.56 & & \\
LHEP  &    0.40 &    0.86 &    0.81 &    0.91 &    0.76 &    0.98 &    0.36 &    0.45 &    0.78 &    0.91 &    0.58 &    0.66 &    0.75 &    0.82 &    0.54 &    0.59 \\
QGSP &         &         &         &         &         &         &         &         &    1.40 &    1.43 &    0.80 &    0.75 &    0.80 &    0.88 &    0.64 &    0.67 \\
FTFP &         &         &         &         &         &         &         &         &         &         &    0.46 &    0.65 &    1.00 &    1.10 &    0.63 &    0.77 \\
MARS &    0.83 &    1.29 &    1.10 &    1.16 &    1.21 &    1.38 &    1.17 &    1.35 &    1.10 &    1.21 &    0.90 &    0.85 &    1.02 &    1.02 &    0.92 &    0.82 \\
\end{tabular}
}
\end{table*}

\begin{figure*}[htbp]
%\vskip 1cm
\begin{center}
  \includegraphics[width=.49\textwidth]{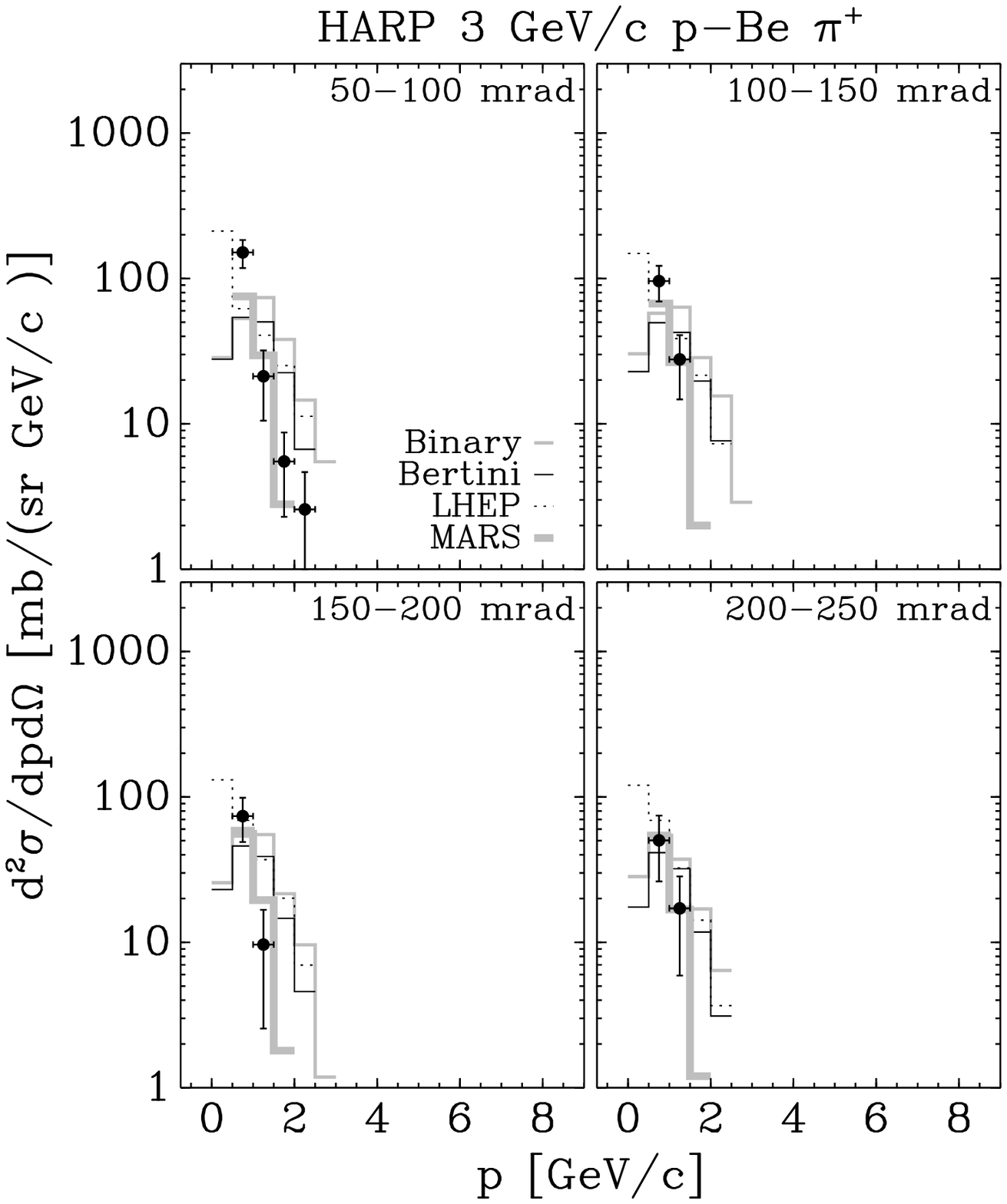}
  \includegraphics[width=.49\textwidth]{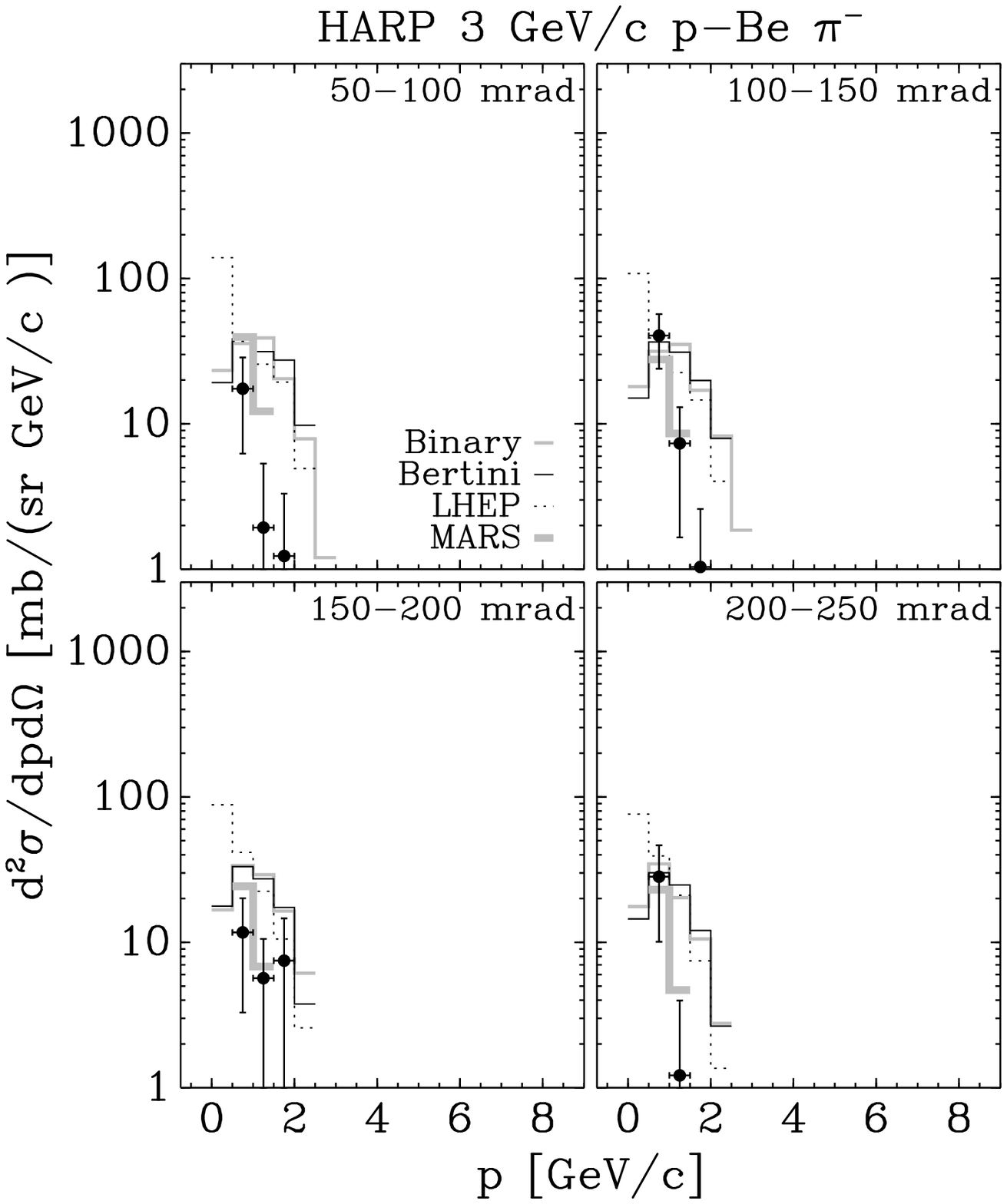}
\end{center}
%\vskip 1cm
\caption{
 Comparison of HARP double-differential $\pi^{\pm}$ cross sections for p--Be at 3 GeV/c with
 GEANT4 and MARS MC predictions, using several generator models 
(see text for details): binary model grey line, Bertini model black 
solid line, LHEP model dotted line, MARS model grey solid line. 
}
\label{fig:G43a}
\end{figure*}

\begin{figure*}[tbp]
%\vskip 1cm
\begin{center}
  \includegraphics[width=.49\textwidth]{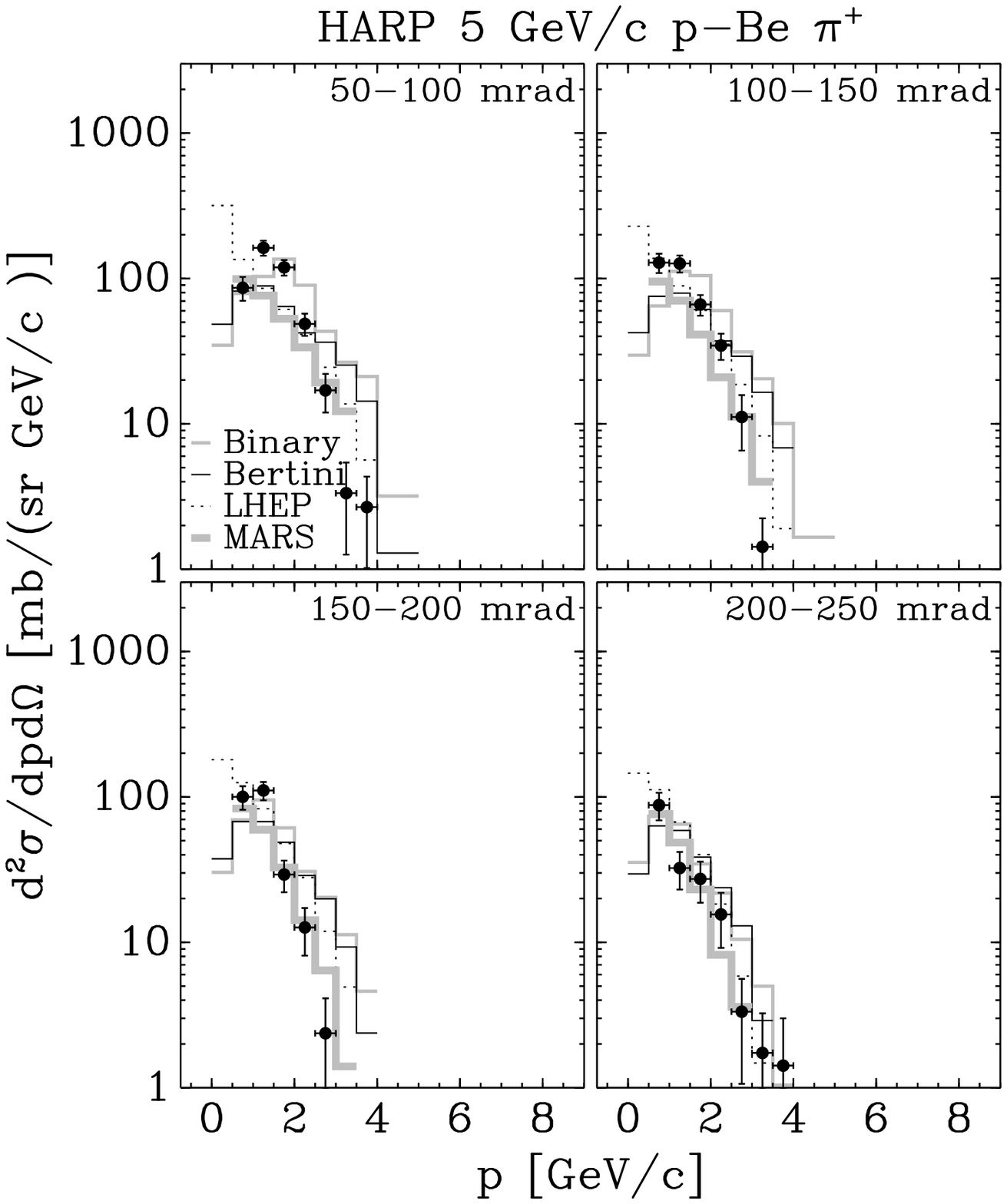}
  \includegraphics[width=.49\textwidth]{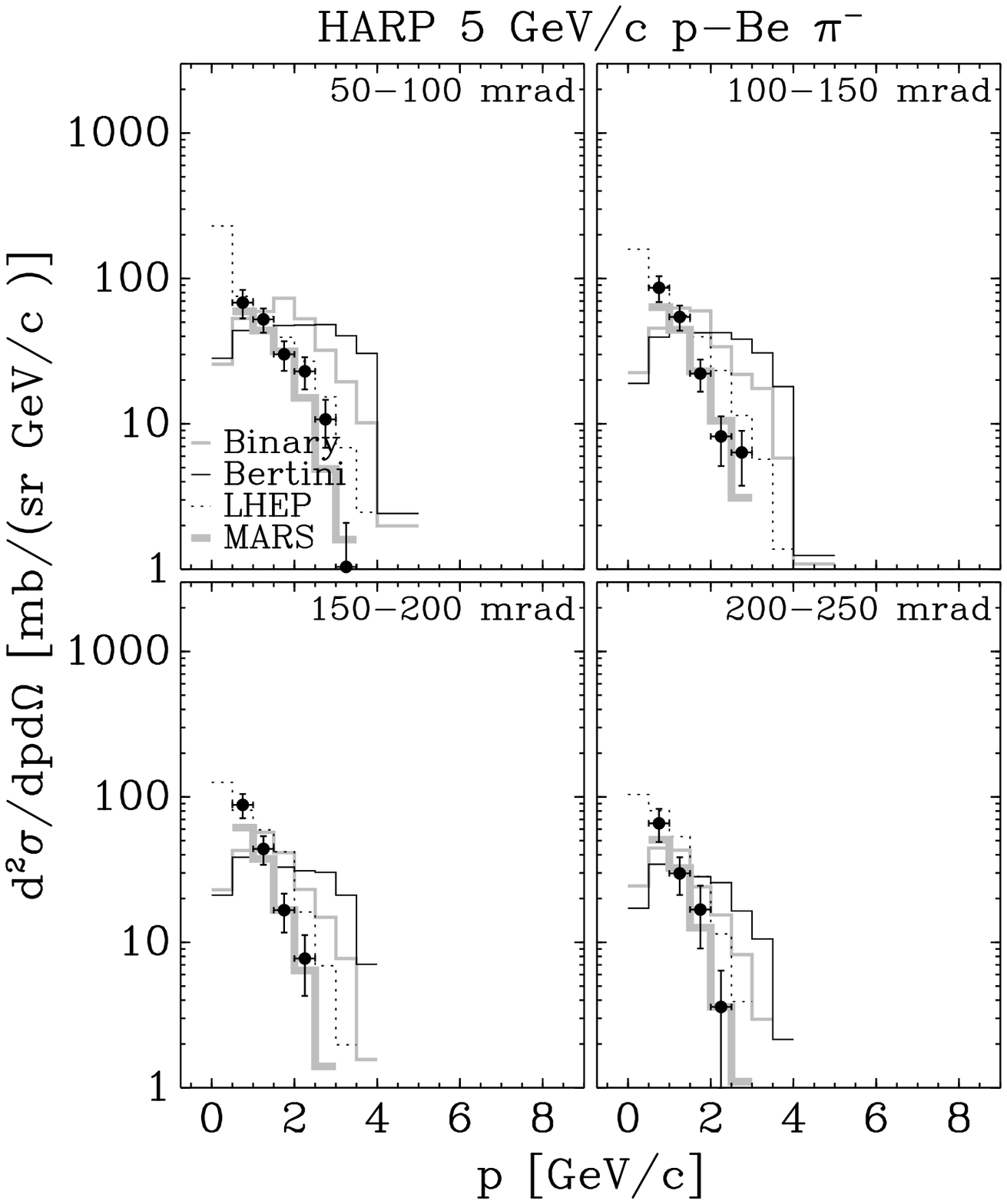}
\end{center}
\caption{
 Comparison of HARP double-differential $\pi^{\pm}$ cross sections for p--Be at 5 GeV/c with
 GEANT4 and MARS MC predictions, using several generator models 
(see text for details): Binary model grey line, Bertini model black solid line, LHEP model dotted line, MARS model grey solid line.
}
\label{fig:G44a}
\end{figure*}
\begin{figure*}[tbp]
%\vskip 1cm
\begin{center}
  \includegraphics[width=.65\textwidth]{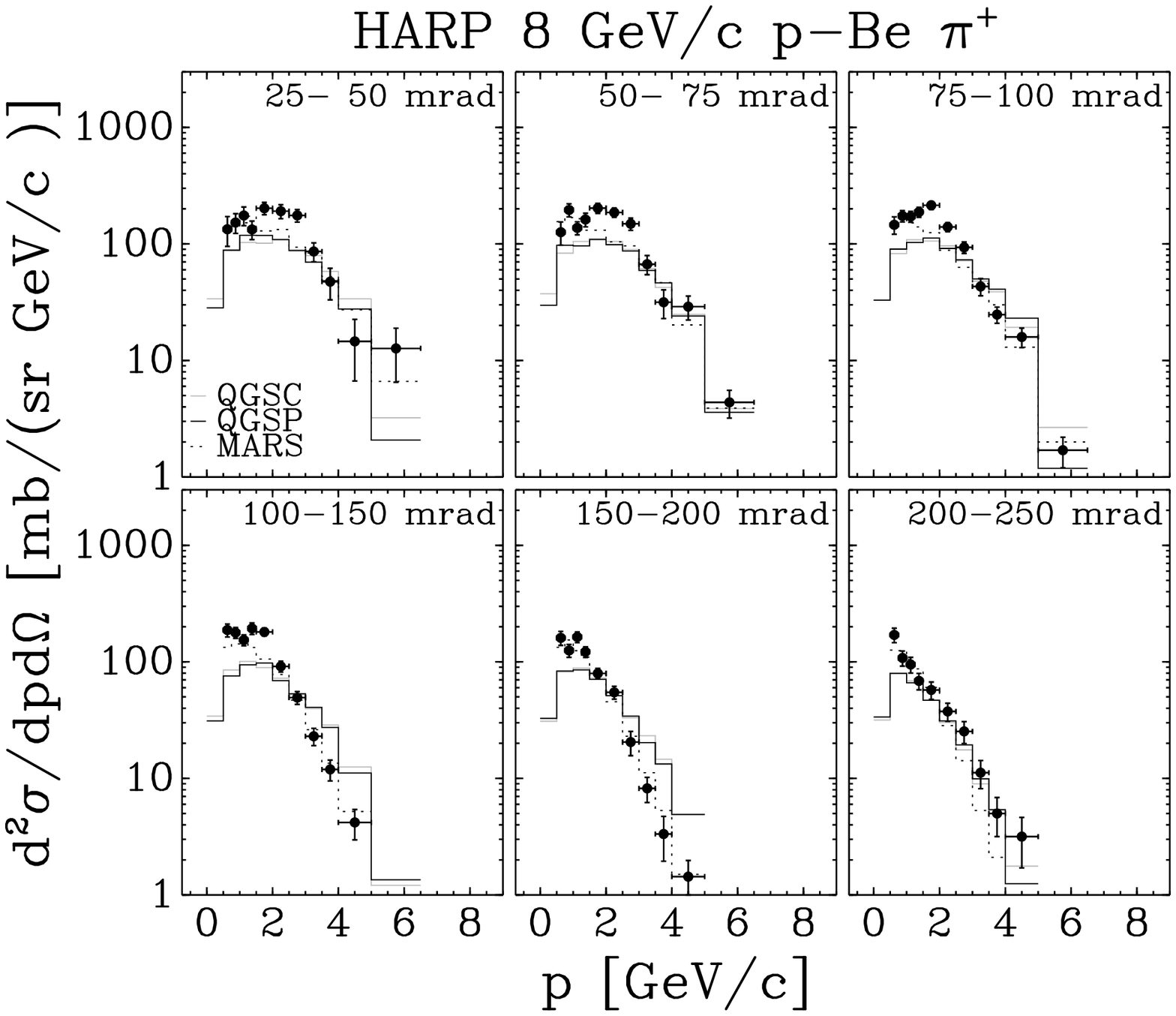}
  \includegraphics[width=.65\textwidth]{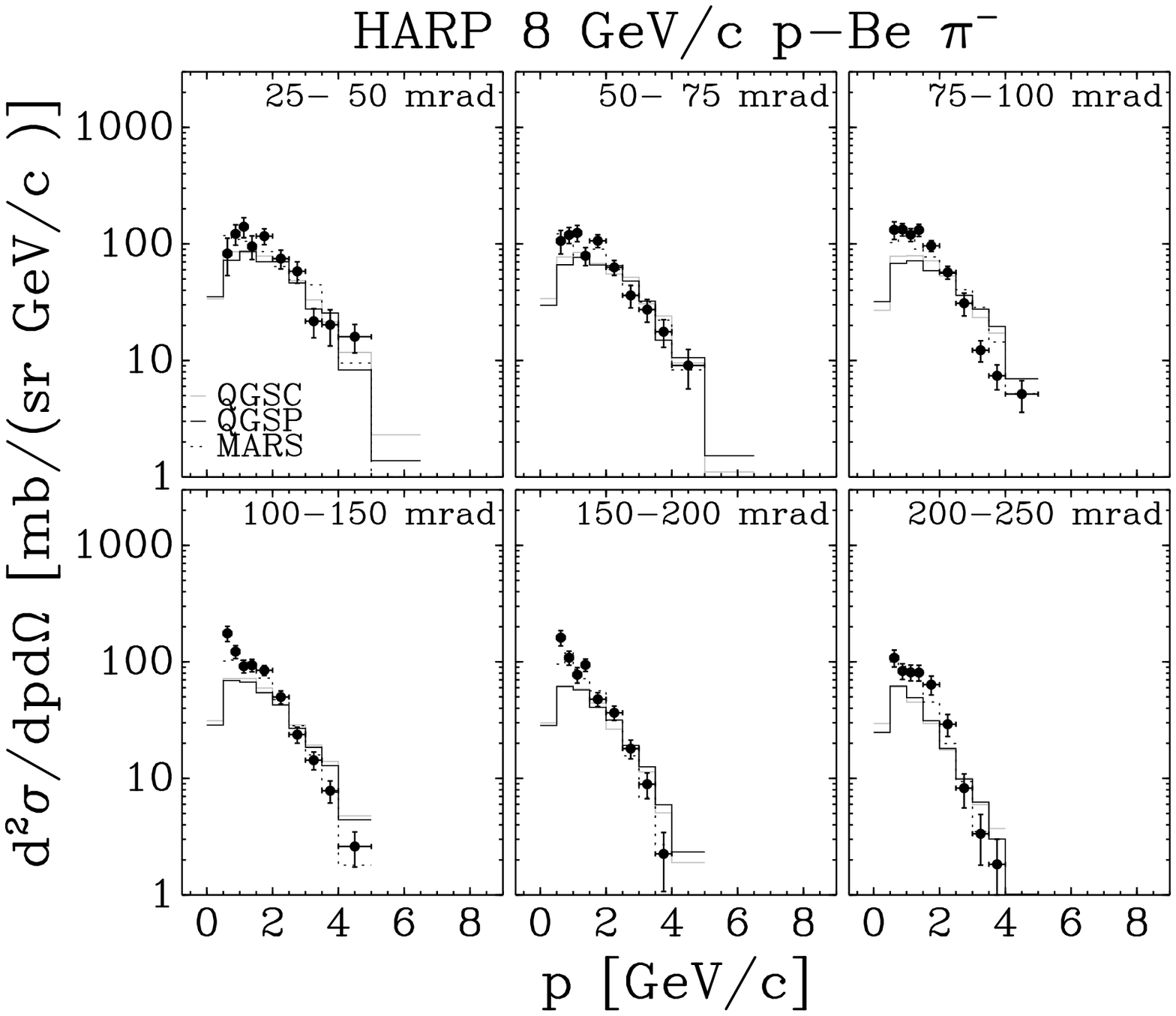}
\end{center}
\caption{
 Comparison of HARP double-differential $\pi^{\pm}$ cross sections for p--Be at 8 GeV/c with
 GEANT4 and MARS MC predictions, using several generator models 
(see text for details): QGSC model dotted line, QGSP model black solid line,
MARS model dashed line.
}
\label{fig:G45a}
\end{figure*}

\begin{figure*}[tbp]
%\vskip 1cm
\begin{center}
  \includegraphics[width=.65\textwidth]{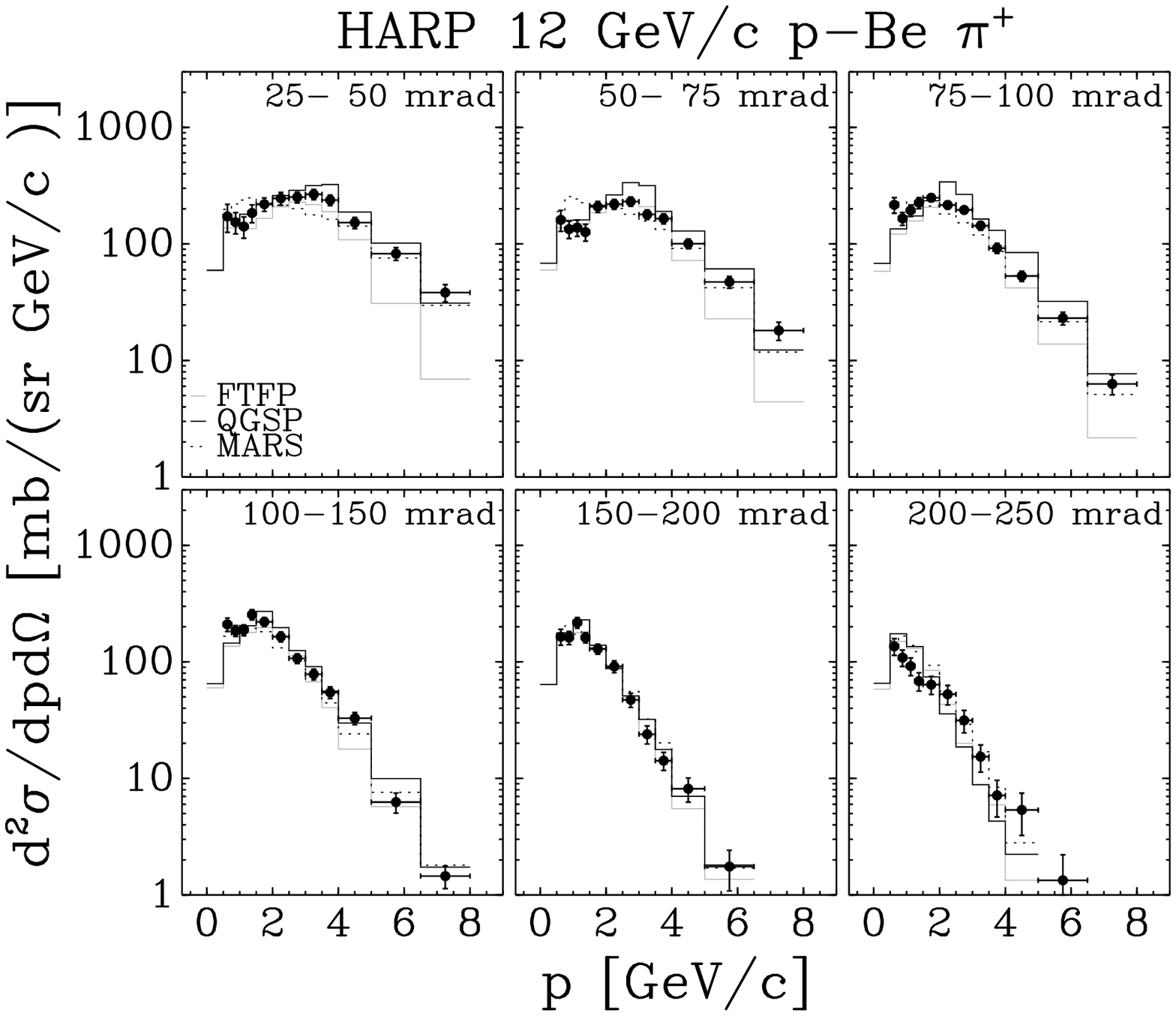}
  \includegraphics[width=.65\textwidth]{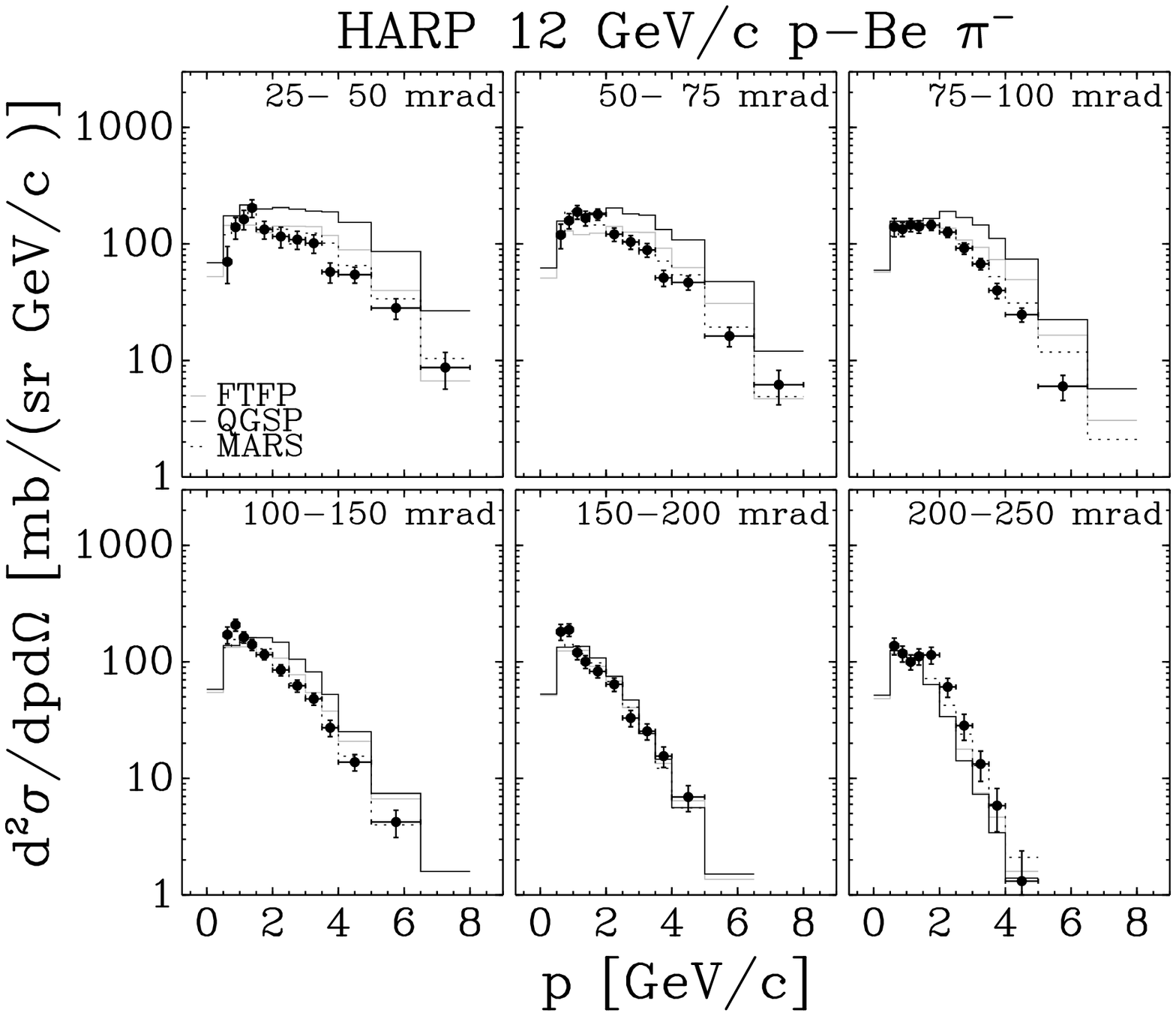}
\end{center}
%\vskip 1cm
\caption{
 Comparison of HARP double-differential $\pi^{\pm}$ cross sections for p--Be at 12 GeV/c with
 GEANT4 and MARS MC predictions, using several generator models 
(see text for details):FTFP model dotted line, QGSP model black solid line, 
MARS dashed line
}
\label{fig:G46}
\end{figure*}

\begin{figure*}[tbp]
%\vskip 1cm
\begin{center}
  \includegraphics[width=.49\textwidth]{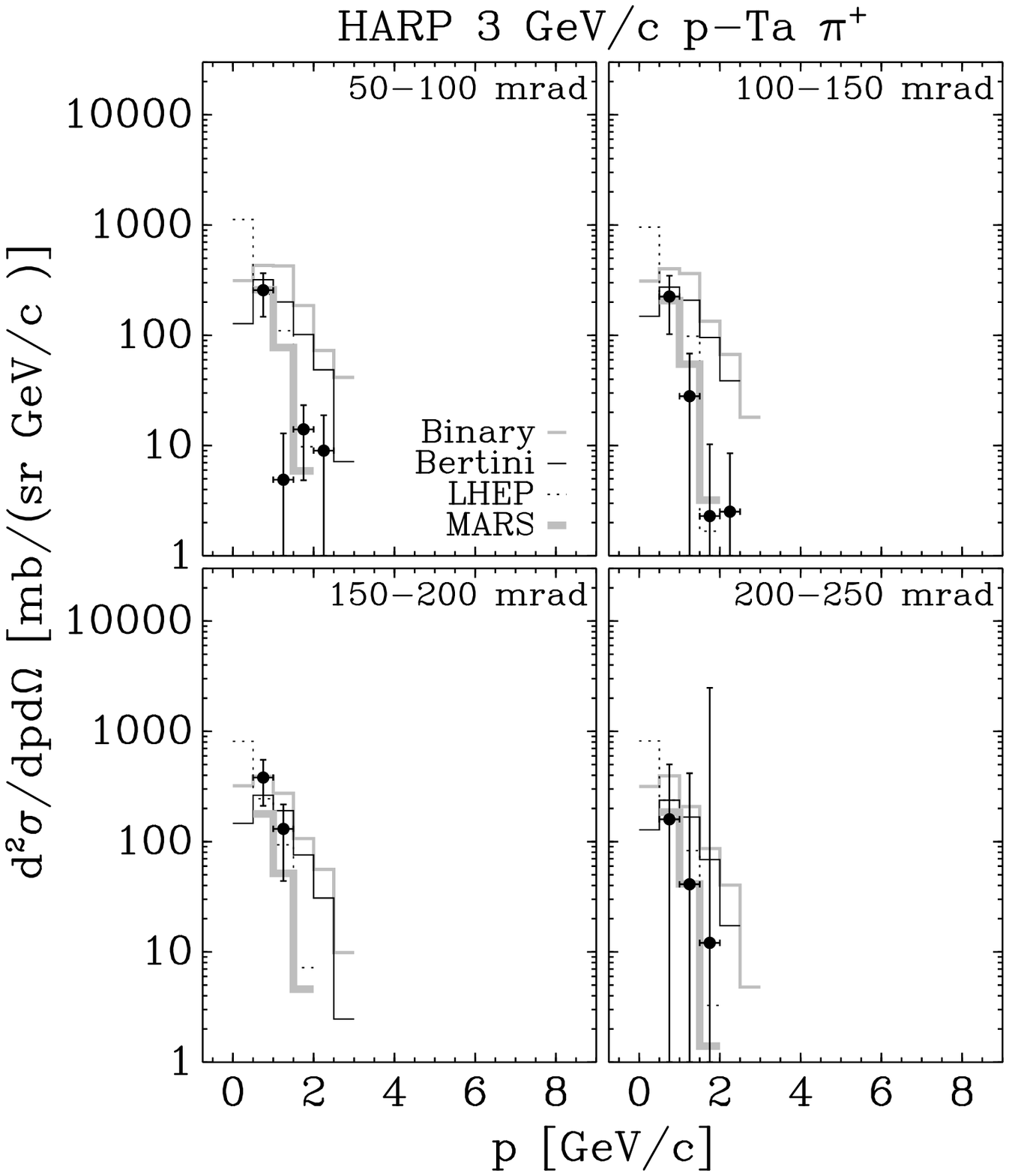}
  \includegraphics[width=.49\textwidth]{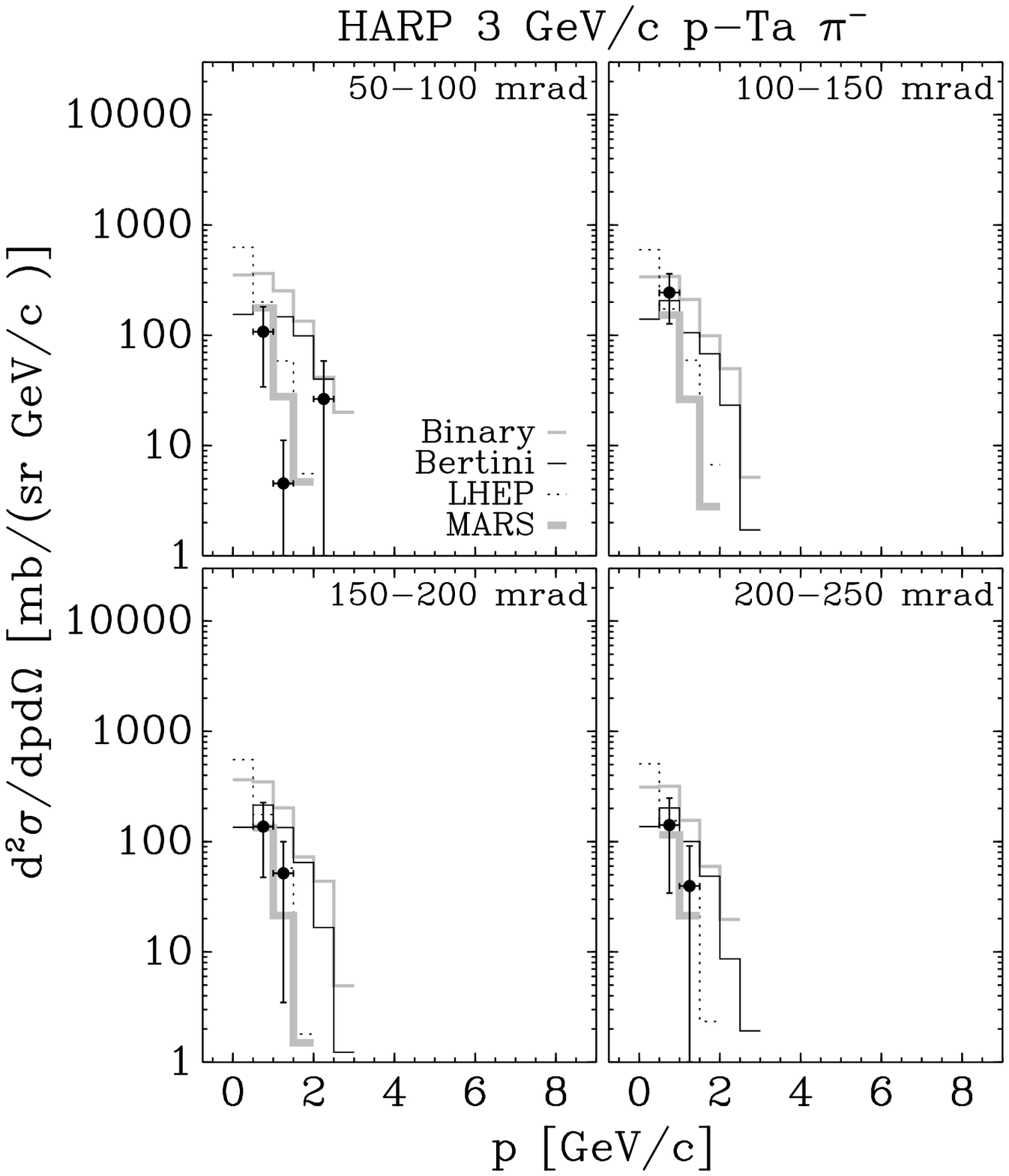}
\end{center}
%\vskip 1cm
\caption{
 Comparison of HARP double-differential $\pi^{\pm}$ cross sections for p--Ta at 3 GeV/c with
 GEANT4 and MARS MC predictions, using several generator models 
(see text for details): Binary model grey line, Bertini model black
solid line, LHEP model dotted line, MARS model grey solid line.
}
\label{fig:G53}
\end{figure*}

\begin{figure*}[tbp]
%\vskip -2cm
\begin{center}
  \includegraphics[width=.49\textwidth]{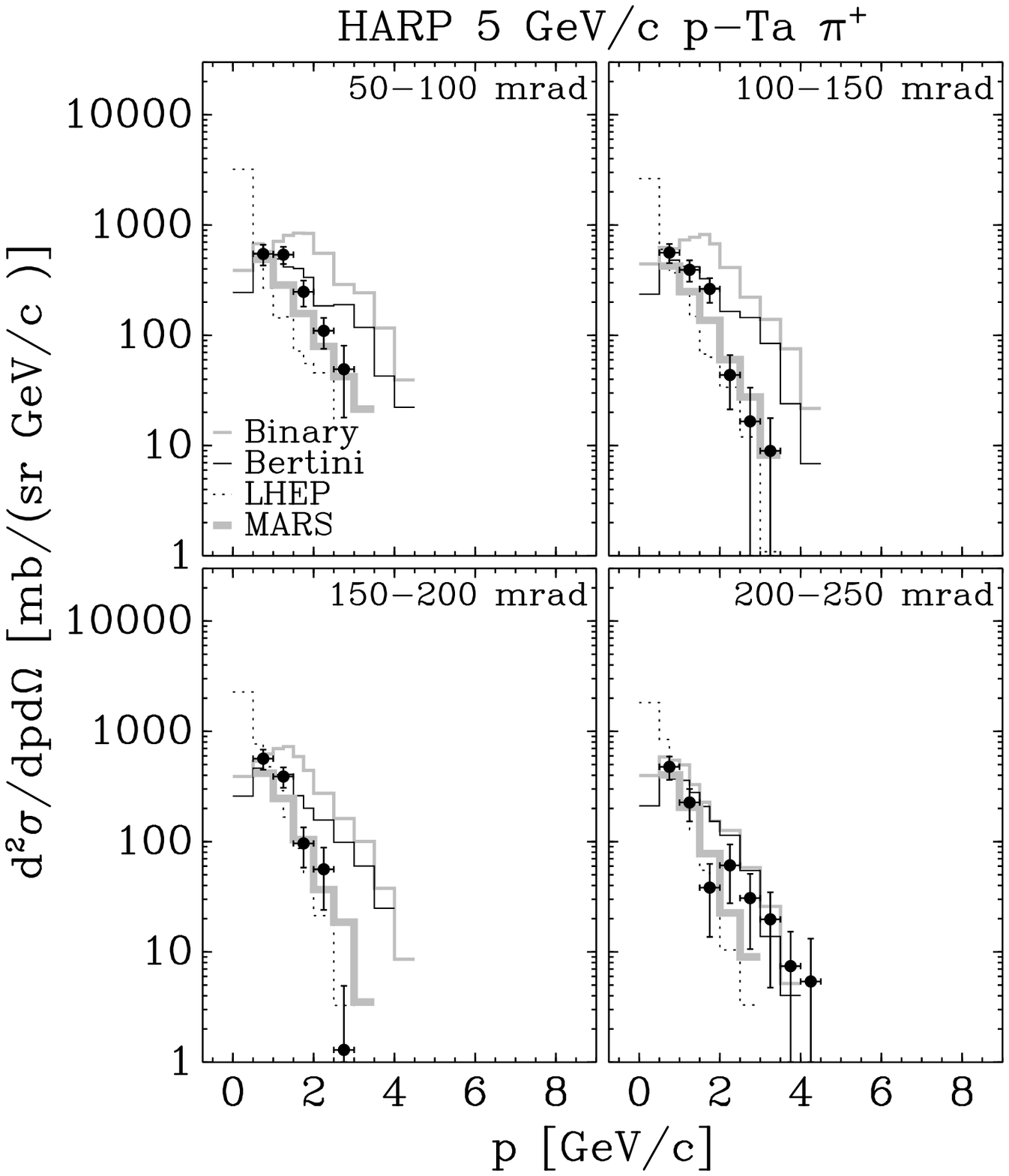}
  \includegraphics[width=.49\textwidth]{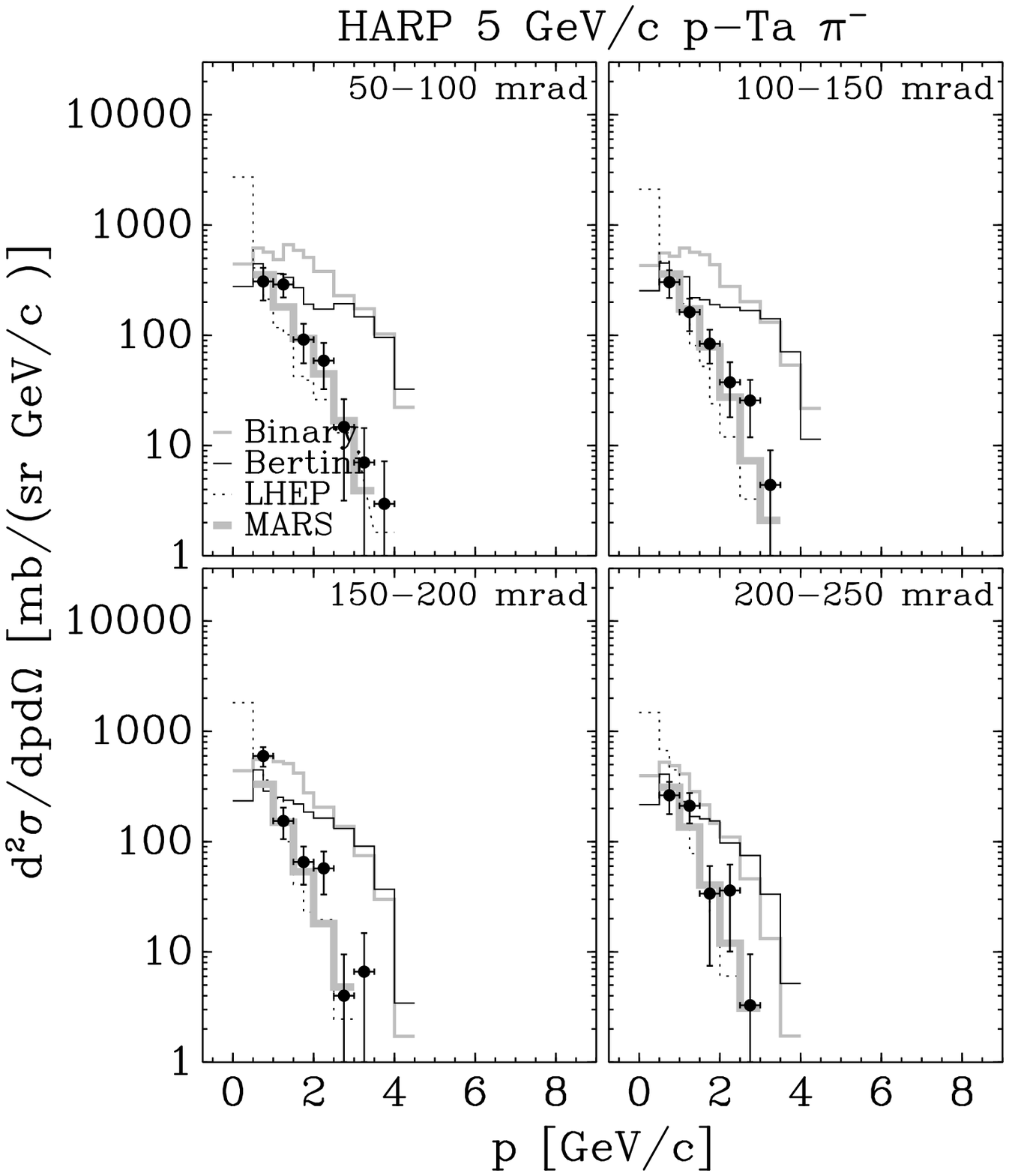}
\end{center}
\caption{
 Comparison of HARP double-differential $\pi^{\pm}$ cross sections for p--Ta at 5 GeV/c with
 GEANT4 and MARS MC predictions, using several generator models 
(see text for details): Binary model grey line, Bertini model black
solid line, LHEP model dotted line, MARS model grey solid line.
}
\label{fig:G54a}
\end{figure*}

\begin{figure*}[tbp]
%\vskip 1cm
\begin{center}
  \includegraphics[width=.65\textwidth]{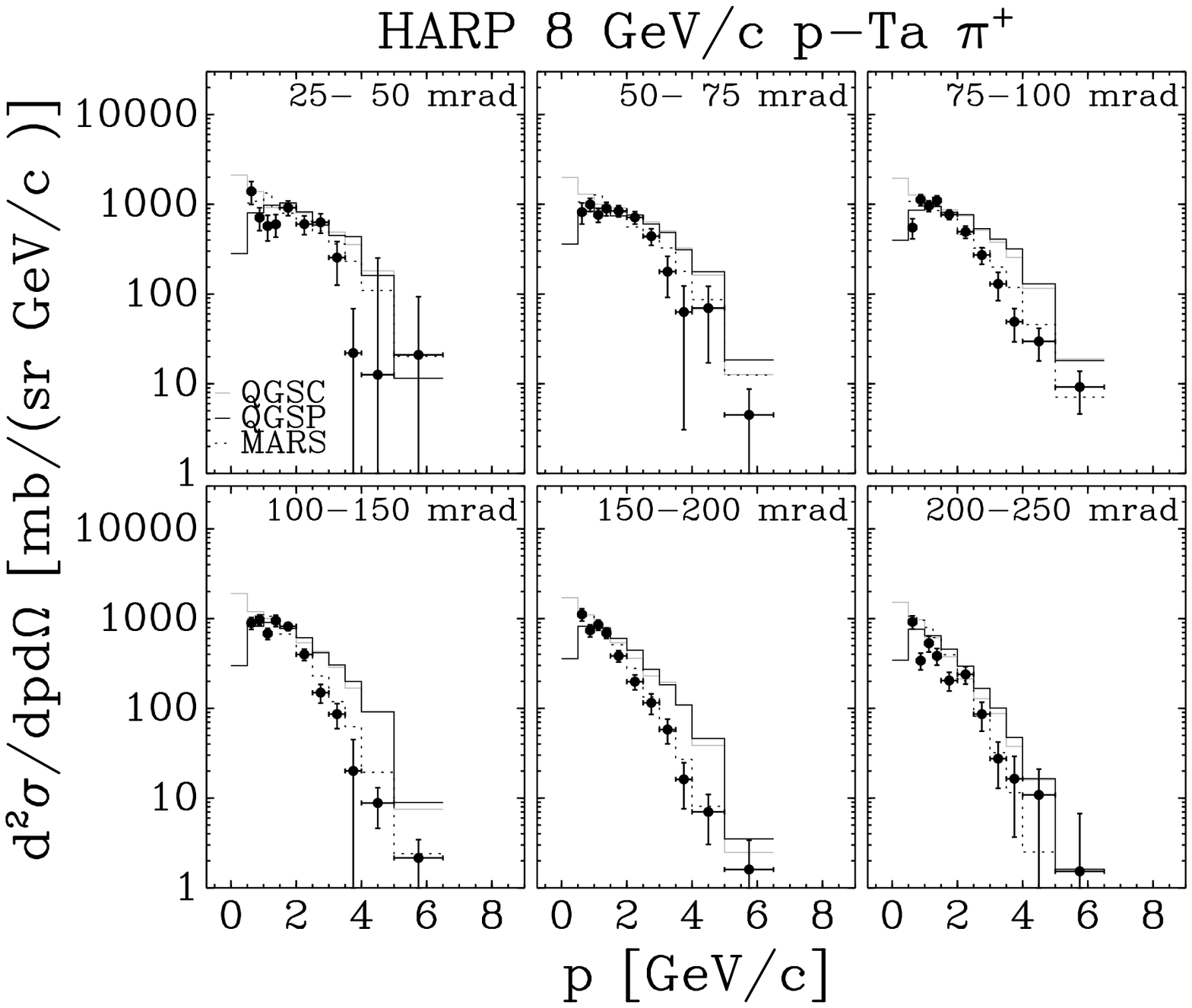}
  \includegraphics[width=.65\textwidth]{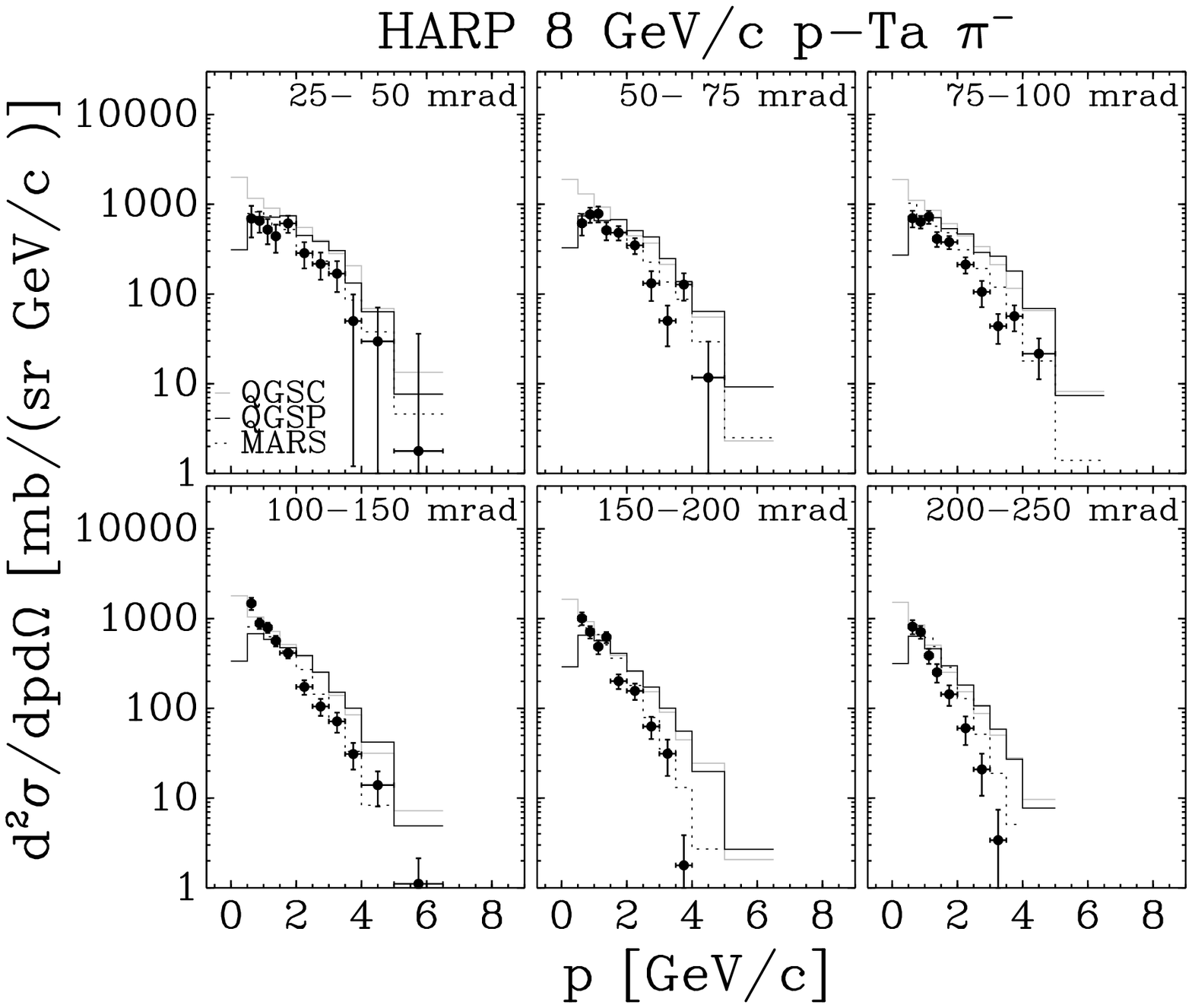}
\end{center}
%\vskip 1cm
\caption{
 Comparison of HARP double-differential $\pi^{\pm}$ cross sections for p--Ta at 8 GeV/c with
 GEANT4 and MARS MC predictions, using several generator models 
(see text for details): QGSC model dotted line, QGSP model black solid line,
MARS dashed line.
}
\label{fig:G55a}
\end{figure*}

\begin{figure*}[htbp]
%\vskip 1cm
\begin{center}
  \includegraphics[width=.65\textwidth]{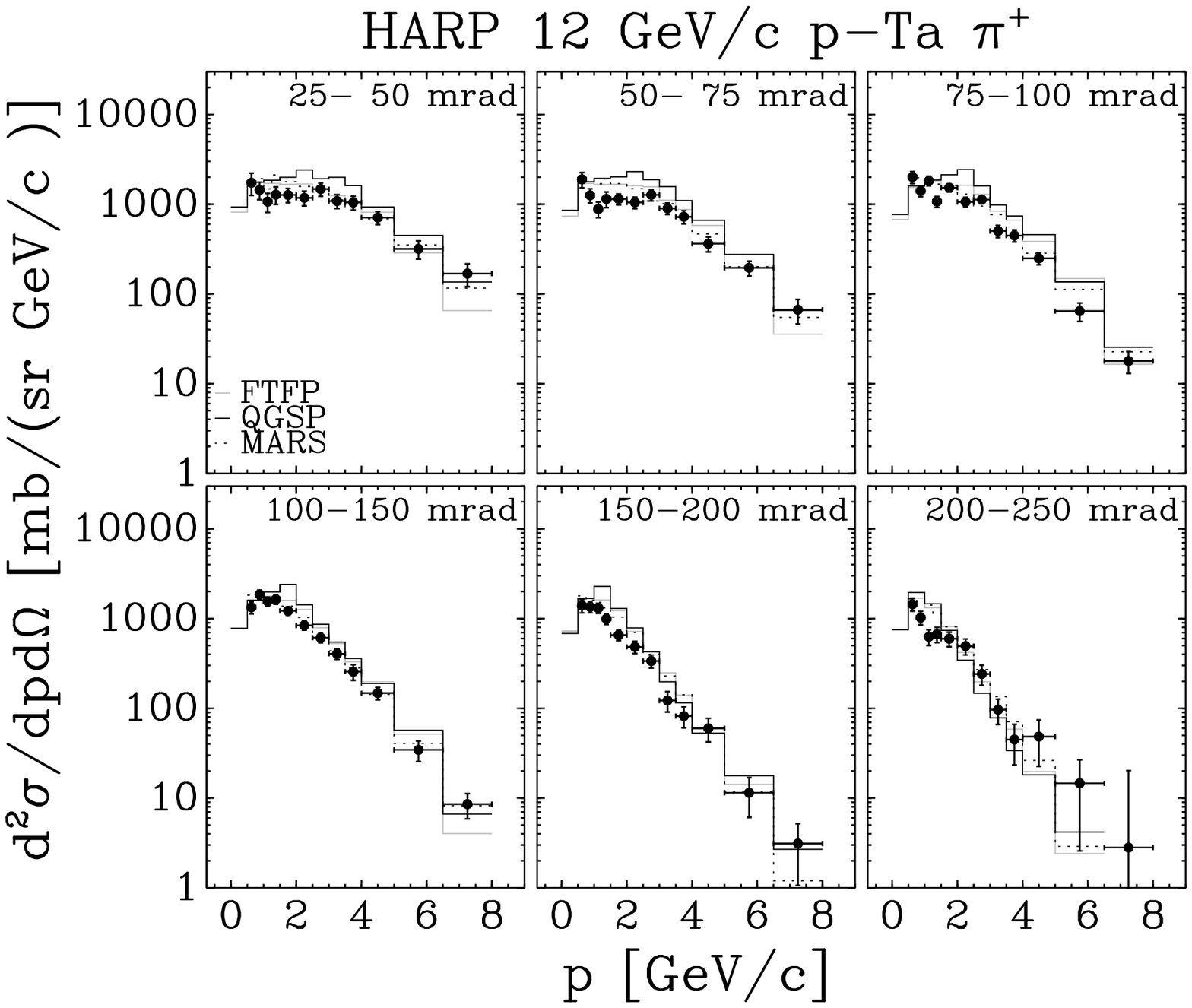}
  \includegraphics[width=.65\textwidth]{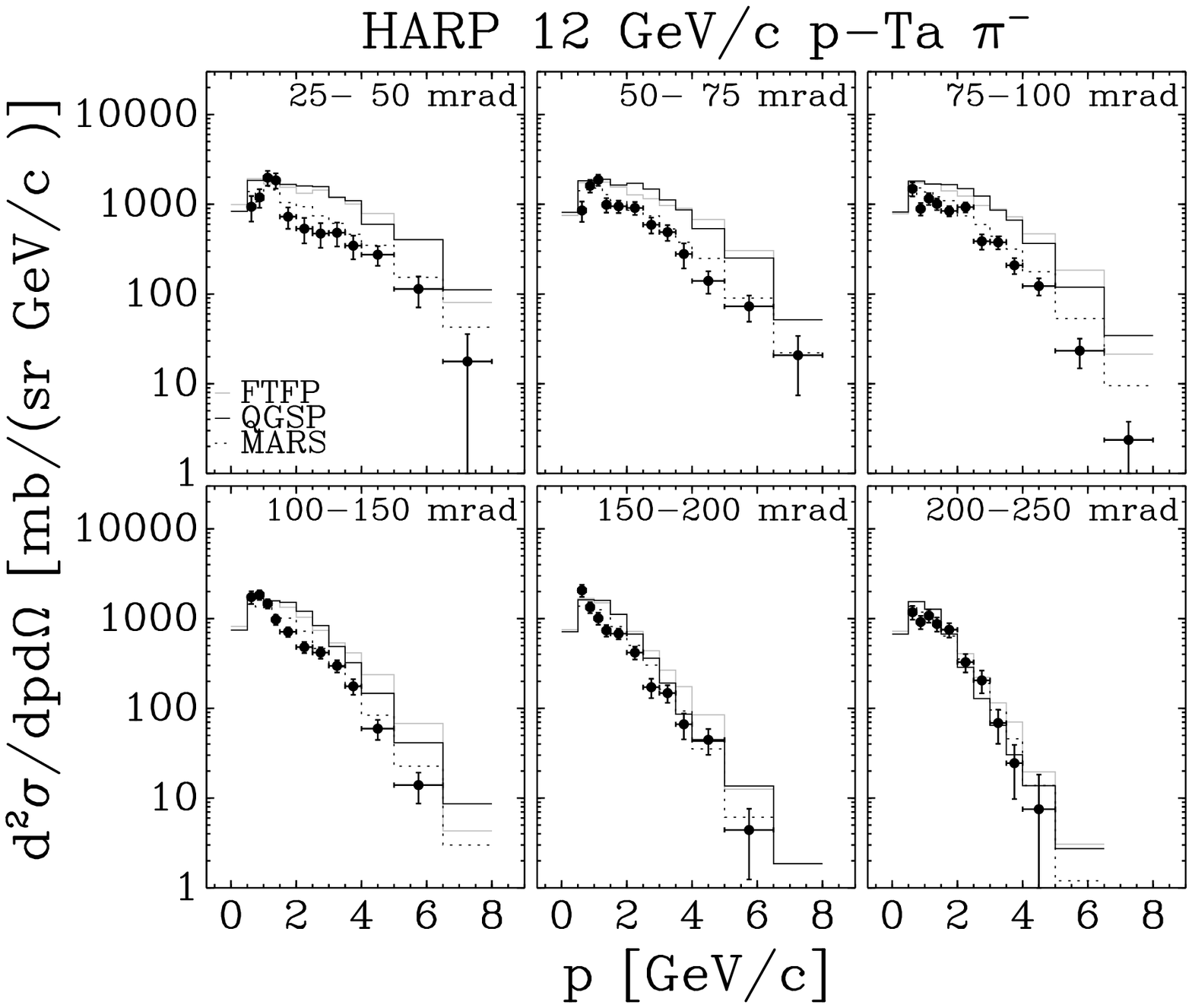}
\end{center}
%\vskip 1cm
\caption{
 Comparison of HARP double-differential $\pi^{\pm}$ cross sections for p--Ta at 12 GeV/c with
 GEANT4 and MARS MC predictions, using several generator models 
(see text for details): FTFP  model dotted grey line, QGSP model black solid line,
MARS  dashed line.
}
\label{fig:G56}
\end{figure*}

The MARS code system~\cite{ref:mars} uses instead as basic model an inclusive
approach multiparticle production originated by R. Feynmann. Above 3~GeV
phenomenological particle production models are used, while below 5~GeV
a cascade-exciton model~\cite{ref:casca} combined with the Fermi break-up model, the coalescence
model, an evaporation model and a multifragmentation extension are used
instead.

In figures~\ref{fig:G43a} to \ref{fig:G56} 
data have been compared to Monte Carlo predictions, using beryllium and 
tantalum as examples of a light and a heavy target. 

Computing the $\chi^2$ between models and data themselves, where a systematic 
error going from $0 \%$ to $50 \%$ have been added to simulation results
we obtain the results shown in Table~\ref{tab_chi1}.
Normalization factors, between data and Monte Carlo simulations, are
shown instead in Table~\ref{tab_norm}.

Over the full energy range covered by the HARP experiment, the best
comparison is obtained with the MARS Monte Carlo.
This may be owing to the fact that MARS is using different models in
different energy regions, equivalent to use a collection of models as
implemented in the ``physics lists'' of GEANT4.

The full set of HARP data, taken with targets
spanning the full periodic table of elements, with small total errors and full
coverage of the solid angle in a single detector may help the validation
of models used in hadronic simulations in the difficult energy range between
3 and 15 GeV/c of incident momentum, as done in reference~\cite{ref:gall}.

%%%% !!!!! HERE INSERT COMPARISONS WITH DATA, IF ANY !!!!!!
\FloatBarrier

\section{Summary and conclusions}\label{sec:conclusions}

In this paper we report our final results on
double-differential cross sections for the production
of positive and negative pions 
in the kinematic range 
0.5 GeV/c$\leq p_\pi \leq 8$ GeV/c 
and 0.025 rad $\leq \theta_\pi \leq$ 0.25 rad
from the collisions
of protons of 3, 5, 8 and 12 GeV/c on   beryllium, aluminium, carbon, copper,
  tin, tantalum targets of 5\% $\lambda_I$.
In addition results at 8.9 \GeVc (Be only), 12.9 \GeVc (Al only) and
12 \GeVc ($N_2$ and $O_2$) are reported.

A parametrization, inspired by the Sanford-Wang formula, of all our
datasets at all energies is also presented. This may be used as a fast
approximation to HARP data, valid within a factor 2-3 of the quoted errors. 
 
The pion yield averaged over different
momentum and angular ranges 
increase smoothly
with the atomic number $A$ of the target and with the
energy of the incoming proton beam.
The $A$-dependence is slightly different for \pim and \pip production,
the latter saturating earlier towards higher $A$, especially at lower
beam momenta.

Comparisons with GEANT4 and MARS generators are presented.

\section{Acknowledgments}

We gratefully acknowledge the help and support of the PS beam staff
and of the numerous technical collaborators who contributed to the
detector design, construction, commissioning and operation.  
In particular, we would like to thank
G.~Barichello,
R.~Brocard,
K.~Burin,
V.~Carassiti,
F.~Chignoli,
D.~Conventi,
G.~Decreuse,
M.~Delattre,
C.~Detraz,  
A.~Domeniconi,
M.~Dwuznik,   
F.~Evangelisti,
B.~Friend,
A.~Iaciofano,
I.~Krasin, 
D.~Lacroix,
J.-C.~Legrand,
M.~Lobello, 
M.~Lollo,
J.~Loquet,
F.~Marinilli,
R.~Mazza,
J.~Mulon,
L.~Musa,
R.~Nicholson,
A.~Pepato,
P.~Petev, 
X.~Pons,
I.~Rusinov,
M.~Scandurra,
E.~Usenko,
R.~van der Vlugt,
for their support in the construction of the detector
and P. Dini for his contribution to Monte Carlo production. 
The collaboration acknowledges the major contributions and advice of
M.~Baldo-Ceolin, 
L.~Linssen, 
M.T.~Muciaccia and A. Pullia
during the construction of the experiment.
The collaboration is indebted to 
V.~Ableev,
%M.~Baldo~Ceolin,
F.~Bergsma,
P.~Binko,
E.~Boter,
M.~Calvi, 
C.~Cavion,
M.Chizov, 
A.~Chukanov,
A.~DeSanto, 
A.~DeMin, 
M.~Doucet,
D.~D\"{u}llmann,
V.~Ermilova, 
W.~Flegel,
Y.~Hayato,
A.~Ichikawa,
%%%A.~Ivanchenko,
O.~Klimov,
T.~Kobayashi,
D.~Kustov, 
M.~Laveder, 
%L.~Linssen,
M.~Mass,
H.~Meinhard,
A.~Menegolli, 
%M.T.~Muciaccia, 
T.~Nakaya,
K.~Nishikawa,
M.~Paganoni,
F.~Paleari,
M.~Pasquali,
M.~Placentino,
%A.~Pullia,
V.~Serdiouk,
S.~Simone,
P.J.~Soler,
S.~Troquereau,
S.~Ueda,
A.~Valassi and
R.~Veenhof
for their contributions to the experiment.

We acknowledge the contributions of 
%V.~Ammosov,
%G.~Chelkov,
%D.~Dedovich,
%A.~De~Min, 
%F.~Dydak and
%M.~Gostkin,
%A.~Guskov, 
%D.~Khartchenko, 
%V.~Koreshev,
%Z.~Kroumchtein,
%I.~Nefedov,
%A.~Semak, 
%E.~Usenko, 
%J.~Wotschack %,
%V.~Zaets, 
%and
%A.~Zhemchugov
V.~Ammosov,
G.~Chelkov,
D.~Dedovich,
F.~Dydak,
M.~Gostkin,
A.~Guskov,
D.~Khartchenko,
V.~Koreshev,
Z.~Kroumchtein,
I.~Nefedov,
A.~Semak,
J.~Wotschack,
V.~Zaets and
A.~Zhemchugov
to the work described in this paper.

 The experiment was made possible by grants from
the Institut Interuniversitaire des Sciences Nucl\'eair\-es and the
Interuniversitair Instituut voor Kernwetenschappen (Belgium), 
Ministerio de Educacion y Ciencia, Grant FPA2003-06921-c02-02 and
Generalitat Valenciana, grant GV00-054-1,
CERN (Geneva, Switzerland), 
the German Bundesministerium f\"ur Bildung und Forschung (Germany), 
the Istituto Na\-zio\-na\-le di Fisica Nucleare (Italy), 
INR RAS (Moscow), the Russian Foundation for Basic Research (grant 08-02-00018)
and the Particle Physics and Astronomy Research Council (UK).
We gratefully acknowledge their support.

\clearpage 
\begin{appendix}
\section{Cross-section data}\label{app:data}
The following tables report the measured differential cross-section
for positive and negative pion production in interactions
of 3, 5, 8 and 12~\GeVc momentum protons
on different types of nuclear targets.
The data are presented in the kinematic range of
0.5~\GeVc$\leq p_\pi \leq 8$~\GeVc
and 0.05~rad $\leq \theta_\pi \leq$ 0.25~rad.
The overall normalization uncertainty ($ \leq 2 \%$) is not included
in the reported errors.

Results at higher incoming beam momenta, from 8~\GeVc to 12~\GeVc,
are also presented
extending to a lower value of the polar angle
$\theta$ (0.025 rad), using a finer binning.

The cross-section values in some bins of the following tables are
omitted and replaced by the symbol $* \pm *$, due to instabilities
in the unfolding procedure coming from the low statistics available.
This appears especially evident for the 3 GeV/c data where 
the total negative final state pion statistics is often well below one 
hundred events.

\begin{table*}[!ht]

  \caption{\label{tab:xsec_results_Be}
    HARP results for the double-differential $\pi^+$  production
    cross-section in the laboratory system,
    $d^2\sigma^{\pi}/(dpd\Omega)$, for p--Be interactions at 3, 5, 8, 8.9, 12~\GeVc.
    Each row refers to a
    different $(p_{\hbox{\small min}} \le p<p_{\hbox{\small max}},
    \theta_{\hbox{\small min}} \le \theta<\theta_{\hbox{\small max}})$ bin,
    where $p$ and $\theta$ are the pion momentum and polar angle, respectively.
    The central value as well as the square-root of the diagonal elements
    of the covariance matrix are given.}

%\begin{tabular}{rrrr|r@{$\pm$}lr{$\pm$}lr{$\pm$}lr{$\pm$}l}
\small{
\begin{tabular}{rrrr|r@{$\pm$}lr@{$\pm$}lr@{$\pm$}lr@{$\pm$}lr@{$\pm$}l}
\hline
$\theta_{\hbox{\small min}}$ &
$\theta_{\hbox{\small max}}$ &
$p_{\hbox{\small min}}$ &
$p_{\hbox{\small max}}$ &
\multicolumn{10}{c}{$d^2\sigma^{\pi^+}/(dpd\Omega)$}
\\
(rad) & (rad) & (\GeVc) & (\GeVc) &
\multicolumn{10}{c}{(barn/(sr \GeVc))}
\\
  &  &  &
&\multicolumn{2}{c}{$ \bf{3 \ \GeVc}$}
&\multicolumn{2}{c}{$ \bf{5 \ \GeVc}$}
&\multicolumn{2}{c}{$ \bf{8 \ \GeVc}$}
&\multicolumn{2}{c}{$ \bf{8.9 \ \GeVc}$}
&\multicolumn{2}{c}{$ \bf{12 \ \GeVc}$}
\\
\hline

0.050 &0.100 & 0.50 & 1.00&  0.15 &   0.03&  0.09 &   0.02&  0.16 &   0.02&  0.17 &   0.02&  0.17 &   0.02\\ 
      &      & 1.00 & 1.50& 0.021 &  0.011&  0.16 &   0.02& 0.167 &  0.013& 0.167 &  0.012&  0.18 &   0.02\\ 
      &      & 1.50 & 2.00& 0.006 &  0.003& 0.119 &  0.015& 0.210 &  0.015& 0.203 &  0.014&  0.23 &   0.02\\ 
      &      & 2.00 & 2.50& 0.003 &  0.002& 0.049 &  0.009& 0.159 &  0.011& 0.205 &  0.012&  0.22 &   0.02\\ 
      &      & 2.50 & 3.00&       &       & 0.017 &  0.005& 0.116 &  0.011& 0.154 &  0.010& 0.210 &  0.013\\ 
      &      & 3.00 & 3.50&       &       & 0.003 &  0.002& 0.053 &  0.007& 0.088 &  0.008& 0.158 &  0.011\\ 
      &      & 3.50 & 4.00&       &       & 0.003 &  0.002& 0.028 &  0.005& 0.065 &  0.006& 0.122 &  0.010\\ 
      &      & 4.00 & 5.00&       &       & 0.001 &  0.001& 0.021 &  0.004& 0.036 &  0.004& 0.073 &  0.006\\ 
      &      & 5.00 & 6.50&       &       &       &       & 0.003 &  0.001& 0.009 &  0.002& 0.033 &  0.003\\ 
      &      & 6.50 & 8.00&       &       &       &       &       &       &       &       & 0.011 &  0.002\\ 
0.100 &0.150 & 0.50 & 1.00&  0.10 &   0.03&  0.13 &   0.02&  0.18 &   0.02&  0.17 &   0.02&  0.20 &   0.02\\ 
      &      & 1.00 & 1.50& 0.028 &  0.013&  0.13 &   0.02&  0.17 &   0.02&  0.19 &   0.02&  0.22 &   0.02\\ 
      &      & 1.50 & 2.00& 0.001 &  0.001& 0.066 &  0.011& 0.181 &  0.015& 0.168 &  0.013&  0.22 &   0.02\\ 
      &      & 2.00 & 2.50&    *  &    *  & 0.035 &  0.007& 0.092 &  0.010& 0.130 &  0.010&  0.16 &   0.02\\ 
      &      & 2.50 & 3.00&       &       & 0.011 &  0.005& 0.049 &  0.006& 0.075 &  0.007& 0.107 &  0.010\\ 
      &      & 3.00 & 3.50&       &       & 0.001 &  0.001& 0.023 &  0.004& 0.040 &  0.005& 0.078 &  0.008\\ 
      &      & 3.50 & 4.00&       &       &     * &     * & 0.012 &  0.002& 0.024 &  0.003& 0.055 &  0.006\\ 
      &      & 4.00 & 5.00&       &       &     * &     * & 0.004 &  0.001& 0.010 &  0.002& 0.033 &  0.004\\ 
      &      & 5.00 & 6.50&       &       &       &       &    *  &    *  &    *  &    *  & 0.006 &  0.001\\ 
      &      & 6.50 & 8.00&       &       &       &       &       &       &       &       &     * &      *\\ 
0.150 &0.200 & 0.50 & 1.00&  0.07 &   0.02&  0.10 &   0.02&  0.14 &   0.02&  0.17 &   0.02&  0.16 &   0.02\\ 
      &      & 1.00 & 1.50& 0.010 &  0.007&  0.11 &   0.02& 0.143 &  0.013& 0.155 &  0.012&  0.19 &   0.02\\ 
      &      & 1.50 & 2.00&    *  &    *  & 0.029 &  0.007& 0.080 &  0.008& 0.102 &  0.008& 0.129 &  0.013\\ 
      &      & 2.00 & 2.50&    *  &    *  & 0.013 &  0.005& 0.055 &  0.007& 0.065 &  0.007& 0.091 &  0.011\\ 
      &      & 2.50 & 3.00&       &       & 0.002 &  0.002& 0.020 &  0.005& 0.034 &  0.004& 0.047 &  0.007\\ 
      &      & 3.00 & 3.50&       &       & 0.001 &  0.001& 0.008 &  0.002& 0.020 &  0.002& 0.024 &  0.004\\ 
      &      & 3.50 & 4.00&       &       &     * &     * & 0.003 &  0.001& 0.011 &  0.002& 0.014 &  0.003\\ 
      &      & 4.00 & 5.00&       &       &     * &     * & 0.001 &  0.001& 0.005 &  0.001& 0.008 &  0.002\\ 
      &      & 5.00 & 6.50&       &       &       &       &  *    &    *  &    *  &     * & 0.002 &  0.001\\ 
      &      & 6.50 & 8.00&       &       &       &       &       &       &       &       &   *   &     * \\ 
0.200 &0.250 & 0.50 & 1.00&  0.05 &   0.02&  0.09 &   0.02&  0.14 &   0.02& 0.120 &  0.015&  0.12 &   0.02\\ 
      &      & 1.00 & 1.50& 0.017 &  0.011& 0.032 &  0.009& 0.082 &  0.012& 0.071 &  0.010& 0.080 &  0.013\\ 
      &      & 1.50 & 2.00&    *  &    *  & 0.027 &  0.009& 0.057 &  0.010& 0.049 &  0.007& 0.064 &  0.011\\ 
      &      & 2.00 & 2.50&    *  &    *  & 0.016 &  0.006& 0.038 &  0.007& 0.033 &  0.006& 0.053 &  0.010\\ 
      &      & 2.50 & 3.00&       &       & 0.003 &  0.002& 0.025 &  0.005& 0.025 &  0.003& 0.031 &  0.007\\ 
      &      & 3.00 & 3.50&       &       & 0.002 &  0.002& 0.011 &  0.003& 0.014 &  0.002& 0.015 &  0.004\\ 
      &      & 3.50 & 4.00&       &       & 0.001 &  0.002& 0.005 &  0.002& 0.006 &  0.001& 0.007 &  0.002\\ 
      &      & 4.00 & 5.00&       &       & 0.001 &  0.001& 0.003 &  0.001& 0.004 &  0.001& 0.005 &  0.002\\ 
      &      & 5.00 & 6.50&       &       &       &       & 0.001 &  0.001&     * &     * & 0.001 &  0.001\\ 
      &      & 6.50 & 8.00&       &       &       &       &       &       &       &       & 0.000 &  0.002\\ 

%%%%%%%%%%%%%%%%%%%%%%%%%%%%%%%%%%%%%%%%%%%%%%%%%%%%%%%%%%%%%%%%%%%%%%%%%%%%%
\hline
\end{tabular}
}
\end{table*}
\begin{table*}[!ht]
  \caption{\label{tab:xsec_results_Be}
    HARP results for the double-differential  $\pi^-$ production
    cross-section in the laboratory system,
    $d^2\sigma^{\pi}/(dpd\Omega)$, for p--Be interactions at 3, 5, 8, 8.9, 12~\GeVc.
    Each row refers to a
    different $(p_{\hbox{\small min}} \le p<p_{\hbox{\small max}},
    \theta_{\hbox{\small min}} \le \theta<\theta_{\hbox{\small max}})$ bin,
    where $p$ and $\theta$ are the pion momentum and polar angle, respectively.
    The central value as well as the square-root of the diagonal elements
    of the covariance matrix are given.}

%\begin{tabular}{rrrr|r@{$\pm$}lr{$\pm$}lr{$\pm$}lr{$\pm$}l}
\small{
\begin{tabular}{rrrr|r@{$\pm$}lr@{$\pm$}lr@{$\pm$}lr@{$\pm$}lr@{$\pm$}l}
\hline
$\theta_{\hbox{\small min}}$ &
$\theta_{\hbox{\small max}}$ &
$p_{\hbox{\small min}}$ &
$p_{\hbox{\small max}}$ &
\multicolumn{10}{c}{$d^2\sigma^{\pi^-}/(dpd\Omega)$}
\\
(rad) & (rad) & (\GeVc) & (\GeVc) &
\multicolumn{10}{c}{(barn/(sr \GeVc))}
\\
  &  &  &
&\multicolumn{2}{c}{$ \bf{3 \ \GeVc}$}
&\multicolumn{2}{c}{$ \bf{5 \ \GeVc}$}
&\multicolumn{2}{c}{$ \bf{8 \ \GeVc}$}
&\multicolumn{2}{c}{$ \bf{8.9 \ \GeVc}$}
&\multicolumn{2}{c}{$ \bf{12 \ \GeVc}$}
\\
\hline

0.050 &0.100 & 0.50 & 1.00& 0.017 &  0.011&  0.07 &   0.02&  0.12 &   0.02& 0.118 &  0.014&  0.14 &   0.02\\ 
      &      & 1.00 & 1.50& 0.002 &  0.003& 0.052 &  0.010& 0.116 &  0.011& 0.141 &  0.010& 0.158 &  0.014\\ 
      &      & 1.50 & 2.00& 0.001 &  0.002& 0.030 &  0.007& 0.101 &  0.009& 0.118 &  0.008& 0.159 &  0.013\\ 
      &      & 2.00 & 2.50&     * &     * & 0.023 &  0.006& 0.060 &  0.006& 0.082 &  0.006& 0.124 &  0.011\\ 
      &      & 2.50 & 3.00&       &       & 0.011 &  0.004& 0.033 &  0.006& 0.049 &  0.005& 0.097 &  0.009\\ 
      &      & 3.00 & 3.50&       &       & 0.001 &  0.001& 0.019 &  0.003& 0.037 &  0.003& 0.076 &  0.007\\ 
      &      & 3.50 & 4.00&       &       &    *  &   *   & 0.012 &  0.002& 0.024 &  0.004& 0.045 &  0.005\\ 
      &      & 4.00 & 5.00&       &       &    *  &   *   & 0.007 &  0.002& 0.013 &  0.002& 0.034 &  0.004\\ 
      &      & 5.00 & 6.50&       &       &       &       &     * &      *& 0.002 &  0.001& 0.010 &  0.002\\ 
      &      & 6.50 & 8.00&       &       &       &       &       &       &       &       & 0.003 &  0.001\\ 
0.100 &0.150 & 0.50 & 1.00&  0.04 &   0.02&  0.09 &   0.02&  0.15 &   0.02&  0.15 &   0.02&  0.19 &   0.02\\ 
      &      & 1.00 & 1.50& 0.007 &  0.006& 0.054 &  0.011& 0.093 &  0.010& 0.126 &  0.011& 0.152 &  0.015\\ 
      &      & 1.50 & 2.00& 0.001 &  0.002& 0.022 &  0.006& 0.085 &  0.008& 0.085 &  0.008& 0.115 &  0.012\\ 
      &      & 2.00 & 2.50&    *  &     * & 0.008 &  0.003& 0.050 &  0.006& 0.056 &  0.006& 0.085 &  0.009\\ 
      &      & 2.50 & 3.00&       &       & 0.006 &  0.003& 0.024 &  0.004& 0.036 &  0.004& 0.062 &  0.008\\ 
      &      & 3.00 & 3.50&       &       & 0.001 &  0.001& 0.014 &  0.002& 0.020 &  0.002& 0.048 &  0.006\\ 
      &      & 3.50 & 4.00&       &       &    *  &    *  & 0.008 &  0.002& 0.015 &  0.002& 0.027 &  0.004\\ 
      &      & 4.00 & 5.00&       &       &    *  &    *  & 0.003 &  0.001& 0.005 &  0.001& 0.014 &  0.002\\ 
      &      & 5.00 & 6.50&       &       &       &       &    *  &     * & 0.001 &  0.000& 0.004 &  0.001\\ 
      &      & 6.50 & 8.00&       &       &       &       &       &       &       &       &   *   &   *  \\ 
0.150 &0.200 & 0.50 & 1.00& 0.012 &  0.008&  0.09 &   0.02&  0.13 &   0.02&  0.16 &   0.02&  0.18 &   0.02\\ 
      &      & 1.00 & 1.50& 0.006 &  0.005& 0.044 &  0.010& 0.086 &  0.010& 0.093 &  0.009& 0.110 &  0.013\\ 
      &      & 1.50 & 2.00& 0.007 &  0.007& 0.017 &  0.005& 0.048 &  0.007& 0.060 &  0.007& 0.083 &  0.010\\ 
      &      & 2.00 & 2.50&    *  &     * & 0.008 &  0.003& 0.037 &  0.005& 0.038 &  0.005& 0.064 &  0.008\\ 
      &      & 2.50 & 3.00&       &       &    *  &    *  & 0.018 &  0.003& 0.018 &  0.003& 0.033 &  0.005\\ 
      &      & 3.00 & 3.50&       &       &    *  &    *  & 0.009 &  0.002& 0.010 &  0.002& 0.025 &  0.004\\ 
      &      & 3.50 & 4.00&       &       &    *  &    *  & 0.002 &  0.001& 0.005 &  0.001& 0.016 &  0.003\\ 
      &      & 4.00 & 5.00&       &       &    *  &    *  &     * &    *  & 0.002 &  0.001& 0.007 &  0.002\\ 
      &      & 5.00 & 6.50&       &       &       &       &     * &    *  &    *  &   *   &   *   &   *    \\ 
      &      & 6.50 & 8.00&       &       &       &       &       &       &       &       &   *   &   *   \\ 
0.200 &0.250 & 0.50 & 1.00&  0.03 &   0.02&  0.07 &   0.02& 0.096 &  0.014& 0.102 &  0.014&  0.13 &   0.02\\ 
      &      & 1.00 & 1.50& 0.001 &  0.003& 0.030 &  0.009& 0.081 &  0.011& 0.090 &  0.011& 0.106 &  0.015\\ 
      &      & 1.50 & 2.00&    *  &    *  & 0.017 &  0.008& 0.064 &  0.012& 0.077 &  0.012&  0.11 &   0.02\\ 
      &      & 2.00 & 2.50&    *  &    *  & 0.004 &  0.003& 0.029 &  0.006& 0.040 &  0.007& 0.061 &  0.011\\ 
      &      & 2.50 & 3.00&       &       & 0.000 &  0.001& 0.008 &  0.003& 0.019 &  0.004& 0.028 &  0.007\\ 
      &      & 3.00 & 3.50&       &       &    *  &     * & 0.003 &  0.002& 0.008 &  0.002& 0.013 &  0.004\\ 
      &      & 3.50 & 4.00&       &       &    *  &     * & 0.002 &  0.001& 0.003 &  0.001& 0.006 &  0.002\\ 
      &      & 4.00 & 5.00&       &       &    *  &     * & 0.001 &  0.001& 0.001 &  0.000& 0.001 &  0.001\\ 
      &      & 5.00 & 6.50&       &       &       &       & 0.000 &  0.001&   *   &    *  & *   &   *   \\ 
      &      & 6.50 & 8.00&       &       &       &       &       &       &       &       & 0.000 &  0.001\\ 
%%%%%%%%%%%%%%%%%%%%%%%%%%%%%%%%%%%%%%%%%%%%%%%%%%%%%%%%%%%%%%%%%%%%%%%%%%%%%
\hline
\end{tabular}
}
\end{table*}

\begin{table*}[!ht]
  \caption{\label{tab:xsec_results_C}
    HARP results for the double-differential $\pi^+$  production
    cross-section in the laboratory system,
    $d^2\sigma^{\pi}/(dpd\Omega)$, for p--C interactions at 3, 5, 8, 12~\GeVc.
    Each row refers to a
    different $(p_{\hbox{\small min}} \le p<p_{\hbox{\small max}},
    \theta_{\hbox{\small min}} \le \theta<\theta_{\hbox{\small max}})$ bin,
    where $p$ and $\theta$ are the pion momentum and polar angle, respectively.
    The central value as well as the square-root of the diagonal elements
    of the covariance matrix are given.}

%\begin{tabular}{rrrr|r@{$\pm$}lr{$\pm$}lr{$\pm$}lr{$\pm$}l}
\small{
\begin{tabular}{rrrr|r@{$\pm$}lr@{$\pm$}lr@{$\pm$}lr@{$\pm$}l}
\hline
$\theta_{\hbox{\small min}}$ &
$\theta_{\hbox{\small max}}$ &
$p_{\hbox{\small min}}$ &
$p_{\hbox{\small max}}$ &
\multicolumn{8}{c}{$d^2\sigma^{\pi^+}/(dpd\Omega)$}
\\
(rad) & (rad) & (\GeVc) & (\GeVc) &
\multicolumn{8}{c}{(barn/(sr \GeVc ))}
\\
  &  &  &
&\multicolumn{2}{c}{$ \bf{3 \ \GeVc}$}
&\multicolumn{2}{c}{$ \bf{5 \ \GeVc}$}
&\multicolumn{2}{c}{$ \bf{8 \ \GeVc}$}
&\multicolumn{2}{c}{$ \bf{12 \ \GeVc}$}
\\
\hline

0.050 &0.100 & 0.50 & 1.00&  0.11 &   0.03&  0.12 &   0.02&  0.18 &   0.02&  0.19 &   0.03\\ 
      &      & 1.00 & 1.50& 0.015 &  0.012&  0.18 &   0.02&  0.20 &   0.02&  0.24 &   0.02\\ 
      &      & 1.50 & 2.00& 0.013 &  0.007& 0.113 &  0.014&  0.25 &   0.02&  0.28 &   0.02\\ 
      &      & 2.00 & 2.50& 0.001 &  0.001& 0.044 &  0.008& 0.185 &  0.014&  0.27 &   0.02\\ 
      &      & 2.50 & 3.00&       &       & 0.023 &  0.005& 0.135 &  0.013&  0.23 &   0.02\\ 
      &      & 3.00 & 3.50&       &       & 0.013 &  0.005& 0.060 &  0.010& 0.167 &  0.014\\ 
      &      & 3.50 & 4.00&       &       & 0.003 &  0.002& 0.030 &  0.007& 0.145 &  0.014\\ 
      &      & 4.00 & 5.00&       &       &    *  &    *  & 0.021 &  0.004& 0.084 &  0.007\\ 
      &      & 5.00 & 6.50&       &       &       &       & 0.004 &  0.001& 0.037 &  0.004\\ 
      &      & 6.50 & 8.00&       &       &       &       &       &       & 0.016 &  0.003\\ 
0.100 &0.150 & 0.50 & 1.00&  0.12 &   0.04&  0.16 &   0.02&  0.21 &   0.02&  0.21 &   0.03\\ 
      &      & 1.00 & 1.50& 0.024 &  0.015&  0.15 &   0.02&  0.22 &   0.02&  0.23 &   0.02\\ 
      &      & 1.50 & 2.00&    *  &    *  & 0.094 &  0.013&  0.20 &   0.02&  0.26 &   0.03\\ 
      &      & 2.00 & 2.50&    *  &    *  & 0.022 &  0.005& 0.123 &  0.014&  0.21 &   0.02\\ 
      &      & 2.50 & 3.00&       &       & 0.012 &  0.004& 0.049 &  0.007& 0.135 &  0.014\\ 
      &      & 3.00 & 3.50&       &       & 0.002 &  0.001& 0.027 &  0.006& 0.093 &  0.011\\ 
      &      & 3.50 & 4.00&       &       &    *  &    *  & 0.012 &  0.003& 0.061 &  0.008\\ 
      &      & 4.00 & 5.00&       &       &    *  &    *  & 0.004 &  0.001& 0.037 &  0.005\\ 
      &      & 5.00 & 6.50&       &       &       &       &    *  &     * & 0.010 &  0.002\\ 
      &      & 6.50 & 8.00&       &       &       &       &       &       & 0.003 &  0.001\\ 
0.150 &0.200 & 0.50 & 1.00&  0.09 &   0.03&  0.16 &   0.02&  0.22 &   0.03&  0.24 &   0.03\\ 
      &      & 1.00 & 1.50& 0.019 &  0.012& 0.094 &  0.014&  0.19 &   0.02&  0.21 &   0.02\\ 
      &      & 1.50 & 2.00&    *  &    *  & 0.042 &  0.008& 0.108 &  0.011&  0.15 &   0.02\\ 
      &      & 2.00 & 2.50&    *  &    *  & 0.019 &  0.006& 0.060 &  0.009& 0.091 &  0.012\\ 
      &      & 2.50 & 3.00&       &       & 0.005 &  0.003& 0.032 &  0.005& 0.050 &  0.008\\ 
      &      & 3.00 & 3.50&       &       &   *   &    *   & 0.019 &  0.004& 0.031 &  0.006\\ 
      &      & 3.50 & 4.00&       &       &   *   &    *   & 0.007 &  0.002& 0.024 &  0.005\\ 
      &      & 4.00 & 5.00&       &       &   *   &    *   & 0.002 &  0.001& 0.011 &  0.003\\ 
      &      & 5.00 & 6.50&       &       &       &       &    *   &    *   & 0.004 &  0.001\\ 
      &      & 6.50 & 8.00&       &       &       &       &       &       &   *    & *      \\ 
0.200 &0.250 & 0.50 & 1.00&  0.07 &   0.03&  0.12 &   0.02&  0.17 &   0.02&  0.17 &   0.03\\ 
      &      & 1.00 & 1.50& 0.012 &  0.021& 0.053 &  0.012&  0.12 &   0.02&  0.08 &   0.02\\ 
      &      & 1.50 & 2.00& 0.001 &  0.010& 0.024 &  0.007& 0.076 &  0.013&  0.09 &   0.02\\ 
      &      & 2.00 & 2.50&    *  &    *  & 0.010 &  0.004& 0.055 &  0.010& 0.044 &  0.010\\ 
      &      & 2.50 & 3.00&       &       & 0.007 &  0.003& 0.032 &  0.007& 0.028 &  0.008\\ 
      &      & 3.00 & 3.50&       &       & 0.002 &  0.001& 0.017 &  0.005& 0.018 &  0.005\\ 
      &      & 3.50 & 4.00&       &       & 0.000 &  0.001& 0.007 &  0.003& 0.008 &  0.004\\ 
      &      & 4.00 & 5.00&       &       &    *  &    *   & 0.003 &  0.002& 0.005 &  0.003\\ 
      &      & 5.00 & 6.50&       &       &       &       & 0.001 &  0.001& 0.002 &  0.001\\ 
      &      & 6.50 & 8.00&       &       &       &       &       &       & 0.001 &  0.003\\ 

%%%%%%%%%%%%%%%%%%%%%%%%%%%%%%%%%%%%%%%%%%%%%%%%%%%%%%%%%%%%%%%%%%%%%%%%%%%%%
\hline
\end{tabular}
}
\end{table*}

\begin{table*}[!ht]
  \caption{\label{tab:xsec_results_C}
    HARP results for the double-differential  $\pi^-$ production
    cross-section in the laboratory system,
    $d^2\sigma^{\pi}/(dpd\Omega)$, for p--C interactions at 3, 5, 8, 12~\GeVc.
    Each row refers to a
    different $(p_{\hbox{\small min}} \le p<p_{\hbox{\small max}},
    \theta_{\hbox{\small min}} \le \theta<\theta_{\hbox{\small max}})$ bin,
    where $p$ and $\theta$ are the pion momentum and polar angle, respectively.
    The central value as well as the square-root of the diagonal elements
    of the covariance matrix are given.}

%\begin{tabular}{rrrr|r@{$\pm$}lr{$\pm$}lr{$\pm$}lr{$\pm$}l}
\small{
\begin{tabular}{rrrr|r@{$\pm$}lr@{$\pm$}lr@{$\pm$}lr@{$\pm$}l}
\hline
$\theta_{\hbox{\small min}}$ &
$\theta_{\hbox{\small max}}$ &
$p_{\hbox{\small min}}$ &
$p_{\hbox{\small max}}$ &
\multicolumn{8}{c}{$d^2\sigma^{\pi^-}/(dpd\Omega)$}
\\
(rad) & (rad) & (\GeVc) & (\GeVc) &
\multicolumn{8}{c}{(barn/(sr \GeVc ))}
\\
  &  &  &
&\multicolumn{2}{c}{$ \bf{3 \ \GeVc}$}
&\multicolumn{2}{c}{$ \bf{5 \ \GeVc}$}
&\multicolumn{2}{c}{$ \bf{8 \ \GeVc}$}
&\multicolumn{2}{c}{$ \bf{12 \ \GeVc}$}
\\
\hline

0.050 &0.100 & 0.50 & 1.00&  0.02 &   0.02&  0.07 &   0.02&  0.13 &   0.02&  0.13 &   0.02\\ 
      &      & 1.00 & 1.50& 0.006 &  0.008& 0.062 &  0.010&  0.16 &   0.02&  0.20 &   0.02\\ 
      &      & 1.50 & 2.00& 0.002 &  0.003& 0.038 &  0.008& 0.108 &  0.011&  0.19 &   0.02\\ 
      &      & 2.00 & 2.50&    *  &     * & 0.015 &  0.004& 0.071 &  0.008& 0.148 &  0.015\\ 
      &      & 2.50 & 3.00&       &       & 0.005 &  0.002& 0.032 &  0.007& 0.102 &  0.013\\ 
      &      & 3.00 & 3.50&       &       &   *    &   *    & 0.024 &  0.005& 0.077 &  0.009\\ 
      &      & 3.50 & 4.00&       &       &   *    &   *    & 0.025 &  0.005& 0.056 &  0.008\\ 
      &      & 4.00 & 5.00&       &       &   *    &   *    & 0.008 &  0.002& 0.041 &  0.005\\ 
      &      & 5.00 & 6.50&       &       &       &       &   *    &   *    & 0.009 &  0.002\\ 
      &      & 6.50 & 8.00&       &       &       &       &       &       & 0.002 &  0.001\\ 
0.100 &0.150 & 0.50 & 1.00&  0.01 &   0.01&  0.10 &   0.02&  0.20 &   0.03&  0.24 &   0.03\\ 
      &      & 1.00 & 1.50& 0.002 &  0.004& 0.047 &  0.010& 0.144 &  0.015&  0.21 &   0.02\\ 
      &      & 1.50 & 2.00&   *    &  *     & 0.019 &  0.005& 0.080 &  0.009&  0.15 &   0.02\\ 
      &      & 2.00 & 2.50&   *    &  *     & 0.013 &  0.004& 0.050 &  0.007& 0.095 &  0.012\\ 
      &      & 2.50 & 3.00&       &       & 0.006 &  0.002& 0.028 &  0.005& 0.079 &  0.010\\ 
      &      & 3.00 & 3.50&       &       & 0.001 &  0.001& 0.015 &  0.003& 0.050 &  0.008\\ 
      &      & 3.50 & 4.00&       &       &   *    &  *     & 0.009 &  0.002& 0.028 &  0.004\\ 
      &      & 4.00 & 5.00&       &       &   *    &  *     & 0.003 &  0.001& 0.017 &  0.004\\ 
      &      & 5.00 & 6.50&       &       &       &       &    *   &     *  & 0.003 &  0.001\\ 
      &      & 6.50 & 8.00&       &       &       &       &       &       &    *   &    *    \\ 
0.150 &0.200 & 0.50 & 1.00&  0.03 &   0.02& 0.079 &  0.015&  0.16 &   0.02&  0.22 &   0.03\\ 
      &      & 1.00 & 1.50& 0.006 &  0.007& 0.041 &  0.009& 0.092 &  0.012&  0.13 &   0.02\\ 
      &      & 1.50 & 2.00& 0.001 &  0.003& 0.018 &  0.004& 0.059 &  0.009&  0.13 &   0.02\\ 
      &      & 2.00 & 2.50&   *    &  *     & 0.012 &  0.003& 0.036 &  0.006& 0.074 &  0.011\\ 
      &      & 2.50 & 3.00&       &       & 0.004 &  0.002& 0.025 &  0.005& 0.051 &  0.009\\ 
      &      & 3.00 & 3.50&       &       & 0.001 &  0.001& 0.008 &  0.002& 0.023 &  0.005\\ 
      &      & 3.50 & 4.00&       &       &   *    &  *     & 0.005 &  0.002& 0.011 &  0.003\\ 
      &      & 4.00 & 5.00&       &       &   *    &  *     &   *    &   *    & 0.007 &  0.002\\ 
      &      & 5.00 & 6.50&       &       &       &       &     *  &    *   &    *   &   *     \\ 
      &      & 6.50 & 8.00&       &       &       &       &       &       &   *    &  *  \\ 
0.200 &0.250 & 0.50 & 1.00&  0.05 &   0.03& 0.054 &  0.014&  0.13 &   0.02&  0.15 &   0.02\\ 
      &      & 1.00 & 1.50& 0.004 &  0.007& 0.041 &  0.010& 0.096 &  0.014&  0.12 &   0.02\\ 
      &      & 1.50 & 2.00&   *    &  *     & 0.011 &  0.005&  0.08 &   0.02&  0.11 &   0.02\\ 
      &      & 2.00 & 2.50&   *    & *      & 0.007 &  0.003& 0.024 &  0.007& 0.060 &  0.014\\ 
      &      & 2.50 & 3.00&       &       & 0.001 &  0.001& 0.009 &  0.003& 0.024 &  0.007\\ 
      &      & 3.00 & 3.50&       &       & 0.001 &  0.001& 0.003 &  0.002& 0.010 &  0.004\\ 
      &      & 3.50 & 4.00&       &       &   *    &   *    & 0.001 &  0.001& 0.005 &  0.003\\ 
      &      & 4.00 & 5.00&       &       &   *    &   *    & 0.000 &  0.001& 0.001 &  0.001\\ 
      &      & 5.00 & 6.50&       &       &       &       & 0.000 &  0.001& 0.001 &  0.001\\ 
      &      & 6.50 & 8.00&       &       &       &       &       &       & 0.000 &  0.001\\ 
%%%%%%%%%%%%%%%%%%%%%%%%%%%%%%%%%%%%%%%%%%%%%%%%%%%%%%%%%%%%%%%%%%%%%%%%%%%%%
\hline
\end{tabular}
}
\end{table*}

\begin{table*}[!ht]
  \caption{\label{tab:xsec_results_N2}
    HARP results for the double-differential $\pi^{\pm}$  production
    cross-section in the laboratory system,
    $d^2\sigma^{\pi}/(dpd\Omega)$, for $p-N_2$ interactions at 12~\GeVc.
    Each row refers to a
    different $(p_{\hbox{\small min}} \le p<p_{\hbox{\small max}},
    \theta_{\hbox{\small min}} \le \theta<\theta_{\hbox{\small max}})$ bin,
    where $p$ and $\theta$ are the pion momentum and polar angle, respectively.
    The central value as well as the square-root of the diagonal elements
    of the covariance matrix are given.}

\begin{center}
%\begin{tabular}{rrrr|r@{$\pm$}lr{$\pm$}lr{$\pm$}lr{$\pm$}l}
\small{
\begin{tabular}{rrrr|r@{$\pm$}lr@{$\pm$}lr@{$\pm$}lr@{$\pm$}lr@{$\pm$}lr@{$\pm$}l}
\hline
$\theta_{\hbox{\small min}}$ &
$\theta_{\hbox{\small max}}$ &
$p_{\hbox{\small min}}$ &
$p_{\hbox{\small max}}$ &
\multicolumn{2}{c}{$d^2\sigma^{\pi^+}/(dpd\Omega)$} &
\multicolumn{2}{c}{$d^2\sigma^{\pi^-}/(dpd\Omega)$}
\\
(rad) & (rad) & (\GeVc) & (\GeVc) &
\multicolumn{2}{c}{(barn/(sr \GeVc))} &
\multicolumn{2}{c}{(barn/(sr \GeVc))}
\\
  &  &  &
&\multicolumn{2}{c}{$ \bf{12 \ \GeVc}$} 
&\multicolumn{2}{c}{$ \bf{12 \ \GeVc}$}
\\
\hline

0.05 & 0.10 & 0.50 & 1.00&  0.24 &   0.04  & 0.20 & 0.03\\ 
      &      & 1.00 & 1.50&  0.32 &   0.03 & 0.26 & 0.03 \\ 
      &      & 1.50 & 2.00&  0.29 &   0.03 & 0.20 & 0.02 \\ 
      &      & 2.00 & 2.50&  0.31 &   0.03 & 0.17 & 0.02 \\ 
      &      & 2.50 & 3.00&  0.28 &   0.02 & 0.127 & 0.015 \\ 
      &      & 3.00 & 3.50&  0.18 &   0.02 & 0.071 & 0.010 \\ 
      &      & 3.50 & 4.00&  0.15 &   0.02 & 0.065 & 0.009 \\ 
      &      & 4.00 & 5.00& 0.098 &  0.009 & 0.037 & 0.006 \\ 
      &      & 5.00 & 6.50& 0.046 &  0.006 & 0.014 & 0.003 \\ 
      &      & 6.50 & 8.00& 0.014 &  0.003 & 0.002 & 0.001 \\ 
 0.10 & 0.15 & 0.50 & 1.00&  0.24 &   0.04 & 0.30  & 0.04 \\ 
      &      & 1.00 & 1.50&  0.34 &   0.03 & 0.22  & 0.02 \\ 
      &      & 1.50 & 2.00&  0.25 &   0.03 & 0.18  & 0.02 \\ 
      &      & 2.00 & 2.50&  0.23 &   0.03 & 0.08  & 0.02 \\ 
      &      & 2.50 & 3.00&  0.15 &   0.02 & 0.079 & 0.012 \\ 
      &      & 3.00 & 3.50& 0.102 &  0.013 & 0.063 & 0.009 \\ 
      &      & 3.50 & 4.00& 0.062 &  0.010 & 0.039 & 0.006 \\ 
      &      & 4.00 & 5.00& 0.034 &  0.005 & 0.014 & 0.003 \\ 
      &      & 5.00 & 6.50& 0.011 &  0.002 & 0.004 & 0.001 \\ 
      &      & 6.50 & 8.00& 0.002 &  0.001 &     * & * \\ 
 0.15 & 0.20 & 0.50 & 1.00&  0.34 &   0.05 & 0.27  & 0.04 \\ 
      &      & 1.00 & 1.50&  0.24 &   0.03 & 0.15  & 0.02 \\ 
      &      & 1.50 & 2.00&  0.15 &   0.02 & 0.12  & 0.02 \\ 
      &      & 2.00 & 2.50&  0.11 &   0.02 & 0.084 & 0.015 \\ 
      &      & 2.50 & 3.00& 0.076 &  0.014 & 0.042 & 0.011 \\ 
      &      & 3.00 & 3.50& 0.036 &  0.008 & 0.017 & 0.007 \\ 
      &      & 3.50 & 4.00& 0.028 &  0.005 & 0.020 & 0.005 \\ 
      &      & 4.00 & 5.00& 0.018 &  0.003 & 0.006 & 0.002 \\ 
      &      & 5.00 & 6.50& 0.005 &  0.002 &       & \\ 
      &      & 6.50 & 8.00& 0.001 &  0.000 &       & \\ 
 0.20 & 0.25 & 0.50 & 1.00&  0.21 &   0.04 & 0.15  & 0.03 \\ 
      &      & 1.00 & 1.50&  0.08 &   0.02 & 0.19  & 0.03 \\ 
      &      & 1.50 & 2.00&  0.10 &   0.02 & 0.12  & 0.03 \\ 
      &      & 2.00 & 2.50& 0.063 &  0.014 & 0.08  & 0.02  \\ 
      &      & 2.50 & 3.00& 0.048 &  0.011 & 0.050 & 0.012 \\ 
      &      & 3.00 & 3.50& 0.022 &  0.006 & 0.015 & 0.005 \\ 
      &      & 3.50 & 4.00& 0.009 &  0.004 & 0.006 & 0.004 \\ 
      &      & 4.00 & 5.00& 0.008 &  0.003 & 0.001 & 0.002 \\ 
      &      & 5.00 & 6.50& 0.002 &  0.001 & 0.000 & 0.001 \\ 
      &      & 6.50 & 8.00& 0.001 &  0.002 & 0.000 & 0.001 \\ 

%%%%%%%%%%%%%%%%%%%%%%%%%%%%%%%%%%%%%%%%%%%%%%%%%%%%%%%%%%%%%%%%%%%%%%%%%%%%%
\hline
\end{tabular}
}
\end{center}
\end{table*}

\begin{table*}[!ht]
  \caption{\label{tab:xsec_results_O2}
    HARP results for the double-differential $\pi^{\pm}$  production
    cross-section in the laboratory system,
    $d^2\sigma^{\pi}/(dpd\Omega)$, for $p-O_2$ interactions at 12~\GeVc.
    Each row refers to a
    different $(p_{\hbox{\small min}} \le p<p_{\hbox{\small max}},
    \theta_{\hbox{\small min}} \le \theta<\theta_{\hbox{\small max}})$ bin,
    where $p$ and $\theta$ are the pion momentum and polar angle, respectively.
    The central value as well as the square-root of the diagonal elements
    of the covariance matrix are given.}

\begin{center}
%\begin{tabular}{rrrr|r@{$\pm$}l}
\small{
\begin{tabular}{rrrr|r@{$\pm$}lr@{$\pm$}l}
\hline
$\theta_{\hbox{\small min}}$ &
$\theta_{\hbox{\small max}}$ &
$p_{\hbox{\small min}}$ &
$p_{\hbox{\small max}}$ &
\multicolumn{2}{c}{$d^2\sigma^{\pi^+}/(dpd\Omega)$} &
\multicolumn{2}{c}{$d^2\sigma^{\pi^-}/(dpd\Omega)$}
\\
(rad) & (rad) & (\GeVc) & (\GeVc) &
\multicolumn{2}{c}{(barn/(sr \GeVc)} &
\multicolumn{2}{c}{(barn/(sr \GeVc)}
\\
  &  &  &
&\multicolumn{2}{c}{$ \bf{12 \ \GeVc}$} &
\multicolumn{2}{c}{$ \bf{12 \ \GeVc}$}
\\
\hline 
0.05 & 0.10 & 0.50 & 1.00&  0.29 &   0.06  & 0.29 & 0.06\\ 
      &      & 1.00 & 1.50&  0.42 &   0.06 & 0.31 & 0.05 \\ 
      &      & 1.50 & 2.00&  0.44 &   0.05 & 0.24 & 0.04 \\ 
      &      & 2.00 & 2.50&  0.37 &   0.05 & 0.17 & 0.03 \\ 
      &      & 2.50 & 3.00&  0.31 &   0.04 & 0.14 & 0.03 \\ 
      &      & 3.00 & 3.50&  0.27 &   0.04 & 0.12 & 0.02  \\ 
      &      & 3.50 & 4.00&  0.18 &   0.03 & 0.07 & 0.02 \\ 
      &      & 4.00 & 5.00&  0.13 &   0.02 & 0.042 & 0.010\\ 
      &      & 5.00 & 6.50& 0.063 &  0.010 & 0.021 & 0.006 \\ 
      &      & 6.50 & 8.00& 0.025 &  0.007 & 0.007 & 0.003 \\ 
 0.10 & 0.15 & 0.50 & 1.00&  0.33 &   0.06 & 0.46  & 0.08 \\ 
      &      & 1.00 & 1.50&  0.49 &   0.07 & 0.33  & 0.05 \\ 
      &      & 1.50 & 2.00&  0.32 &   0.05 & 0.24  & 0.04 \\ 
      &      & 2.00 & 2.50&  0.33 &   0.05 & 0.11  & 0.03 \\ 
      &      & 2.50 & 3.00&  0.22 &   0.04 & 0.14  & 0.03 \\ 
      &      & 3.00 & 3.50&  0.11 &   0.02 & 0.07  & 0.02 \\ 
      &      & 3.50 & 4.00&  0.10 &   0.02 & 0.06  & 0.02 \\ 
      &      & 4.00 & 5.00& 0.052 &  0.011 & 0.024 & 0.008 \\ 
      &      & 5.00 & 6.50& 0.015 &  0.005 & 0.004 & 0.002 \\ 
      &      & 6.50 & 8.00& 0.002 &  0.001 & 0.001 & 0.001 \\ 
 0.15 & 0.20 & 0.50 & 1.00&  0.34 &   0.07 & 0.24  & 0.06 \\ 
      &      & 1.00 & 1.50&  0.31 &   0.05 & 0.20  & 0.04 \\ 
      &      & 1.50 & 2.00&  0.23 &   0.04 & 0.11  & 0.03 \\ 
      &      & 2.00 & 2.50&  0.14 &   0.03 & 0.10  & 0.03 \\ 
      &      & 2.50 & 3.00&  0.08 &   0.02 & 0.05  & 0.02 \\ 
      &      & 3.00 & 3.50&  0.06 &   0.02 & 0.015 & 0.010 \\ 
      &      & 3.50 & 4.00& 0.029 &  0.011 & 0.027 & 0.013 \\ 
      &      & 4.00 & 5.00& 0.020 &  0.008 & 0.011 & 0.007 \\ 
      &      & 5.00 & 6.50& 0.006 &  0.004 & 0.001 & 0.002 \\ 
      &      & 6.50 & 8.00& 0.001 &  0.001 & 0.000 & 0.001 \\ 
 0.20 & 0.25 & 0.50 & 1.00&  0.23 &   0.06 & 0.19  & 0.05 \\ 
      &      & 1.00 & 1.50&  0.14 &   0.04 & 0.20  & 0.06 \\ 
      &      & 1.50 & 2.00&  0.13 &   0.04 & 0.12  & 0.04 \\ 
      &      & 2.00 & 2.50&  0.07 &   0.03 & 0.12  & 0.04 \\ 
      &      & 2.50 & 3.00&  0.04 &   0.02 & 0.06  & 0.03 \\ 
      &      & 3.00 & 3.50&  0.04 &   0.02 & 0.022 & 0.015 \\ 
      &      & 3.50 & 4.00& 0.018 &  0.013 & 0.015 & 0.015 \\ 
      &      & 4.00 & 5.00& 0.011 &  0.009 & 0.005 & 0.006 \\ 
      &      & 5.00 & 6.50& 0.003 &  0.005 & 0.002 & 0.007 \\ 
      &      & 6.50 & 8.00& 0.002 &  0.005 & 0.001 & 0.007 \\ 

\hline
\end{tabular}
}
\end{center}
\end{table*}

\begin{table*}[!ht]
  \caption{\label{tab:xsec_results_Al}
    HARP results for the double-differential $\pi^+$  production
    cross-section in the laboratory system,
    $d^2\sigma^{\pi}/(dpd\Omega)$, for p--Al interactions at 3, 5, 8, 12, 12.9~\GeVc.
    Each row refers to a
    different $(p_{\hbox{\small min}} \le p<p_{\hbox{\small max}},
    \theta_{\hbox{\small min}} \le \theta<\theta_{\hbox{\small max}})$ bin,
    where $p$ and $\theta$ are the pion momentum and polar angle, respectively.
    The central value as well as the square-root of the diagonal elements
    of the covariance matrix are given.}

%\begin{tabular}{rrrr|r@{$\pm$}lr{$\pm$}lr{$\pm$}lr{$\pm$}l}
\small{
\begin{tabular}{rrrr|r@{$\pm$}lr@{$\pm$}lr@{$\pm$}lr@{$\pm$}lr@{$\pm$}l}
\hline
$\theta_{\hbox{\small min}}$ &
$\theta_{\hbox{\small max}}$ &
$p_{\hbox{\small min}}$ &
$p_{\hbox{\small max}}$ &
\multicolumn{10}{c}{$d^2\sigma^{\pi^+}/(dpd\Omega)$}
\\
(rad) & (rad) & (\GeVc) & (\GeVc) &
\multicolumn{10}{c}{(barn/(sr \GeVc))}
\\
  &  &  &
&\multicolumn{2}{c}{$ \bf{3 \ \GeVc}$}
&\multicolumn{2}{c}{$ \bf{5 \ \GeVc}$}
&\multicolumn{2}{c}{$ \bf{8 \ \GeVc}$}
&\multicolumn{2}{c}{$ \bf{12 \ \GeVc}$}
&\multicolumn{2}{c}{$ \bf{12.9 \ \GeVc}$}
\\
\hline

0.050 &0.100 & 0.50 & 1.00&  0.19 &   0.06&  0.17 &   0.03&  0.34 &   0.04&  0.47 &   0.06&  0.49 &   0.04\\ 
      &      & 1.00 & 1.50&  0.02 &   0.02&  0.27 &   0.03&  0.35 &   0.03&  0.42 &   0.04&  0.46 &   0.03\\ 
      &      & 1.50 & 2.00&     * &      *&  0.20 &   0.03&  0.36 &   0.03&  0.48 &   0.04&  0.52 &   0.03\\ 
      &      & 2.00 & 2.50&     * &      *& 0.064 &  0.012&  0.31 &   0.02&  0.47 &   0.04&  0.47 &   0.02\\ 
      &      & 2.50 & 3.00&       &       & 0.038 &  0.010&  0.23 &   0.02&  0.44 &   0.04&  0.41 &   0.02\\ 
      &      & 3.00 & 3.50&       &       & 0.006 &  0.005&  0.18 &   0.02&  0.28 &   0.03&  0.36 &   0.02\\ 
      &      & 3.50 & 4.00&       &       & 0.004 &  0.003& 0.109 &  0.014&  0.23 &   0.03&  0.27 &   0.02\\ 
      &      & 4.00 & 5.00&       &       & 0.001 &  0.001& 0.077 &  0.010& 0.143 &  0.014& 0.169 &  0.011\\ 
      &      & 5.00 & 6.50&       &       &       &       & 0.006 &  0.002& 0.048 &  0.007& 0.077 &  0.005\\ 
      &      & 6.50 & 8.00&       &       &       &       &       &       & 0.021 &  0.004& 0.029 &  0.003\\ 
0.100 &0.150 & 0.50 & 1.00&  0.13 &   0.05&  0.24 &   0.04&  0.36 &   0.04&  0.44 &   0.06&  0.49 &   0.04\\ 
      &      & 1.00 & 1.50&  0.01 &   0.01&  0.22 &   0.03&  0.38 &   0.04&  0.50 &   0.05&  0.53 &   0.04\\ 
      &      & 1.50 & 2.00&     * &     * &  0.15 &   0.02&  0.32 &   0.03&  0.46 &   0.05&  0.47 &   0.03\\ 
      &      & 2.00 & 2.50&     * &     * & 0.048 &  0.011&  0.19 &   0.02&  0.32 &   0.04&  0.35 &   0.03\\ 
      &      & 2.50 & 3.00&       &       & 0.015 &  0.007&  0.13 &   0.02&  0.22 &   0.03&  0.28 &   0.02\\ 
      &      & 3.00 & 3.50&       &       & 0.003 &  0.001& 0.091 &  0.013&  0.18 &   0.02&  0.18 &   0.02\\ 
      &      & 3.50 & 4.00&       &       & 0.001 &  0.001& 0.053 &  0.008&  0.09 &   0.02& 0.128 &  0.011\\ 
      &      & 4.00 & 5.00&       &       &     * &      *& 0.020 &  0.005& 0.050 &  0.008& 0.070 &  0.007\\ 
      &      & 5.00 & 6.50&       &       &       &       & 0.001 &  0.000& 0.017 &  0.003& 0.021 &  0.003\\ 
      &      & 6.50 & 8.00&       &       &       &       &       &       & 0.004 &  0.001& 0.004 &  0.001\\ 
0.150 &0.200 & 0.50 & 1.00&  0.19 &   0.07&  0.26 &   0.04&  0.39 &   0.05&  0.40 &   0.06&  0.50 &   0.05\\ 
      &      & 1.00 & 1.50&  0.03 &   0.02&  0.17 &   0.03&  0.31 &   0.03&  0.34 &   0.04&  0.35 &   0.03\\ 
      &      & 1.50 & 2.00&     * &      *& 0.054 &  0.013&  0.17 &   0.02&  0.29 &   0.03&  0.26 &   0.02\\ 
      &      & 2.00 & 2.50&     * &     * & 0.023 &  0.010& 0.098 &  0.014&  0.17 &   0.02&  0.17 &   0.02\\ 
      &      & 2.50 & 3.00&       &       & 0.005 &  0.004& 0.058 &  0.010&  0.10 &   0.02& 0.117 &  0.011\\ 
      &      & 3.00 & 3.50&       &       & 0.001 &  0.001& 0.039 &  0.007& 0.046 &  0.011& 0.072 &  0.008\\ 
      &      & 3.50 & 4.00&       &       &     * &     * & 0.023 &  0.006& 0.033 &  0.008& 0.047 &  0.006\\ 
      &      & 4.00 & 5.00&       &       &     * &     * & 0.008 &  0.002& 0.025 &  0.007& 0.021 &  0.004\\ 
      &      & 5.00 & 6.50&       &       &       &       &     * &      *& 0.005 &  0.003& 0.005 &  0.001\\ 
      &      & 6.50 & 8.00&       &       &       &       &       &       &     * &      *&     * &     * \\ 
0.200 &0.250 & 0.50 & 1.00&  0.05 &   0.04&  0.16 &   0.03&  0.24 &   0.03&  0.27 &   0.05&  0.33 &   0.04\\ 
      &      & 1.00 & 1.50&  0.01 &   0.02&  0.08 &   0.02&  0.18 &   0.03&  0.17 &   0.04&  0.21 &   0.02\\ 
      &      & 1.50 & 2.00&     * &     * & 0.039 &  0.014&  0.17 &   0.03&  0.10 &   0.03&  0.15 &   0.02\\ 
      &      & 2.00 & 2.50&     * &     *  & 0.020 &  0.009&  0.07 &   0.02&  0.10 &   0.03& 0.103 &  0.014\\ 
      &      & 2.50 & 3.00&       &       & 0.004 &  0.003& 0.037 &  0.010&  0.05 &   0.02& 0.076 &  0.010\\ 
      &      & 3.00 & 3.50&       &       & 0.001 &  0.001& 0.025 &  0.007& 0.031 &  0.010& 0.038 &  0.006\\ 
      &      & 3.50 & 4.00&       &       & 0.001 &  0.001& 0.013 &  0.005& 0.018 &  0.009& 0.021 &  0.004\\ 
      &      & 4.00 & 5.00&       &       & 0.001 &  0.002& 0.007 &  0.004& 0.012 &  0.008& 0.015 &  0.003\\ 
      &      & 5.00 & 6.50&       &       &       &       & 0.001 &  0.002& 0.005 &  0.004& 0.004 &  0.001\\ 
      &      & 6.50 & 8.00&       &       &       &       &       &       & 0.002 &  0.007& 0.001 &  0.001\\ 
%%%%%%%%%%%%%%%%%%%%%%%%%%%%%%%%%%%%%%%%%%%%%%%%%%%%%%%%%%%%%%%%%%%%%%%%%%%%%
\hline
\end{tabular}
}
\end{table*}
\begin{table*}[!ht]
  \caption{\label{tab:xsec_results_Al}
    HARP results for the double-differential  $\pi^-$ production
    cross-section in the laboratory system,
    $d^2\sigma^{\pi}/(dpd\Omega)$, for p--Al interactions at 3, 5, 8, 12, 12.9~\GeVc.
    Each row refers to a
    different $(p_{\hbox{\small min}} \le p<p_{\hbox{\small max}},
    \theta_{\hbox{\small min}} \le \theta<\theta_{\hbox{\small max}})$ bin,
    where $p$ and $\theta$ are the pion momentum and polar angle, respectively.
    The central value as well as the square-root of the diagonal elements
    of the covariance matrix are given.}

%\begin{tabular}{rrrr|r@{$\pm$}lr{$\pm$}lr{$\pm$}lr{$\pm$}l}
\small{
\begin{tabular}{rrrr|r@{$\pm$}lr@{$\pm$}lr@{$\pm$}lr@{$\pm$}lr@{$\pm$}l}
\hline
$\theta_{\hbox{\small min}}$ &
$\theta_{\hbox{\small max}}$ &
$p_{\hbox{\small min}}$ &
$p_{\hbox{\small max}}$ &
\multicolumn{10}{c}{$d^2\sigma^{\pi^-}/(dpd\Omega)$}
\\
(rad) & (rad) & (\GeVc) & (\GeVc) &
\multicolumn{10}{c}{(barn/(sr \GeVc))}
\\
  &  &  &
&\multicolumn{2}{c}{$ \bf{3 \ \GeVc}$}
&\multicolumn{2}{c}{$ \bf{5 \ \GeVc}$}
&\multicolumn{2}{c}{$ \bf{8 \ \GeVc}$}
&\multicolumn{2}{c}{$ \bf{12 \ \GeVc}$}
&\multicolumn{2}{c}{$ \bf{12.9 \ \GeVc}$}
\\
\hline

0.050 &0.100 & 0.50 & 1.00&  0.03 &   0.03&  0.14 &   0.03&  0.28 &   0.04&  0.38 &   0.05&  0.37 &   0.04\\ 
      &      & 1.00 & 1.50&    *  &     * &  0.10 &   0.02&  0.21 &   0.02&  0.35 &   0.04&  0.40 &   0.03\\ 
      &      & 1.50 & 2.00& 0.001 &  0.003& 0.050 &  0.013&  0.19 &   0.02&  0.34 &   0.04&  0.36 &   0.02\\ 
      &      & 2.00 & 2.50&    *  &     * & 0.023 &  0.008& 0.117 &  0.014&  0.27 &   0.03&  0.29 &   0.02\\ 
      &      & 2.50 & 3.00&       &       & 0.007 &  0.004& 0.066 &  0.013&  0.17 &   0.02&  0.20 &   0.02\\ 
      &      & 3.00 & 3.50&       &       &    *  & *    & 0.040 &  0.007&  0.14 &   0.02& 0.179 &  0.011\\ 
      &      & 3.50 & 4.00&       &       &    *  &  *   & 0.032 &  0.006&  0.11 &   0.02& 0.115 &  0.010\\ 
      &      & 4.00 & 5.00&       &       &       &  *   & 0.011 &  0.004& 0.048 &  0.009& 0.069 &  0.006\\ 
      &      & 5.00 & 6.50&       &       &       &       & 0.001 &  0.001& 0.020 &  0.004& 0.028 &  0.003\\ 
      &      & 6.50 & 8.00&       &       &       &       &       &       & 0.003 &  0.002& 0.007 &  0.002\\ 
0.100 &0.150 & 0.50 & 1.00&  0.03 &   0.03&  0.16 &   0.03&  0.32 &   0.04&  0.48 &   0.07&  0.46 &   0.05\\ 
      &      & 1.00 & 1.50&    *  &     * &  0.08 &   0.02&  0.22 &   0.02&  0.39 &   0.04&  0.38 &   0.03\\ 
      &      & 1.50 & 2.00& 0.002 &  0.005& 0.038 &  0.010&  0.14 &   0.02&  0.24 &   0.03&  0.30 &   0.02\\ 
      &      & 2.00 & 2.50&    *  &     * & 0.017 &  0.006& 0.080 &  0.012&  0.19 &   0.03&  0.23 &   0.02\\ 
      &      & 2.50 & 3.00&       &       & 0.011 &  0.004& 0.038 &  0.007&  0.14 &   0.02& 0.172 &  0.015\\ 
      &      & 3.00 & 3.50&       &       & 0.002 &  0.001& 0.026 &  0.006& 0.095 &  0.015& 0.103 &  0.009\\ 
      &      & 3.50 & 4.00&       &       &    *  &    * & 0.014 &  0.003& 0.066 &  0.012& 0.081 &  0.008\\ 
      &      & 4.00 & 5.00&       &       &    *  &    * & 0.003 &  0.002& 0.032 &  0.007& 0.038 &  0.005\\ 
      &      & 5.00 & 6.50&       &       &       &       &   *  &     * & 0.008 &  0.002& 0.007 &  0.002\\ 
      &      & 6.50 & 8.00&       &       &       &       &       &       & 0.001 &  0.000& 0.001 &  0.000\\ 
0.150 &0.200 & 0.50 & 1.00&  0.07 &   0.04&  0.18 &   0.03&  0.35 &   0.04&  0.49 &   0.07&  0.43 &   0.04\\ 
      &      & 1.00 & 1.50&  0.02 &   0.02&  0.08 &   0.02&  0.17 &   0.02&  0.25 &   0.04&  0.35 &   0.03\\ 
      &      & 1.50 & 2.00&   *   &    *  & 0.027 &  0.008& 0.093 &  0.014&  0.19 &   0.03&  0.22 &   0.02\\ 
      &      & 2.00 & 2.50&   *   &    *  & 0.014 &  0.005& 0.065 &  0.011&  0.12 &   0.02& 0.156 &  0.015\\ 
      &      & 2.50 & 3.00&       &       & 0.005 &  0.004& 0.040 &  0.008& 0.064 &  0.013& 0.084 &  0.010\\ 
      &      & 3.00 & 3.50&       &       & 0.000 &  0.001& 0.013 &  0.004& 0.063 &  0.014& 0.046 &  0.006\\ 
      &      & 3.50 & 4.00&       &       & *     &  *    & 0.005 &  0.002& 0.022 &  0.007& 0.026 &  0.004\\ 
      &      & 4.00 & 5.00&       &       & *     &  *    & 0.001 &  0.001& 0.007 &  0.003& 0.012 &  0.002\\ 
      &      & 5.00 & 6.50&       &       &       &       &    *  &     * & 0.001 &  0.001& 0.001 &  0.001\\ 
      &      & 6.50 & 8.00&       &       &       &       &       &       &    *  &     * &    *  &    *  \\ 
0.200 &0.250 & 0.50 & 1.00&  0.03 &   0.03&  0.12 &   0.03&  0.21 &   0.03&  0.29 &   0.05&  0.34 &   0.04\\ 
      &      & 1.00 & 1.50&  0.02 &   0.03&  0.07 &   0.02&  0.19 &   0.03&  0.26 &   0.04&  0.28 &   0.03\\ 
      &      & 1.50 & 2.00& 0.000 &  0.001& 0.010 &  0.007&  0.14 &   0.03&  0.16 &   0.04&  0.26 &   0.04\\ 
      &      & 2.00 & 2.50&     * &     * & 0.005 &  0.004&  0.06 &   0.02&  0.11 &   0.03&  0.13 &   0.02\\ 
      &      & 2.50 & 3.00&       &       & 0.002 &  0.002& 0.022 &  0.007& 0.028 &  0.012& 0.064 &  0.012\\ 
      &      & 3.00 & 3.50&       &       &     * &     * & 0.009 &  0.004& 0.016 &  0.007& 0.031 &  0.006\\ 
      &      & 3.50 & 4.00&       &       &     * &     * & 0.002 &  0.002& 0.013 &  0.006& 0.013 &  0.003\\ 
      &      & 4.00 & 5.00&       &       &     * &     * & 0.001 &  0.003& 0.004 &  0.004& 0.005 &  0.001\\ 
      &      & 5.00 & 6.50&       &       &       &       & 0.000 &  0.001& 0.000 &  0.001& 0.001 &  0.001\\ 
      &      & 6.50 & 8.00&       &       &       &       &       &       & 0.000 &  0.004& 0.001 &  0.001\\ 
%%%%%%%%%%%%%%%%%%%%%%%%%%%%%%%%%%%%%%%%%%%%%%%%%%%%%%%%%%%%%%%%%%%%%%%%%%%%%
\hline
\end{tabular}
}
\end{table*}

\begin{table*}[!ht]
  \caption{\label{tab:xsec_results_Cu}
    HARP results for the double-differential $\pi^+$  production
    cross-section in the laboratory system,
    $d^2\sigma^{\pi}/(dpd\Omega)$, for p--Cu interactions at 3, 5, 8, 12~\GeVc.
    Each row refers to a
    different $(p_{\hbox{\small min}} \le p<p_{\hbox{\small max}},
    \theta_{\hbox{\small min}} \le \theta<\theta_{\hbox{\small max}})$ bin,
    where $p$ and $\theta$ are the pion momentum and polar angle, respectively.
    The central value as well as the square-root of the diagonal elements
    of the covariance matrix are given.}

%\begin{tabular}{rrrr|r@{$\pm$}lr{$\pm$}lr{$\pm$}lr{$\pm$}l}
\small{
\begin{tabular}{rrrr|r@{$\pm$}lr@{$\pm$}lr@{$\pm$}lr@{$\pm$}l}
\hline
$\theta_{\hbox{\small min}}$ &
$\theta_{\hbox{\small max}}$ &
$p_{\hbox{\small min}}$ &
$p_{\hbox{\small max}}$ &
\multicolumn{8}{c}{$d^2\sigma^{\pi^+}/(dpd\Omega)$}
\\
(rad) & (rad) & (\GeVc) & (\GeVc) &
\multicolumn{8}{c}{(barn/(sr \GeVc))}
\\
  &  &  &
&\multicolumn{2}{c}{$ \bf{3 \ \GeVc}$}
&\multicolumn{2}{c}{$ \bf{5 \ \GeVc}$}
&\multicolumn{2}{c}{$ \bf{8 \ \GeVc}$}
&\multicolumn{2}{c}{$ \bf{12 \ \GeVc}$}
\\
\hline

0.050 &0.100 & 0.50 & 1.00&  0.33 &   0.13&  0.30 &   0.06&  0.55 &   0.07&  0.77 &   0.10\\ 
      &      & 1.00 & 1.50&  0.05 &   0.04&  0.40 &   0.05&  0.59 &   0.05&  0.76 &   0.07\\ 
      &      & 1.50 & 2.00&  0.02 &   0.02&  0.25 &   0.04&  0.56 &   0.04&  0.81 &   0.07\\ 
      &      & 2.00 & 2.50&     * &      *&  0.10 &   0.02&  0.44 &   0.04&  0.75 &   0.06\\ 
      &      & 2.50 & 3.00&       &       & 0.025 &  0.011&  0.26 &   0.03&  0.61 &   0.05\\ 
      &      & 3.00 & 3.50&       &       & 0.025 &  0.013&  0.13 &   0.02&  0.53 &   0.05\\ 
      &      & 3.50 & 4.00&       &       & 0.021 &  0.012&  0.05 &   0.02&  0.36 &   0.04\\ 
      &      & 4.00 & 5.00&       &       & 0.001 &  0.002& 0.036 &  0.010&  0.22 &   0.02\\ 
      &      & 5.00 & 6.50&       &       &       &       & 0.005 &  0.002& 0.087 &  0.012\\ 
      &      & 6.50 & 8.00&       &       &       &       &       &       & 0.028 &  0.006\\ 
0.100 &0.150 & 0.50 & 1.00&  0.19 &   0.10&  0.49 &   0.07&  0.63 &   0.07&  0.90 &   0.10\\ 
      &      & 1.00 & 1.50&  0.07 &   0.06&  0.36 &   0.05&  0.60 &   0.06&  0.84 &   0.08\\ 
      &      & 1.50 & 2.00&  0.002&  0.006&  0.20 &   0.03&  0.48 &   0.04&  0.72 &   0.07\\ 
      &      & 2.00 & 2.50&     * &     * &  0.05 &   0.02&  0.28 &   0.03&  0.53 &   0.06\\ 
      &      & 2.50 & 3.00&       &       & 0.012 &  0.007&  0.12 &   0.02&  0.37 &   0.04\\ 
      &      & 3.00 & 3.50&       &       & 0.007 &  0.003& 0.057 &  0.013&  0.27 &   0.03\\ 
      &      & 3.50 & 4.00&       &       & 0.001 &  0.001& 0.022 &  0.006&  0.18 &   0.02\\ 
      &      & 4.00 & 5.00&       &       &    *  &    *  & 0.014 &  0.004&  0.09 &   0.02\\ 
      &      & 5.00 & 6.50&       &       &       &       & 0.002 &  0.001& 0.015 &  0.004\\ 
      &      & 6.50 & 8.00&       &       &       &       &       &       & 0.003 &  0.001\\ 
0.150 &0.200 & 0.50 & 1.00&  0.26 &   0.13&  0.47 &   0.07&  0.61 &   0.07&  0.86 &   0.11\\ 
      &      & 1.00 & 1.50&  0.05 &   0.05&  0.25 &   0.04&  0.42 &   0.04&  0.64 &   0.07\\ 
      &      & 1.50 & 2.00&  0.00 &   0.01&  0.09 &   0.02&  0.28 &   0.03&  0.39 &   0.05\\ 
      &      & 2.00 & 2.50&    *  &    *  & 0.016 &  0.011&  0.15 &   0.02&  0.31 &   0.04\\ 
      &      & 2.50 & 3.00&       &       & 0.006 &  0.006& 0.076 &  0.015&  0.14 &   0.03\\ 
      &      & 3.00 & 3.50&       &       & 0.004 &  0.003& 0.040 &  0.010&  0.13 &   0.02\\ 
      &      & 3.50 & 4.00&       &       & 0.000 &  0.001& 0.009 &  0.004&  0.07 &   0.02\\ 
      &      & 4.00 & 5.00&       &       &    *  &     * & 0.010 &  0.004& 0.019 &  0.006\\ 
      &      & 5.00 & 6.50&       &       &       &       & 0.001 &  0.001& 0.005 &  0.002\\ 
      &      & 6.50 & 8.00&       &       &       &       &       &       & 0.002 &  0.001\\ 
0.200 &0.250 & 0.50 & 1.00&  0.04 &   0.07&  0.32 &   0.06&  0.47 &   0.06&  0.59 &   0.09\\ 
      &      & 1.00 & 1.50&  0.03 &   0.13&  0.14 &   0.04&  0.24 &   0.04&  0.33 &   0.06\\ 
      &      & 1.50 & 2.00&    *  &     * &  0.04 &   0.02&  0.21 &   0.04&  0.14 &   0.03\\ 
      &      & 2.00 & 2.50&    *  &     * &  0.04 &   0.02&  0.09 &   0.02&  0.12 &   0.03\\ 
      &      & 2.50 & 3.00&       &       & 0.002 &  0.004& 0.028 &  0.011&  0.11 &   0.03\\ 
      &      & 3.00 & 3.50&       &       & 0.001 &  0.001& 0.018 &  0.008&  0.03 &   0.02\\ 
      &      & 3.50 & 4.00&       &       & 0.001 &  0.002& 0.017 &  0.007& 0.010 &  0.011\\ 
      &      & 4.00 & 5.00&       &       & 0.000 &  0.001& 0.015 &  0.007& 0.009 &  0.010\\ 
      &      & 5.00 & 6.50&       &       &       &       & 0.002 &  0.003& 0.001 &  0.002\\ 
      &      & 6.50 & 8.00&       &       &       &       &       &       & 0.001 &  0.009\\ 

%%%%%%%%%%%%%%%%%%%%%%%%%%%%%%%%%%%%%%%%%%%%%%%%%%%%%%%%%%%%%%%%%%%%%%%%%%%%%
\hline
\end{tabular}
}
\end{table*}
\begin{table*}[!ht]
  \caption{\label{tab:xsec_results_Cu}
    HARP results for the double-differential  $\pi^-$ production
    cross-section in the laboratory system,
    $d^2\sigma^{\pi}/(dpd\Omega)$, for p--Cu interactions at 3, 5, 8, 12~\GeVc.
    Each row refers to a
    different $(p_{\hbox{\small min}} \le p<p_{\hbox{\small max}},
    \theta_{\hbox{\small min}} \le \theta<\theta_{\hbox{\small max}})$ bin,
    where $p$ and $\theta$ are the pion momentum and polar angle, respectively.
    The central value as well as the square-root of the diagonal elements
    of the covariance matrix are given.}

%\begin{tabular}{rrrr|r@{$\pm$}lr{$\pm$}lr{$\pm$}lr{$\pm$}l}
\small{
\begin{tabular}{rrrr|r@{$\pm$}lr@{$\pm$}lr@{$\pm$}lr@{$\pm$}l}
\hline
$\theta_{\hbox{\small min}}$ &
$\theta_{\hbox{\small max}}$ &
$p_{\hbox{\small min}}$ &
$p_{\hbox{\small max}}$ &
\multicolumn{8}{c}{$d^2\sigma^{\pi^-}/(dpd\Omega)$}
\\
(rad) & (rad) & (\GeVc) & (\GeVc) &
\multicolumn{8}{c}{(barn/(sr \GeVc))}
\\
  &  &  &
&\multicolumn{2}{c}{$ \bf{3 \ \GeVc}$}
&\multicolumn{2}{c}{$ \bf{5 \ \GeVc}$}
&\multicolumn{2}{c}{$ \bf{8 \ \GeVc}$}
&\multicolumn{2}{c}{$ \bf{12 \ \GeVc}$}
\\
\hline

0.050 &0.100 & 0.50 & 1.00&  0.07 &   0.06&  0.18 &   0.04&  0.45 &   0.06&  0.58 &   0.08\\ 
      &      & 1.00 & 1.50&    *   &      * &  0.18 &   0.03&  0.38 &   0.04&  0.70 &   0.07\\ 
      &      & 1.50 & 2.00&  0.002&  0.008&  0.08 &   0.02&  0.32 &   0.03&  0.57 &   0.06\\ 
      &      & 2.00 & 2.50&    *  &   *   &  0.05 &   0.02&  0.17 &   0.02&  0.41 &   0.05\\ 
      &      & 2.50 & 3.00&       &       & 0.003 &  0.003&  0.08 &   0.02&  0.35 &   0.04\\ 
      &      & 3.00 & 3.50&       &       & 0.001 &  0.002& 0.041 &  0.009&  0.22 &   0.04\\ 
      &      & 3.50 & 4.00&       &       &     * &     * & 0.050 &  0.010&  0.15 &   0.02\\ 
      &      & 4.00 & 5.00&       &       &     * &     * & 0.012 &  0.006& 0.089 &  0.014\\ 
      &      & 5.00 & 6.50&       &       &       &       & 0.000 &  0.001& 0.029 &  0.007\\ 
      &      & 6.50 & 8.00&       &       &       &       &       &       & 0.006 &  0.004\\ 
0.100 &0.150 & 0.50 & 1.00&  0.14 &   0.10&  0.26 &   0.05&  0.57 &   0.07&  0.90 &   0.12\\ 
      &      & 1.00 & 1.50&    *  &    *  &  0.13 &   0.03&  0.42 &   0.04&  0.71 &   0.07\\ 
      &      & 1.50 & 2.00&    *  &    *  & 0.053 &  0.015&  0.22 &   0.03&  0.38 &   0.05\\ 
      &      & 2.00 & 2.50&    *  &    *  & 0.020 &  0.009&  0.14 &   0.02&  0.31 &   0.04\\ 
      &      & 2.50 & 3.00&       &       & 0.008 &  0.004& 0.088 &  0.014&  0.22 &   0.03\\ 
      &      & 3.00 & 3.50&       &       & 0.001 &  0.001& 0.044 &  0.009&  0.14 &   0.02\\ 
      &      & 3.50 & 4.00&       &       &    *  &    *   & 0.020 &  0.006&  0.10 &   0.02\\ 
      &      & 4.00 & 5.00&       &       &    *  &    *   & 0.006 &  0.002& 0.033 &  0.008\\ 
      &      & 5.00 & 6.50&       &       &       &       &    *  &    *   & 0.007 &  0.003\\ 
      &      & 6.50 & 8.00&       &       &       &       &       &       & 0.001 &  0.000\\ 
0.150 &0.200 & 0.50 & 1.00&  0.04 &   0.04&  0.26 &   0.05&  0.60 &   0.07&  0.95 &   0.12\\ 
      &      & 1.00 & 1.50&  0.02 &   0.03&  0.11 &   0.03&  0.31 &   0.04&  0.42 &   0.06\\ 
      &      & 1.50 & 2.00&  0.01 &   0.03&  0.05 &   0.02&  0.19 &   0.03&  0.40 &   0.05\\ 
      &      & 2.00 & 2.50& 0.000 &  0.002& 0.027 &  0.010&  0.10 &   0.02&  0.22 &   0.04\\ 
      &      & 2.50 & 3.00&       &       & 0.018 &  0.010& 0.048 &  0.011&  0.13 &   0.03\\ 
      &      & 3.00 & 3.50&       &       & 0.001 &  0.001& 0.021 &  0.007&  0.08 &   0.02\\ 
      &      & 3.50 & 4.00&       &       &  *    &    *  & 0.010 &  0.004& 0.048 &  0.014\\ 
      &      & 4.00 & 5.00&       &       &  *    &   *   & 0.002 &  0.001& 0.029 &  0.008\\ 
      &      & 5.00 & 6.50&       &       &       &       &    *  &    *  & 0.003 &  0.002\\ 
      &      & 6.50 & 8.00&       &       &       &       &       &       &    *  &    *  \\ 
0.200 &0.250 & 0.50 & 1.00&  0.09 &   0.10&  0.12 &   0.04&  0.42 &   0.06&  0.56 &   0.09\\ 
      &      & 1.00 & 1.50&     * &     * &  0.11 &   0.03&  0.27 &   0.04&  0.34 &   0.06\\ 
      &      & 1.50 & 2.00&     * &     * &  0.03 &   0.02&  0.19 &   0.04&  0.37 &   0.07\\ 
      &      & 2.00 & 2.50&     * &     * & 0.006 &  0.005& 0.051 &  0.015&  0.23 &   0.05\\ 
      &      & 2.50 & 3.00&       &       & 0.001 &  0.001& 0.016 &  0.007&  0.08 &   0.02\\ 
      &      & 3.00 & 3.50&       &       & 0.000 &  0.001& 0.006 &  0.005& 0.033 &  0.012\\ 
      &      & 3.50 & 4.00&       &       & 0.000 &  0.002& 0.003 &  0.003& 0.026 &  0.013\\ 
      &      & 4.00 & 5.00&       &       & 0.000 &  0.002& 0.001 &  0.003& 0.011 &  0.009\\ 
      &      & 5.00 & 6.50&       &       &       &       & 0.000 &  0.001& 0.005 &  0.008\\ 
      &      & 6.50 & 8.00&       &       &       &       &       &       & 0.001 &  0.005\\ 
%%%%%%%%%%%%%%%%%%%%%%%%%%%%%%%%%%%%%%%%%%%%%%%%%%%%%%%%%%%%%%%%%%%%%%%%%%%%%
\hline
\end{tabular}
}
\end{table*}

~
\begin{table*}[!ht]
  \caption{\label{tab:xsec_results_Sn}
    HARP results for the double-differential $\pi^+$  production
    cross-section in the laboratory system,
    $d^2\sigma^{\pi}/(dpd\Omega)$, for p--Sn interactions at 3, 5, 8, 12~\GeVc.
    Each row refers to a
    different $(p_{\hbox{\small min}} \le p<p_{\hbox{\small max}},
    \theta_{\hbox{\small min}} \le \theta<\theta_{\hbox{\small max}})$ bin,
    where $p$ and $\theta$ are the pion momentum and polar angle, respectively.
    The central value as well as the square-root of the diagonal elements
    of the covariance matrix are given.}

%\begin{tabular}{rrrr|r@{$\pm$}lr{$\pm$}lr{$\pm$}lr{$\pm$}l}
\small{
\begin{tabular}{rrrr|r@{$\pm$}lr@{$\pm$}lr@{$\pm$}lr@{$\pm$}l}
\hline
$\theta_{\hbox{\small min}}$ &
$\theta_{\hbox{\small max}}$ &
$p_{\hbox{\small min}}$ &
$p_{\hbox{\small max}}$ &
\multicolumn{8}{c}{$d^2\sigma^{\pi^+}/(dpd\Omega)$}
\\
(rad) & (rad) & (\GeVc) & (\GeVc) &
\multicolumn{8}{c}{(barn/(sr \GeVc))}
\\
  &  &  &
&\multicolumn{2}{c}{$ \bf{3 \ \GeVc}$}
&\multicolumn{2}{c}{$ \bf{5 \ \GeVc}$}
&\multicolumn{2}{c}{$ \bf{8 \ \GeVc}$}
&\multicolumn{2}{c}{$ \bf{12 \ \GeVc}$}
\\
\hline

0.050 &0.100 & 0.50 & 1.00&  0.32 &   0.12&  0.42 &   0.08&  0.80 &   0.10&  1.18 &   0.12\\ 
      &      & 1.00 & 1.50& 0.001 &  0.003&  0.39 &   0.06&  0.83 &   0.07&  1.07 &   0.09\\ 
      &      & 1.50 & 2.00& 0.016 &  0.011&  0.25 &   0.05&  0.74 &   0.06&  1.09 &   0.07\\ 
      &      & 2.00 & 2.50&  0.01 &   0.01&  0.10 &   0.03&  0.68 &   0.06&  1.02 &   0.07\\ 
      &      & 2.50 & 3.00&       &       &  0.05 &   0.02&  0.38 &   0.05&  0.88 &   0.06\\ 
      &      & 3.00 & 3.50&       &       &    *  &    *  &  0.14 &   0.03&  0.71 &   0.05\\ 
      &      & 3.50 & 4.00&       &       &    *  &    *  &  0.06 &   0.02&  0.44 &   0.05\\ 
      &      & 4.00 & 5.00&       &       &    *  &    *  &  0.06 &   0.02&  0.31 &   0.02\\ 
      &      & 5.00 & 6.50&       &       &       &       & 0.008 &  0.003&  0.12 &   0.02\\ 
      &      & 6.50 & 8.00&       &       &       &       &       &       & 0.044 &  0.007\\ 
0.100 &0.150 & 0.50 & 1.00&  0.10 &   0.07&  0.53 &   0.08&  1.03 &   0.11&  1.24 &   0.12\\ 
      &      & 1.00 & 1.50&  0.04 &   0.05&  0.45 &   0.07&  0.71 &   0.07&  1.21 &   0.10\\ 
      &      & 1.50 & 2.00&  0.02 &   0.02&  0.21 &   0.04&  0.66 &   0.06&  0.94 &   0.08\\ 
      &      & 2.00 & 2.50&     * &      *&  0.09 &   0.03&  0.39 &   0.04&  0.75 &   0.07\\ 
      &      & 2.50 & 3.00&       &       &  0.05 &   0.02&  0.18 &   0.03&  0.60 &   0.05\\ 
      &      & 3.00 & 3.50&       &       & 0.008 &  0.004& 0.054 &  0.015&  0.35 &   0.04\\ 
      &      & 3.50 & 4.00&       &       & 0.002 &  0.001& 0.031 &  0.009&  0.20 &   0.02\\ 
      &      & 4.00 & 5.00&       &       & 0.001 &  0.001& 0.016 &  0.005& 0.116 &  0.015\\ 
      &      & 5.00 & 6.50&       &       &       &       & 0.002 &  0.001& 0.029 &  0.006\\ 
      &      & 6.50 & 8.00&       &       &       &       &       &       & 0.005 &  0.001\\ 
0.150 &0.200 & 0.50 & 1.00&  0.19 &   0.12&  0.43 &   0.08&  0.91 &   0.10&  1.16 &   0.13\\ 
      &      & 1.00 & 1.50&  0.03 &   0.05&  0.30 &   0.05&  0.55 &   0.06&  1.01 &   0.09\\ 
      &      & 1.50 & 2.00&     * &      *&  0.11 &   0.03&  0.38 &   0.04&  0.60 &   0.06\\ 
      &      & 2.00 & 2.50&  0.01 &   0.05&  0.04 &   0.02&  0.20 &   0.03&  0.38 &   0.04\\ 
      &      & 2.50 & 3.00&       &       &  0.01 &   0.01&  0.09 &   0.02&  0.22 &   0.03\\ 
      &      & 3.00 & 3.50&       &       & 0.000 &  0.001& 0.043 &  0.011&  0.15 &   0.02\\ 
      &      & 3.50 & 4.00&       &       & 0.002 &  0.002& 0.019 &  0.007& 0.089 &  0.014\\ 
      &      & 4.00 & 5.00&       &       & 0.000 &  0.001& 0.007 &  0.003& 0.042 &  0.009\\ 
      &      & 5.00 & 6.50&       &       &       &       & 0.001 &  0.001& 0.014 &  0.004\\ 
      &      & 6.50 & 8.00&       &       &       &       &       &       & 0.001 &  0.001\\ 
0.200 &0.250 & 0.50 & 1.00&     * &   *   &  0.39 &   0.08&  0.73 &   0.10&  0.85 &   0.11\\ 
      &      & 1.00 & 1.50&     * &   *   &  0.16 &   0.05&  0.48 &   0.07&  0.50 &   0.07\\ 
      &      & 1.50 & 2.00&     * &   *   &  0.07 &   0.03&  0.30 &   0.05&  0.32 &   0.05\\ 
      &      & 2.00 & 2.50&     * &   *   & 0.015 &  0.010&  0.22 &   0.04&  0.21 &   0.04\\ 
      &      & 2.50 & 3.00&       &       & 0.012 &  0.010&  0.08 &   0.02&  0.16 &   0.03\\ 
      &      & 3.00 & 3.50&       &       & 0.013 &  0.009& 0.033 &  0.011&  0.08 &   0.02\\ 
      &      & 3.50 & 4.00&       &       & 0.006 &  0.006& 0.018 &  0.008& 0.039 &  0.011\\ 
      &      & 4.00 & 5.00&       &       & 0.002 &  0.003& 0.013 &  0.007& 0.028 &  0.009\\ 
      &      & 5.00 & 6.50&       &       &       &       & 0.004 &  0.004& 0.006 &  0.003\\ 
      &      & 6.50 & 8.00&       &       &       &       &       &       & 0.003 &  0.006\\

%%%%%%%%%%%%%%%%%%%%%%%%%%%%%%%%%%%%%%%%%%%%%%%%%%%%%%%%%%%%%%%%%%%%%%%%%%%%%
\hline
\end{tabular}
}
\end{table*}
\begin{table*}[!ht]
  \caption{\label{tab:xsec_results_Sn}
    HARP results for the double-differential  $\pi^-$ production
    cross-section in the laboratory system,
    $d^2\sigma^{\pi}/(dpd\Omega)$, for p--Sn interactions at 3, 5, 8, 12~\GeVc.
    Each row refers to a
    different $(p_{\hbox{\small min}} \le p<p_{\hbox{\small max}},
    \theta_{\hbox{\small min}} \le \theta<\theta_{\hbox{\small max}})$ bin,
    where $p$ and $\theta$ are the pion momentum and polar angle, respectively.
    The central value as well as the square-root of the diagonal elements
    of the covariance matrix are given.}

%\begin{tabular}{rrrr|r@{$\pm$}lr{$\pm$}lr{$\pm$}lr{$\pm$}l}
\small{
\begin{tabular}{rrrr|r@{$\pm$}lr@{$\pm$}lr@{$\pm$}lr@{$\pm$}l}
\hline
$\theta_{\hbox{\small min}}$ &
$\theta_{\hbox{\small max}}$ &
$p_{\hbox{\small min}}$ &
$p_{\hbox{\small max}}$ &
\multicolumn{8}{c}{$d^2\sigma^{\pi^-}/(dpd\Omega)$}
\\
(rad) & (rad) & (\GeVc) & (\GeVc) &
\multicolumn{8}{c}{(barn/(sr \GeVc))}
\\
  &  &  &
&\multicolumn{2}{c}{$ \bf{3 \ \GeVc}$}
&\multicolumn{2}{c}{$ \bf{5 \ \GeVc}$}
&\multicolumn{2}{c}{$ \bf{8 \ \GeVc}$}
&\multicolumn{2}{c}{$ \bf{12 \ \GeVc}$}
\\
\hline

0.050 &0.100 & 0.50 & 1.00&  0.11 &   0.07&  0.26 &   0.07&  0.60 &   0.08&  0.92 &   0.10\\ 
      &      & 1.00 & 1.50&  0.004&   0.006&  0.21 &   0.04&  0.50 &   0.05&  1.02 &   0.08\\ 
      &      & 1.50 & 2.00&  0.01 &   0.01&  0.15 &   0.04&  0.41 &   0.04&  0.82 &   0.06\\ 
      &      & 2.00 & 2.50&     * &      *& 0.033 &  0.015&  0.24 &   0.03&  0.50 &   0.05\\ 
      &      & 2.50 & 3.00&       &       & 0.012 &  0.008&  0.13 &   0.02&  0.46 &   0.04\\ 
      &      & 3.00 & 3.50&       &       &   *   &   *   &  0.09 &   0.02&  0.30 &   0.03\\ 
      &      & 3.50 & 4.00&       &       &   *   &   *   & 0.044 &  0.010&  0.20 &   0.02\\ 
      &      & 4.00 & 5.00&       &       &   *   &   *   & 0.024 &  0.009&  0.12 &   0.02\\ 
      &      & 5.00 & 6.50&       &       &       &       & 0.003 &  0.002& 0.030 &  0.007\\ 
      &      & 6.50 & 8.00&       &       &       &       &       &       & 0.006 &  0.003\\ 
0.100 &0.150 & 0.50 & 1.00&  0.10 &   0.07&  0.36 &   0.08&  0.89 &   0.11&  1.40 &   0.15\\ 
      &      & 1.00 & 1.50&     * &      *&  0.11 &   0.03&  0.59 &   0.06&  0.92 &   0.08\\ 
      &      & 1.50 & 2.00&     * &      *&  0.07 &   0.02&  0.32 &   0.04&  0.63 &   0.06\\ 
      &      & 2.00 & 2.50&     * &      *& 0.024 &  0.012&  0.18 &   0.03&  0.43 &   0.04\\ 
      &      & 2.50 & 3.00&       &       & 0.016 &  0.007&  0.11 &   0.02&  0.25 &   0.03\\ 
      &      & 3.00 & 3.50&       &       & 0.007 &  0.005& 0.064 &  0.014&  0.22 &   0.03\\ 
      &      & 3.50 & 4.00&       &       &     * &   *   & 0.020 &  0.006&  0.12 &   0.02\\ 
      &      & 4.00 & 5.00&       &       &     * &  *    & 0.010 &  0.004& 0.058 &  0.011\\ 
      &      & 5.00 & 6.50&       &       &       &       & 0.001 &  0.001& 0.008 &  0.003\\ 
      &      & 6.50 & 8.00&       &       &       &       &       &       & 0.001 &  0.000\\ 
0.150 &0.200 & 0.50 & 1.00&  0.06 &   0.11&  0.34 &   0.07&  0.77 &   0.09&  1.17 &   0.13\\ 
      &      & 1.00 & 1.50&  0.06 &   0.04&  0.15 &   0.04&  0.49 &   0.06&  0.75 &   0.07\\ 
      &      & 1.50 & 2.00&     * &      *&  0.07 &   0.02&  0.21 &   0.03&  0.51 &   0.05\\ 
      &      & 2.00 & 2.50&  0.005&  0.012&  0.06 &   0.02&  0.14 &   0.02&  0.30 &   0.04\\ 
      &      & 2.50 & 3.00&       &       &  0.00 &   0.01&  0.07 &   0.02&  0.18 &   0.03\\ 
      &      & 3.00 & 3.50&       &       &     * &     * & 0.016 &  0.008&  0.10 &   0.02\\ 
      &      & 3.50 & 4.00&       &       &     * &     * & 0.012 &  0.005& 0.061 &  0.013\\ 
      &      & 4.00 & 5.00&       &       &     * &     * & 0.001 &  0.001& 0.033 &  0.007\\ 
      &      & 5.00 & 6.50&       &       &       &       & *     &  *    & 0.005 &  0.002\\ 
      &      & 6.50 & 8.00&       &       &       &       &       &       & *     &  *    \\ 
0.200 &0.250 & 0.50 & 1.00&  0.05 &   0.06&  0.36 &   0.08&  0.59 &   0.08&  0.88 &   0.11\\ 
      &      & 1.00 & 1.50&  0.05 &   0.05&  0.12 &   0.04&  0.42 &   0.06&  0.54 &   0.07\\ 
      &      & 1.50 & 2.00&     * &      *&  0.02 &   0.02&  0.29 &   0.05&  0.55 &   0.09\\ 
      &      & 2.00 & 2.50&     * &      *& 0.005 &  0.005&  0.16 &   0.03&  0.27 &   0.05\\ 
      &      & 2.50 & 3.00&       &       &     * &    *  &  0.06 &   0.02&  0.13 &   0.03\\ 
      &      & 3.00 & 3.50&       &       & 0.000 &  0.001& 0.023 &  0.011&  0.06 &   0.02\\ 
      &      & 3.50 & 4.00&       &       & *     &   *   & 0.004 &  0.004& 0.033 &  0.011\\ 
      &      & 4.00 & 5.00&       &       & *     &   *   & 0.001 &  0.005& 0.008 &  0.004\\ 
      &      & 5.00 & 6.50&       &       &       &       & 0.000 &  0.001& 0.002 &  0.003\\ 
      &      & 6.50 & 8.00&       &       &       &       &       &       & 0.000 &  0.003\\ 
%%%%%%%%%%%%%%%%%%%%%%%%%%%%%%%%%%%%%%%%%%%%%%%%%%%%%%%%%%%%%%%%%%%%%%%%%%%%%
\hline
\end{tabular}
}
\end{table*}

\begin{table*}[!ht]
  \caption{\label{tab:xsec_results_Ta}
    HARP results for the double-differential $\pi^+$  production
    cross-section in the laboratory system,
    $d^2\sigma^{\pi}/(dpd\Omega)$, for p--Ta interactions at 3, 5, 8, 12~\GeVc.
    Each row refers to a
    different $(p_{\hbox{\small min}} \le p<p_{\hbox{\small max}},
    \theta_{\hbox{\small min}} \le \theta<\theta_{\hbox{\small max}})$ bin,
    where $p$ and $\theta$ are the pion momentum and polar angle, respectively.
    The central value as well as the square-root of the diagonal elements
    of the covariance matrix are given.}

%\begin{tabular}{rrrr|r@{$\pm$}lr{$\pm$}lr{$\pm$}lr{$\pm$}l}
\small{
\begin{tabular}{rrrr|r@{$\pm$}lr@{$\pm$}lr@{$\pm$}lr@{$\pm$}l}
\hline
$\theta_{\hbox{\small min}}$ &
$\theta_{\hbox{\small max}}$ &
$p_{\hbox{\small min}}$ &
$p_{\hbox{\small max}}$ &
\multicolumn{8}{c}{$d^2\sigma^{\pi^+}/(dpd\omega)$}
\\
(rad) & (rad) & (\GeVc) & (\GeVc) &
\multicolumn{8}{c}{(barn/(sr \GeVc))}
\\
  &  &  &
&\multicolumn{2}{c}{$ \bf{3 \ \GeVc}$}
&\multicolumn{2}{c}{$ \bf{5 \ \GeVc}$}
&\multicolumn{2}{c}{$ \bf{8 \ \GeVc}$}
&\multicolumn{2}{c}{$ \bf{12 \ \GeVc}$}
\\
\hline

0.050 &0.100 & 0.50 & 1.00&  0.26 &   0.11&  0.55 &   0.12&  0.87 &   0.12&  1.65 &   0.20\\ 
      &      & 1.00 & 1.50& 0.004 &  0.007&  0.54 &   0.10&  0.95 &   0.09&  1.27 &   0.13\\ 
      &      & 1.50 & 2.00& 0.014 &  0.009&  0.25 &   0.07&  0.80 &   0.08&  1.37 &   0.12\\ 
      &      & 2.00 & 2.50&  0.01 &   0.01&  0.11 &   0.03&  0.59 &   0.07&  1.05 &   0.11\\ 
      &      & 2.50 & 3.00&       &       &  0.05 &   0.03&  0.34 &   0.06&  1.19 &   0.11\\ 
      &      & 3.00 & 3.50&       &       &     *  &   *    &  0.15 &   0.05&  0.67 &   0.08\\ 
      &      & 3.50 & 4.00&       &       &     *  &   *    &  0.05 &   0.03&  0.56 &   0.07\\ 
      &      & 4.00 & 5.00&       &       &     *  &   *    &  0.05 &   0.02&  0.30 &   0.04\\ 
      &      & 5.00 & 6.50&       &       &       &       & 0.007 &  0.003&  0.12 &   0.02\\ 
      &      & 6.50 & 8.00&       &       &       &       &       &       & 0.038 &  0.009\\ 
0.100 &0.150 & 0.50 & 1.00&  0.22 &   0.12&  0.56 &   0.11&  0.93 &   0.11&  1.60 &   0.19\\ 
      &      & 1.00 & 1.50&  0.03 &   0.04&  0.39 &   0.09&  0.82 &   0.10&  1.60 &   0.16\\ 
      &      & 1.50 & 2.00& 0.002 &  0.008&  0.26 &   0.07&  0.82 &   0.09&  1.22 &   0.13\\ 
      &      & 2.00 & 2.50& 0.002 &  0.007&  0.04 &   0.02&  0.40 &   0.06&  0.84 &   0.10\\ 
      &      & 2.50 & 3.00&       &       &  0.02 &   0.02&  0.15 &   0.04&  0.61 &   0.07\\ 
      &      & 3.00 & 3.50&       &       & 0.009 &  0.009&  0.09 &   0.03&  0.41 &   0.05\\ 
      &      & 3.50 & 4.00&       &       &   *    &   *    &  0.02 &   0.02&  0.26 &   0.05\\ 
      &      & 4.00 & 5.00&       &       & 0.000 &  0.001& 0.009 &  0.004&  0.15 &   0.02\\ 
      &      & 5.00 & 6.50&       &       &       &       & 0.002 &  0.001& 0.034 &  0.009\\ 
      &      & 6.50 & 8.00&       &       &       &       &       &       & 0.009 &  0.003\\ 
0.150 &0.200 & 0.50 & 1.00&  0.38 &   0.17&  0.57 &   0.12&  0.93 &   0.12&  1.37 &   0.19\\ 
      &      & 1.00 & 1.50&  0.13 &   0.09&  0.39 &   0.08&  0.77 &   0.09&  1.15 &   0.13\\ 
      &      & 1.50 & 2.00&   *   &    *   &  0.10 &   0.04&  0.38 &   0.06&  0.66 &   0.09\\ 
      &      & 2.00 & 2.50&   *   &    *   &  0.06 &   0.03&  0.20 &   0.04&  0.48 &   0.08\\ 
      &      & 2.50 & 3.00&       &       &  0.00 &   0.00&  0.12 &   0.03&  0.34 &   0.06\\ 
      &      & 3.00 & 3.50&       &       &  0.00 &   0.00&  0.06 &   0.02&  0.12 &   0.03\\ 
      &      & 3.50 & 4.00&       &       & 0.000 &  0.001& 0.016 &  0.009&  0.08 &   0.02\\ 
      &      & 4.00 & 5.00&       &       &    *   &   *    & 0.007 &  0.004&  0.06 &   0.02\\ 
      &      & 5.00 & 6.50&       &       &       &       & 0.002 &  0.002& 0.011 &  0.005\\ 
      &      & 6.50 & 8.00&       &       &       &       &       &       & 0.003 &  0.002\\ 
0.200 &0.250 & 0.50 & 1.00&  0.16 &   0.34&  0.48 &   0.11&  0.63 &   0.10&  1.24 &   0.18\\ 
      &      & 1.00 & 1.50&  0.04 &   0.38&  0.23 &   0.07&  0.46 &   0.08&  0.65 &   0.11\\ 
      &      & 1.50 & 2.00&    *  &   *   &  0.04 &   0.02&  0.20 &   0.05&  0.60 &   0.11\\ 
      &      & 2.00 & 2.50&    *  &   *   &  0.06 &   0.03&  0.24 &   0.05&  0.49 &   0.10\\ 
      &      & 2.50 & 3.00&       &       &  0.03 &   0.02&  0.09 &   0.03&  0.24 &   0.06\\ 
      &      & 3.00 & 3.50&       &       & 0.020 &  0.015& 0.027 &  0.015&  0.10 &   0.03\\ 
      &      & 3.50 & 4.00&       &       & 0.007 &  0.008& 0.016 &  0.013&  0.04 &   0.02\\ 
      &      & 4.00 & 5.00&       &       & 0.005 &  0.008& 0.011 &  0.010&  0.05 &   0.03\\ 
      &      & 5.00 & 6.50&       &       &       &       & 0.002 &  0.005& 0.015 &  0.012\\ 
      &      & 6.50 & 8.00&       &       &       &       &       &       & 0.003 &  0.017\\ 

%%%%%%%%%%%%%%%%%%%%%%%%%%%%%%%%%%%%%%%%%%%%%%%%%%%%%%%%%%%%%%%%%%%%%%%%%%%%%
\hline
\end{tabular}
}
\end{table*}
\begin{table*}[!ht]
  \caption{\label{tab:xsec_results_Ta}
    HARP results for the double-differential  $\pi^-$ production
    cross-section in the laboratory system,
    $d^2\sigma^{\pi}/(dpd\Omega)$, for p--Ta interactions at 3, 5, 8, 12~\GeVc.
    Each row refers to a
    different $(p_{\hbox{\small min}} \le p<p_{\hbox{\small max}},
    \theta_{\hbox{\small min}} \le \theta<\theta_{\hbox{\small max}})$ bin,
    where $p$ and $\theta$ are the pion momentum and polar angle, respectively.
    The central value as well as the square-root of the diagonal elements
    of the covariance matrix are given.}

%\begin{tabular}{rrrr|r@{$\pm$}lr{$\pm$}lr{$\pm$}lr{$\pm$}l}
\small{
\begin{tabular}{rrrr|r@{$\pm$}lr@{$\pm$}lr@{$\pm$}lr@{$\pm$}l}
\hline
$\theta_{\hbox{\small min}}$ &
$\theta_{\hbox{\small max}}$ &
$p_{\hbox{\small min}}$ &
$p_{\hbox{\small max}}$ &
\multicolumn{8}{c}{$d^2\sigma^{\pi^-}/(dpd\Omega)$}
\\
(rad) & (rad) & (\GeVc) & (\GeVc) &
\multicolumn{8}{c}{(barn/(sr \GeVc ))}
\\
  &  &  &
&\multicolumn{2}{c}{$ \bf{3 \ \GeVc}$}
&\multicolumn{2}{c}{$ \bf{5 \ \GeVc}$}
&\multicolumn{2}{c}{$ \bf{8 \ \GeVc}$}
&\multicolumn{2}{c}{$ \bf{12 \ \GeVc}$}
\\
\hline

0.050 &0.100 & 0.50 & 1.00&  0.11 &   0.07&  0.31 &   0.10&  0.68 &   0.10&  1.21 &   0.16\\ 
      &      & 1.00 & 1.50& 0.004 &  0.007&  0.29 &   0.07&  0.60 &   0.08&  1.23 &   0.12\\ 
      &      & 1.50 & 2.00&    *  &   *    &  0.09 &   0.04&  0.42 &   0.06&  0.89 &   0.10\\ 
      &      & 2.00 & 2.50&  0.03 &   0.03&  0.06 &   0.03&  0.27 &   0.04&  0.92 &   0.10\\ 
      &      & 2.50 & 3.00&       &       & 0.015 &  0.012&  0.12 &   0.03&  0.47 &   0.08\\ 
      &      & 3.00 & 3.50&       &       & 0.007 &  0.007& 0.046 &  0.015&  0.42 &   0.06\\ 
      &      & 3.50 & 4.00&       &       &   *    &  *   &  0.09 &   0.02&  0.24 &   0.05\\ 
      &      & 4.00 & 5.00&       &       & 0.000 &  0.001& 0.017 &  0.011&  0.13 &   0.02\\ 
      &      & 5.00 & 6.50&       &       &       &       &   *    & *      & 0.044 &  0.012\\ 
      &      & 6.50 & 8.00&       &       &       &       &       &       & 0.010 &  0.006\\ 
0.100 &0.150 & 0.50 & 1.00&  0.24 &   0.12&  0.30 &   0.09&  1.18 &   0.15&  1.79 &   0.22\\ 
      &      & 1.00 & 1.50&    *   &   *    &  0.16 &   0.05&  0.68 &   0.08&  1.22 &   0.13\\ 
      &      & 1.50 & 2.00&    *   &   *    &  0.08 &   0.03&  0.41 &   0.05&  0.71 &   0.09\\ 
      &      & 2.00 & 2.50&    *   &   *    &  0.04 &   0.02&  0.17 &   0.03&  0.48 &   0.07\\ 
      &      & 2.50 & 3.00&       &       & 0.026 &  0.014&  0.10 &   0.02&  0.42 &   0.06\\ 
      &      & 3.00 & 3.50&       &       &  *     &   *    &  0.07 &   0.02&  0.30 &   0.05\\ 
      &      & 3.50 & 4.00&       &       &  *     &   *    & 0.031 &  0.010&  0.18 &   0.03\\ 
      &      & 4.00 & 5.00&       &       &  *     &   *    & 0.014 &  0.006& 0.059 &  0.015\\ 
      &      & 5.00 & 6.50&       &       &       &       & 0.001 &  0.001& 0.014 &  0.005\\ 
      &      & 6.50 & 8.00&       &       &       &       &       &       & 0.000 &  0.001\\ 
0.150 &0.200 & 0.50 & 1.00&  0.14 &   0.09&  0.60 &   0.12&  0.86 &   0.12&  1.70 &   0.22\\ 
      &      & 1.00 & 1.50&  0.05 &   0.05&  0.15 &   0.05&  0.55 &   0.07&  0.87 &   0.11\\ 
      &      & 1.50 & 2.00&   *    &   *    &  0.07 &   0.02&  0.20 &   0.04&  0.68 &   0.09\\ 
      &      & 2.00 & 2.50&   *    &   *    &  0.06 &   0.02&  0.16 &   0.03&  0.42 &   0.07\\ 
      &      & 2.50 & 3.00&       &       &  0.00 &   0.01&  0.06 &   0.02&  0.17 &   0.04\\ 
      &      & 3.00 & 3.50&       &       & 0.007 &  0.008& 0.031 &  0.014&  0.15 &   0.03\\ 
      &      & 3.50 & 4.00&       &       &  *     &  *     &  *     &   *    &  0.07 &   0.02\\ 
      &      & 4.00 & 5.00&       &       &  *     &  *     &   *    &   *    & 0.045 &  0.014\\ 
      &      & 5.00 & 6.50&       &       &       &       & 0.000 &  0.001& 0.004 &  0.003\\ 
      &      & 6.50 & 8.00&       &       &       &       &       &       & 0.000 &  0.001\\ 
0.200 &0.250 & 0.50 & 1.00&  0.14 &   0.11&  0.26 &   0.09&  0.76 &   0.11&  1.05 &   0.16\\ 
      &      & 1.00 & 1.50&  0.04 &   0.05&  0.21 &   0.06&  0.32 &   0.06&  0.97 &   0.15\\ 
      &      & 1.50 & 2.00&   *    &   *    &  0.03 &   0.03&  0.14 &   0.04&  0.75 &   0.14\\ 
      &      & 2.00 & 2.50&   *    &   *    &  0.04 &   0.03&  0.06 &   0.02&  0.33 &   0.08\\ 
      &      & 2.50 & 3.00&       &       & 0.003 &  0.006& 0.021 &  0.010&  0.20 &   0.06\\ 
      &      & 3.00 & 3.50&       &       &    *   &   *    &    *   &  *     &  0.07 &   0.03\\ 
      &      & 3.50 & 4.00&       &       &    *   &   *    & 0.001 &  0.002& 0.024 &  0.015\\ 
      &      & 4.00 & 5.00&       &       &    *   &   *    & 0.000 &  0.005& 0.008 &  0.011\\ 
      &      & 5.00 & 6.50&       &       &       &       & 0.000 &  0.002& 0.001 &  0.002\\ 
      &      & 6.50 & 8.00&       &       &       &       &       &       & 0.000 &  0.009\\ 
%%%%%%%%%%%%%%%%%%%%%%%%%%%%%%%%%%%%%%%%%%%%%%%%%%%%%%%%%%%%%%%%%%%%%%%%%%%%%
\hline
\end{tabular}
}
\end{table*}
\begin{table*}[!ht]
  \caption{\label{tab:xsec_results_Pb}
    HARP results for the double-differential $\pi^+$  production
    cross-section in the laboratory system,
    $d^2\sigma^{\pi}/(dpd\Omega)$, for p--Pb interactions at 3, 5, 8, 12~\GeVc.
    Each row refers to a
    different $(p_{\hbox{\small min}} \le p<p_{\hbox{\small max}},
    \theta_{\hbox{\small min}} \le \theta<\theta_{\hbox{\small max}})$ bin,
    where $p$ and $\theta$ are the pion momentum and polar angle, respectively.
    The central value as well as the square-root of the diagonal elements
    of the covariance matrix are given.}

%\begin{tabular}{rrrr|r@{$\pm$}lr{$\pm$}lr{$\pm$}lr{$\pm$}l}
\small{
\begin{tabular}{rrrr|r@{$\pm$}lr@{$\pm$}lr@{$\pm$}lr@{$\pm$}l}
\hline
$\theta_{\hbox{\small min}}$ &
$\theta_{\hbox{\small max}}$ &
$p_{\hbox{\small min}}$ &
$p_{\hbox{\small max}}$ &
\multicolumn{8}{c}{$d^2\sigma^{\pi^+}/(dpd\omega)$}
\\
(rad) & (rad) & (\GeVc) & (\GeVc) &
\multicolumn{8}{c}{(barn/(sr \GeVc ))}
\\
  &  &  &
&\multicolumn{2}{c}{$ \bf{3 \ \GeVc}$}
&\multicolumn{2}{c}{$ \bf{5 \ \GeVc}$}
&\multicolumn{2}{c}{$ \bf{8 \ \GeVc}$}
&\multicolumn{2}{c}{$ \bf{12 \ \GeVc}$}
\\
\hline

0.050 &0.100 & 0.50 & 1.00&  0.09 &   0.05&  0.45 &   0.11&  0.99 &   0.14&  1.78 &   0.27\\ 
      &      & 1.00 & 1.50&  *    &   *   &  0.63 &   0.11&  0.86 &   0.09&  1.21 &   0.16\\ 
      &      & 1.50 & 2.00& 0.012 &  0.009&  0.46 &   0.09&  0.81 &   0.08&  1.47 &   0.19\\ 
      &      & 2.00 & 2.50&  0.01 &   0.01&  0.06 &   0.02&  0.66 &   0.08&  1.40 &   0.16\\ 
      &      & 2.50 & 3.00&       &       &  0.06 &   0.04&  0.41 &   0.07&  1.02 &   0.13\\ 
      &      & 3.00 & 3.50&       &       &  0.00 &   0.01&  0.16 &   0.05&  0.83 &   0.11\\ 
      &      & 3.50 & 4.00&       &       &  0.01 &   0.02&  0.03 &   0.06&  0.56 &   0.09\\ 
      &      & 4.00 & 5.00&       &       &   *   &    *  &  0.06 &   0.03&  0.39 &   0.06\\ 
      &      & 5.00 & 6.50&       &       &       &       &  0.01 &   0.01&  0.16 &   0.03\\ 
      &      & 6.50 & 8.00&       &       &       &       &       &       & 0.034 &  0.011\\ 
0.100 &0.150 & 0.50 & 1.00&  0.35 &   0.17&  0.74 &   0.14&  1.18 &   0.14&  1.77 &   0.25\\ 
      &      & 1.00 & 1.50&  0.04 &   0.05&  0.42 &   0.10&  0.77 &   0.10&  1.45 &   0.19\\ 
      &      & 1.50 & 2.00&  0.01 &   0.02&  0.18 &   0.05&  0.80 &   0.08&  1.42 &   0.19\\ 
      &      & 2.00 & 2.50& 0.003 &  0.007&  0.07 &   0.03&  0.45 &   0.07&  0.84 &   0.13\\ 
      &      & 2.50 & 3.00&       &       &  0.06 &   0.03&  0.15 &   0.04&  0.61 &   0.10\\ 
      &      & 3.00 & 3.50&       &       & 0.014 &  0.009&  0.05 &   0.02&  0.45 &   0.08\\ 
      &      & 3.50 & 4.00&       &       & 0.001 &  0.001& 0.031 &  0.012&  0.26 &   0.05\\ 
      &      & 4.00 & 5.00&       &       &    *  &     * & 0.011 &  0.006&  0.13 &   0.03\\ 
      &      & 5.00 & 6.50&       &       &       &       & 0.000 &  0.000& 0.044 &  0.013\\ 
      &      & 6.50 & 8.00&       &       &       &       &       &       & 0.008 &  0.004\\ 
0.150 &0.200 & 0.50 & 1.00&  0.14 &   0.13&  0.66 &   0.14&  1.07 &   0.14&  1.32 &   0.23\\ 
      &      & 1.00 & 1.50&  0.08 &   0.07&  0.30 &   0.08&  0.68 &   0.08&  1.08 &   0.16\\ 
      &      & 1.50 & 2.00&  0.02 &   0.04&  0.14 &   0.05&  0.40 &   0.06&  0.75 &   0.12\\ 
      &      & 2.00 & 2.50&  0.00 &   0.01&  0.04 &   0.04&  0.24 &   0.05&  0.56 &   0.10\\ 
      &      & 2.50 & 3.00&       &       &  0.01 &   0.01&  0.08 &   0.02&  0.39 &   0.08\\ 
      &      & 3.00 & 3.50&       &       & 0.000 &  0.001& 0.040 &  0.014&  0.12 &   0.04\\ 
      &      & 3.50 & 4.00&       &       &     * &     * & 0.018 &  0.008&  0.13 &   0.04\\ 
      &      & 4.00 & 5.00&       &       &     * &     * & 0.009 &  0.005&  0.05 &   0.02\\ 
      &      & 5.00 & 6.50&       &       &       &       & 0.002 &  0.002& 0.015 &  0.008\\ 
      &      & 6.50 & 8.00&       &       &       &       &       &       & 0.015 &  0.010\\ 
0.200 &0.250 & 0.50 & 1.00&  *    &   *   &  0.49 &   0.13&  0.54 &   0.09&  0.97 &   0.20\\ 
      &      & 1.00 & 1.50&  0.08 &   0.10&  0.21 &   0.08&  0.41 &   0.08&  0.61 &   0.15\\ 
      &      & 1.50 & 2.00&  0.01 &   0.06&  0.06 &   0.04&  0.22 &   0.06&  0.46 &   0.13\\ 
      &      & 2.00 & 2.50&   *   &    *  &  0.02 &   0.02&  0.13 &   0.03&  0.33 &   0.09\\ 
      &      & 2.50 & 3.00&       &       &  0.02 &   0.02&  0.05 &   0.02&  0.33 &   0.11\\ 
      &      & 3.00 & 3.50&       &       &  0.01 &   0.02&  0.03 &   0.02&  0.17 &   0.06\\ 
      &      & 3.50 & 4.00&       &       & 0.005 &  0.011& 0.020 &  0.014&  0.11 &   0.06\\ 
      &      & 4.00 & 5.00&       &       &     * &      *&  0.01 &   0.01&  0.04 &   0.04\\ 
      &      & 5.00 & 6.50&       &       &       &       &  0.00 &   0.02&  0.01 &   0.01\\ 
      &      & 6.50 & 8.00&       &       &       &       &       &       &  0.00 &   0.03\\

%%%%%%%%%%%%%%%%%%%%%%%%%%%%%%%%%%%%%%%%%%%%%%%%%%%%%%%%%%%%%%%%%%%%%%%%%%%%%
\hline
\end{tabular}
}
\end{table*}
\begin{table*}[!ht]
  \caption{\label{tab:xsec_results_Pb}
    HARP results for the double-differential  $\pi^-$ production
    cross-section in the laboratory system,
    $d^2\sigma^{\pi}/(dpd\Omega)$, for p--Pb interactions at 3, 5, 8, 12~\GeVc.
    Each row refers to a
    different $(p_{\hbox{\small min}} \le p<p_{\hbox{\small max}},
    \theta_{\hbox{\small min}} \le \theta<\theta_{\hbox{\small max}})$ bin,
    where $p$ and $\theta$ are the pion momentum and polar angle, respectively.
    The central value as well as the square-root of the diagonal elements
    of the covariance matrix are given.}

%\begin{tabular}{rrrr|r@{$\pm$}lr{$\pm$}lr{$\pm$}lr{$\pm$}l}
\small{
\begin{tabular}{rrrr|r@{$\pm$}lr@{$\pm$}lr@{$\pm$}lr@{$\pm$}l}
\hline
$\theta_{\hbox{\small min}}$ &
$\theta_{\hbox{\small max}}$ &
$p_{\hbox{\small min}}$ &
$p_{\hbox{\small max}}$ &
\multicolumn{8}{c}{$d^2\sigma^{\pi^-}/(dpd\Omega)$}
\\
(rad) & (rad) & (\GeVc) & (\GeVc) &
\multicolumn{8}{c}{(barn/(sr \GeVc))}
\\
  &  &  &
&\multicolumn{2}{c}{$ \bf{3 \ \GeVc}$}
&\multicolumn{2}{c}{$ \bf{5 \ \GeVc}$}
&\multicolumn{2}{c}{$ \bf{8 \ \GeVc}$}
&\multicolumn{2}{c}{$ \bf{12 \ \GeVc}$}
\\
\hline

0.050 &0.100 & 0.50 & 1.00&  0.06 &   0.06&  0.37 &   0.11&  0.72 &   0.11&  1.35 &   0.22\\ 
      &      & 1.00 & 1.50&     * &    *  &  0.30 &   0.07&  0.54 &   0.08&  1.37 &   0.18\\ 
      &      & 1.50 & 2.00&  0.01 &   0.02&  0.15 &   0.05&  0.47 &   0.06&  0.92 &   0.14\\ 
      &      & 2.00 & 2.50&     * &    *  &  0.06 &   0.03&  0.27 &   0.04&  0.77 &   0.12\\ 
      &      & 2.50 & 3.00&       &       &  0.04 &   0.02&  0.15 &   0.04&  0.43 &   0.09\\ 
      &      & 3.00 & 3.50&       &       &    *  &    *  &  0.05 &   0.02&  0.47 &   0.12\\ 
      &      & 3.50 & 4.00&       &       &    *  &    *   &  0.06 &   0.02&  0.29 &   0.17\\ 
      &      & 4.00 & 5.00&       &       &    *  &    *   & 0.034 &  0.012&  0.13 &   0.09\\ 
      &      & 5.00 & 6.50&       &       &       &       &    *   &   *    &  0.03 &   0.04\\ 
      &      & 6.50 & 8.00&       &       &       &       &       &       & 0.008 &  0.043\\ 
0.100 &0.150 & 0.50 & 1.00&  0.24 &   0.13&  0.43 &   0.11&  1.01 &   0.14&  2.23 &   0.32\\ 
      &      & 1.00 & 1.50&  0.004&  0.008&  0.19 &   0.06&  0.58 &   0.07&  1.39 &   0.18\\ 
      &      & 1.50 & 2.00&  0.003&  0.007&  0.05 &   0.02&  0.37 &   0.05&  0.86 &   0.13\\ 
      &      & 2.00 & 2.50&    *  &     * &  0.07 &   0.03&  0.17 &   0.03&  0.50 &   0.09\\ 
      &      & 2.50 & 3.00&       &       &  0.03 &   0.02&  0.14 &   0.03&  0.31 &   0.07\\ 
      &      & 3.00 & 3.50&       &       & 0.007 &  0.006&  0.07 &   0.02&  0.21 &   0.05\\ 
      &      & 3.50 & 4.00&       &       & 0.003 &  0.006& 0.037 &  0.011&  0.19 &   0.05\\ 
      &      & 4.00 & 5.00&       &       & 0.000 &  0.002& 0.013 &  0.006&  0.09 &   0.03\\ 
      &      & 5.00 & 6.50&       &       &       &       & 0.002 &  0.002& 0.024 &  0.011\\ 
      &      & 6.50 & 8.00&       &       &       &       &       &       & 0.002 &  0.002\\ 
0.150 &0.200 & 0.50 & 1.00&  0.11 &   0.08&  0.39 &   0.10&  0.98 &   0.13&  1.98 &   0.30\\ 
      &      & 1.00 & 1.50&  0.07 &   0.06&  0.14 &   0.05&  0.50 &   0.07&  0.90 &   0.15\\ 
      &      & 1.50 & 2.00&    *  &     * &  0.05 &   0.02&  0.27 &   0.05&  0.58 &   0.11\\ 
      &      & 2.00 & 2.50&    *  &     * &  0.04 &   0.02&  0.17 &   0.03&  0.33 &   0.08\\ 
      &      & 2.50 & 3.00&       &       &  0.01 &   0.01&  0.07 &   0.02&  0.24 &   0.06\\ 
      &      & 3.00 & 3.50&       &       & 0.007 &  0.010& 0.033 &  0.012&  0.19 &   0.05\\ 
      &      & 3.50 & 4.00&       &       & 0.002 &  0.006& 0.011 &  0.007&  0.08 &   0.03\\ 
      &      & 4.00 & 5.00&       &       &    *  &    *  &   *   &    *  &  0.05 &   0.02\\ 
      &      & 5.00 & 6.50&       &       &       &       & 0.000 &  0.001& 0.019 &  0.012\\ 
      &      & 6.50 & 8.00&       &       &       &       &       &       & 0.002 &  0.002\\ 
0.200 &0.250 & 0.50 & 1.00&  0.19 &   0.14&  0.41 &   0.12&  0.68 &   0.11&  1.41 &   0.25\\ 
      &      & 1.00 & 1.50&  0.03 &   0.04&  0.16 &   0.06&  0.37 &   0.07&  0.85 &   0.16\\ 
      &      & 1.50 & 2.00&  0.01 &   0.04&  0.02 &   0.03&  0.17 &   0.05&  0.81 &   0.19\\ 
      &      & 2.00 & 2.50&    *  &     * &  0.02 &   0.02&  0.07 &   0.02&  0.40 &   0.12\\ 
      &      & 2.50 & 3.00&       &       & 0.021 &  0.026& 0.014 &  0.008&  0.16 &   0.06\\ 
      &      & 3.00 & 3.50&       &       &     * &     * &    *   &   *    &  0.15 &   0.08\\ 
      &      & 3.50 & 4.00&       &       &     * &     * &  0.00 &   0.01&  0.04 &   0.04\\ 
      &      & 4.00 & 5.00&       &       &     * &     *  &  0.00 &   0.01&  0.01 &   0.02\\ 
      &      & 5.00 & 6.50&       &       &       &       &     *  &   *    &  0.00 &   0.01\\ 
      &      & 6.50 & 8.00&       &       &       &       &       &       &  0.00 &   0.02\\ 
%%%%%%%%%%%%%%%%%%%%%%%%%%%%%%%%%%%%%%%%%%%%%%%%%%%%%%%%%%%%%%%%%%%%%%%%%%%%%
\hline
\end{tabular}
}
\end{table*}

\begin{table*}[!ht]
  \caption{\label{tab:xsec_results_Befine1}
    HARP results for the double-differential $\pi^+$  production
    cross-section in the laboratory system,
    $d^2\sigma^{\pi}/(dpd\Omega)$, for p--Be interactions at 8,8.9,12~\GeVc.
    Each row refers to a
    different $(p_{\hbox{\small min}} \le p<p_{\hbox{\small max}},
    \theta_{\hbox{\small min}} \le \theta<\theta_{\hbox{\small max}})$ bin,
    where $p$ and $\theta$ are the pion momentum and polar angle, respectively.
    The central value as well as the square-root of the diagonal elements
    of the covariance matrix are given.}
%\begin{tabular}{rrrr|r@{$\pm$}lr{$\pm$}lr{$\pm$}lr{$\pm$}l}
\small{
\begin{tabular}{rrrr|r@{$\pm$}lr@{$\pm$}lr@{$\pm$}l}
\hline
$\theta_{\hbox{\small min}}$ &
$\theta_{\hbox{\small max}}$ &
$p_{\hbox{\small min}}$ &
$p_{\hbox{\small max}}$ &
\multicolumn{6}{c}{$d^2\sigma^{\pi^+}/(dpd\Omega)$}
\\
(rad) & (rad) & (\GeVc) & (\GeVc) &
\multicolumn{6}{c}{(barn/(sr \GeVc))}
\\
  &  &  &
&\multicolumn{2}{c}{$ \bf{8 \ \GeVc}$}
&\multicolumn{2}{c}{$ \bf{8.9 \ \GeVc}$}
&\multicolumn{2}{c}{$ \bf{12 \ \GeVc}$}
\\
\hline

0.025 &0.050 & 0.50 & 0.75&  0.13 &   0.04&  0.12 &   0.03&  0.17 &   0.05\\ 
      &      & 0.75 & 1.00&  0.15 &   0.03&  0.13 &   0.03&  0.15 &   0.03\\ 
      &      & 1.00 & 1.25&  0.18 &   0.03&  0.12 &   0.03&  0.14 &   0.03\\ 
      &      & 1.25 & 1.50&  0.13 &   0.02&  0.13 &   0.03&  0.18 &   0.03\\ 
      &      & 1.50 & 2.00&  0.20 &   0.02&  0.17 &   0.02&  0.22 &   0.03\\ 
      &      & 2.00 & 2.50&  0.19 &   0.03&  0.19 &   0.02&  0.25 &   0.03\\ 
      &      & 2.50 & 3.00&  0.18 &   0.02&  0.18 &   0.02&  0.25 &   0.03\\ 
      &      & 3.00 & 3.50&  0.09 &   0.02&  0.15 &   0.02&  0.27 &   0.03\\ 
      &      & 3.50 & 4.00& 0.047 &  0.014&  0.08 &   0.02&  0.24 &   0.03\\ 
      &      & 4.00 & 5.00& 0.015 &  0.008& 0.069 &  0.012&  0.15 &   0.02\\ 
      &      & 5.00 & 6.50& 0.013 &  0.006& 0.022 &  0.010& 0.083 &  0.010\\ 
      &      & 6.50 & 8.00&       &       &       &       & 0.038 &  0.006\\ 
0.050 &0.075 & 0.50 & 0.75&  0.13 &   0.03&  0.16 &   0.03&  0.16 &   0.03\\ 
      &      & 0.75 & 1.00&  0.20 &   0.02&  0.14 &   0.02&  0.13 &   0.02\\ 
      &      & 1.00 & 1.25&  0.14 &   0.02&  0.17 &   0.02&  0.14 &   0.02\\ 
      &      & 1.25 & 1.50&  0.16 &   0.02&  0.15 &   0.02&  0.13 &   0.02\\ 
      &      & 1.50 & 2.00&  0.20 &   0.02&  0.19 &   0.02&  0.21 &   0.02\\ 
      &      & 2.00 & 2.50&  0.19 &   0.02&  0.22 &   0.02&  0.22 &   0.02\\ 
      &      & 2.50 & 3.00&  0.15 &   0.02&  0.18 &   0.02&  0.23 &   0.02\\ 
      &      & 3.00 & 3.50& 0.067 &  0.013& 0.104 &  0.013&  0.18 &   0.02\\ 
      &      & 3.50 & 4.00& 0.032 &  0.009& 0.096 &  0.011&  0.16 &   0.02\\ 
      &      & 4.00 & 5.00& 0.029 &  0.007& 0.050 &  0.008& 0.101 &  0.010\\ 
      &      & 5.00 & 6.50& 0.004 &  0.001& 0.015 &  0.003& 0.047 &  0.006\\ 
      &      & 6.50 & 8.00&       &       &       &       & 0.018 &  0.003\\ 
0.075 &0.100 & 0.50 & 0.75&  0.15 &   0.02&  0.20 &   0.03&  0.22 &   0.03\\ 
      &      & 0.75 & 1.00&  0.17 &   0.02&  0.18 &   0.02&  0.17 &   0.02\\ 
      &      & 1.00 & 1.25&  0.17 &   0.02&  0.17 &   0.02&  0.19 &   0.02\\ 
      &      & 1.25 & 1.50&  0.19 &   0.02&  0.17 &   0.02&  0.23 &   0.02\\ 
      &      & 1.50 & 2.00&  0.21 &   0.02&  0.21 &   0.02&  0.25 &   0.02\\ 
      &      & 2.00 & 2.50& 0.140 &  0.012& 0.190 &  0.013&  0.22 &   0.02\\ 
      &      & 2.50 & 3.00& 0.093 &  0.011& 0.138 &  0.009& 0.196 &  0.015\\ 
      &      & 3.00 & 3.50& 0.043 &  0.007& 0.077 &  0.008& 0.144 &  0.013\\ 
      &      & 3.50 & 4.00& 0.025 &  0.004& 0.044 &  0.005& 0.092 &  0.009\\ 
      &      & 4.00 & 5.00& 0.016 &  0.003& 0.027 &  0.003& 0.053 &  0.005\\ 
      &      & 5.00 & 6.50& 0.002 &  0.000& 0.005 &  0.001& 0.023 &  0.003\\ 
      &      & 6.50 & 8.00&       &       &       &       & 0.006 &  0.001\\

%%%%%%%%%%%%%%%%%%%%%%%%%%%%%%%%%%%%%%%%%%%%%%%%%%%%%%%%%%%%%%%%%%%%%%%%%%%%%
\hline
\end{tabular}
}
\end{table*}
\clearpage
\begin{table*}[!ht]
  \caption{\label{tab:xsec_results_Befine2}
    HARP results for the double-differential  $\pi^-$ production
    cross-section in the laboratory system,
    $d^2\sigma^{\pi}/(dpd\Omega)$, for p--Be interactions at 8,8.9,12~\GeVc.
    Each row refers to a
    different $(p_{\hbox{\small min}} \le p<p_{\hbox{\small max}},
    \theta_{\hbox{\small min}} \le \theta<\theta_{\hbox{\small max}})$ bin,
    where $p$ and $\theta$ are the pion momentum and polar angle, respectively.
    The central value as well as the square-root of the diagonal elements
    of the covariance matrix are given.}

%\begin{tabular}{rrrr|r@{$\pm$}lr{$\pm$}lr{$\pm$}lr{$\pm$}l}
\small{
\begin{tabular}{rrrr|r@{$\pm$}lr@{$\pm$}lr@{$\pm$}l}
\hline
$\theta_{\hbox{\small min}}$ &
$\theta_{\hbox{\small max}}$ &
$p_{\hbox{\small min}}$ &
$p_{\hbox{\small max}}$ &
\multicolumn{6}{c}{$d^2\sigma^{\pi^-}/(dpd\Omega)$}
\\
(rad) & (rad) & (\GeVc) & (\GeVc) &
\multicolumn{6}{c}{(barn/(sr \GeVc))}
\\
  &  &  &
&\multicolumn{2}{c}{$ \bf{8 \ \GeVc}$}
&\multicolumn{2}{c}{$ \bf{8.9 \ \GeVc}$}
&\multicolumn{2}{c}{$ \bf{12 \ \GeVc}$}
\\
\hline

0.025 &0.050 & 0.50 & 0.75&  0.08 &   0.03&  0.11 &   0.02&  0.07 &   0.02\\ 
      &      & 0.75 & 1.00&  0.12 &   0.02&  0.10 &   0.02&  0.14 &   0.03\\ 
      &      & 1.00 & 1.25&  0.14 &   0.03&  0.17 &   0.03&  0.16 &   0.03\\ 
      &      & 1.25 & 1.50&  0.10 &   0.02&  0.15 &   0.02&  0.20 &   0.04\\ 
      &      & 1.50 & 2.00&  0.12 &   0.02&  0.11 &   0.02&  0.13 &   0.02\\ 
      &      & 2.00 & 2.50& 0.075 &  0.014&  0.09 &   0.02&  0.12 &   0.02\\ 
      &      & 2.50 & 3.00& 0.058 &  0.012& 0.074 &  0.013&  0.11 &   0.02\\ 
      &      & 3.00 & 3.50& 0.022 &  0.006& 0.057 &  0.011&  0.10 &   0.02\\ 
      &      & 3.50 & 4.00& 0.020 &  0.007& 0.042 &  0.008& 0.057 &  0.011\\ 
      &      & 4.00 & 5.00& 0.016 &  0.004& 0.028 &  0.004& 0.055 &  0.008\\ 
      &      & 5.00 & 6.50& 0.001 &  0.001& 0.006 &  0.002& 0.028 &  0.006\\ 
      &      & 6.50 & 8.00&       &       &       &       & 0.009 &  0.003\\ 
0.050 &0.075 & 0.50 & 0.75&  0.11 &   0.02&  0.08 &   0.02&  0.12 &   0.03\\ 
      &      & 0.75 & 1.00&  0.12 &   0.02&  0.11 &   0.02&  0.16 &   0.02\\ 
      &      & 1.00 & 1.25&  0.12 &   0.02&  0.14 &   0.02&  0.19 &   0.03\\ 
      &      & 1.25 & 1.50& 0.079 &  0.014& 0.126 &  0.015&  0.17 &   0.02\\ 
      &      & 1.50 & 2.00& 0.106 &  0.013& 0.121 &  0.013&  0.18 &   0.02\\ 
      &      & 2.00 & 2.50& 0.063 &  0.009& 0.092 &  0.009&  0.12 &   0.02\\ 
      &      & 2.50 & 3.00& 0.036 &  0.008& 0.053 &  0.007& 0.104 &  0.014\\ 
      &      & 3.00 & 3.50& 0.027 &  0.006& 0.045 &  0.006& 0.089 &  0.012\\ 
      &      & 3.50 & 4.00& 0.018 &  0.005& 0.034 &  0.006& 0.051 &  0.008\\ 
      &      & 4.00 & 5.00& 0.009 &  0.003& 0.021 &  0.003& 0.047 &  0.006\\ 
      &      & 5.00 & 6.50& 0.001 &  0.001& 0.003 &  0.001& 0.016 &  0.003\\ 
      &      & 6.50 & 8.00&       &       &       &       & 0.006 &  0.002\\ 
0.075 &0.100 & 0.50 & 0.75&  0.13 &   0.02&  0.11 &   0.02&  0.14 &   0.03\\ 
      &      & 0.75 & 1.00&  0.13 &   0.02&  0.15 &   0.02&  0.13 &   0.02\\ 
      &      & 1.00 & 1.25&  0.12 &   0.02& 0.162 &  0.015&  0.15 &   0.02\\ 
      &      & 1.25 & 1.50&  0.13 &   0.02& 0.136 &  0.013&  0.14 &   0.02\\ 
      &      & 1.50 & 2.00& 0.096 &  0.010& 0.116 &  0.009& 0.145 &  0.014\\ 
      &      & 2.00 & 2.50& 0.057 &  0.007& 0.075 &  0.007& 0.126 &  0.013\\ 
      &      & 2.50 & 3.00& 0.031 &  0.007& 0.046 &  0.005& 0.092 &  0.010\\ 
      &      & 3.00 & 3.50& 0.012 &  0.003& 0.032 &  0.003& 0.067 &  0.008\\ 
      &      & 3.50 & 4.00& 0.007 &  0.002& 0.018 &  0.004& 0.040 &  0.006\\ 
      &      & 4.00 & 5.00& 0.005 &  0.002& 0.006 &  0.001& 0.025 &  0.003\\ 
      &      & 5.00 & 6.50&     * &      *&     * &      *& 0.006 &  0.001\\ 
      &      & 6.50 & 8.00&       &       &       &       &     * &      * \\ 
%%%%%%%%%%%%%%%%%%%%%%%%%%%%%%%%%%%%%%%%%%%%%%%%%%%%%%%%%%%%%%%%%%%%%%%%%%%%%
\hline
\end{tabular}
}
\end{table*}

\clearpage
\begin{table*}[hbt]
  \caption{\label{tab:xsec_results_C7} 
    HARP results for the double-differential $\pi^+$  and $\pi^-$ production
    cross-section in the laboratory system,
    $d^2\sigma^{\pi}/(dpd\Omega)$, for $p$--C 
    interactions at 8 and 12~\GeVc.
    Each row refers to a
    different $(p_{\hbox{\small min}} \le p<p_{\hbox{\small max}},
    \theta_{\hbox{\small min}} \le \theta<\theta_{\hbox{\small max}})$ bin,
    where $p$ and $\theta$ are the pion momentum and polar angle, respectively.
    A finer angular binning than in the previous set of tables is used.
    The central value as well as the square-root of the diagonal elements
    of the covariance matrix are given.}

%\begin{tabular}{rrrr|r@{$\pm$}lr{$\pm$}lr{$\pm$}lr{$\pm$}l}
\small{
\begin{tabular}{rrrr|r@{$\pm$}lr@{$\pm$}l|r@{$\pm$}lr@{$\pm$}l}
\hline
$\theta_{\hbox{\small min}}$ &
$\theta_{\hbox{\small max}}$ &
$p_{\hbox{\small min}}$ &
$p_{\hbox{\small max}}$ &
\multicolumn{4}{c}{$d^2\sigma^{\pi^+}/(dpd\Omega)$} &
\multicolumn{4}{c}{$d^2\sigma^{\pi^-}/(dpd\Omega)$}
\\
(rad) & (rad) & (\GeVc) & (\GeVc) &
\multicolumn{4}{c}{(barn/(sr \GeVc))} &
\multicolumn{4}{c}{(barn/(sr \GeVc))}
\\
  &  &  &
&\multicolumn{2}{c}{$ \bf{8 \ \GeVc}$}
&\multicolumn{2}{c}{$ \bf{12 \ \GeVc}$}
&\multicolumn{2}{c}{$ \bf{8 \ \GeVc}$}
&\multicolumn{2}{c}{$ \bf{12 \ \GeVc}$}
\\
\hline
0.025 &0.050 & 0.50 & 0.75&  0.13 &   0.04&  0.22 &   0.07&  0.07 &   0.03&  0.14 &   0.05\\ 
      &      & 0.75 & 1.00&  0.15 &   0.04&  0.16 &   0.04&  0.16 &   0.04&  0.15 &   0.04\\ 
      &      & 1.00 & 1.25&  0.16 &   0.04&  0.20 &   0.05&  0.19 &   0.04&  0.20 &   0.04\\ 
      &      & 1.25 & 1.50&  0.24 &   0.04&  0.25 &   0.05&  0.08 &   0.02&  0.22 &   0.05\\ 
      &      & 1.50 & 2.00&  0.22 &   0.03&  0.24 &   0.04&  0.16 &   0.03&  0.26 &   0.04\\ 
      &      & 2.00 & 2.50&  0.19 &   0.03&  0.25 &   0.04&  0.08 &   0.02&  0.10 &   0.03\\ 
      &      & 2.50 & 3.00&  0.21 &   0.03&  0.34 &   0.04&  0.07 &   0.02&  0.11 &   0.03\\ 
      &      & 3.00 & 3.50&  0.09 &   0.02&  0.27 &   0.04& 0.034 &  0.010&  0.13 &   0.03\\ 
      &      & 3.50 & 4.00&  0.05 &   0.02&  0.29 &   0.04& 0.020 &  0.009&  0.09 &   0.02\\ 
      &      & 4.00 & 5.00& 0.019 &  0.011&  0.15 &   0.02& 0.017 &  0.005& 0.062 &  0.012\\ 
      &      & 5.00 & 6.50& 0.021 &  0.010&  0.12 &   0.02& 0.001 &  0.001& 0.031 &  0.008\\ 
      &      & 6.50 & 8.00&       &       & 0.055 &  0.010&       &       & 0.010 &  0.005\\ 
0.050 &0.075 & 0.50 & 0.75&  0.14 &   0.03&  0.23 &   0.05&  0.10 &   0.03&  0.10 &   0.03\\ 
      &      & 0.75 & 1.00&  0.18 &   0.03&  0.21 &   0.04&  0.11 &   0.02&  0.16 &   0.03\\ 
      &      & 1.00 & 1.25&  0.18 &   0.03&  0.18 &   0.03&  0.15 &   0.03&  0.21 &   0.03\\ 
      &      & 1.25 & 1.50&  0.17 &   0.02&  0.22 &   0.04&  0.17 &   0.03&  0.20 &   0.03\\ 
      &      & 1.50 & 2.00&  0.23 &   0.02&  0.25 &   0.03&  0.12 &   0.02&  0.17 &   0.02\\ 
      &      & 2.00 & 2.50&  0.19 &   0.02&  0.29 &   0.03& 0.080 &  0.013&  0.15 &   0.02\\ 
      &      & 2.50 & 3.00&  0.19 &   0.02&  0.28 &   0.03& 0.037 &  0.010&  0.12 &   0.02\\ 
      &      & 3.00 & 3.50&  0.07 &   0.02&  0.20 &   0.02& 0.031 &  0.008&  0.11 &   0.02\\ 
      &      & 3.50 & 4.00& 0.039 &  0.013&  0.19 &   0.02& 0.036 &  0.009& 0.075 &  0.014\\ 
      &      & 4.00 & 5.00& 0.035 &  0.008& 0.105 &  0.013& 0.010 &  0.005& 0.053 &  0.009\\ 
      &      & 5.00 & 6.50& 0.007 &  0.002& 0.055 &  0.007&    *  &   *   & 0.012 &  0.004\\ 
      &      & 6.50 & 8.00&       &       & 0.029 &  0.005&       &       & 0.004 &  0.002\\ 
0.075 &0.100 & 0.50 & 0.75&  0.13 &   0.03&  0.15 &   0.03&  0.19 &   0.03&  0.14 &   0.03\\ 
      &      & 0.75 & 1.00&  0.26 &   0.03&  0.19 &   0.03&  0.12 &   0.02&  0.12 &   0.02\\ 
      &      & 1.00 & 1.25&  0.22 &   0.02&  0.26 &   0.03&  0.15 &   0.02&  0.18 &   0.03\\ 
      &      & 1.25 & 1.50&  0.22 &   0.02&  0.26 &   0.03&  0.15 &   0.02&  0.20 &   0.03\\ 
      &      & 1.50 & 2.00&  0.27 &   0.02&  0.30 &   0.03& 0.100 &  0.013&  0.21 &   0.02\\ 
      &      & 2.00 & 2.50&  0.18 &   0.02&  0.25 &   0.02& 0.065 &  0.010&  0.15 &   0.02\\ 
      &      & 2.50 & 3.00& 0.095 &  0.013&  0.20 &   0.02& 0.029 &  0.008&  0.09 &   0.02\\ 
      &      & 3.00 & 3.50& 0.054 &  0.010& 0.143 &  0.014& 0.019 &  0.004& 0.056 &  0.009\\ 
      &      & 3.50 & 4.00& 0.023 &  0.005& 0.112 &  0.013& 0.017 &  0.004& 0.042 &  0.007\\ 
%%%%%%%%%%%%%%%%%%%%%%%%%%%%%%%%%%%%%%%%%%%%%%%%%%%%%%%%%%%%%%%%%%%%%%%%%%%%%
\hline
\end{tabular}
}
\end{table*}

\begin{table*}[!ht]
  \caption{\label{tab:xsec_results_Alfine1}
    HARP results for the double-differential $\pi^+$  production
    cross-section in the laboratory system,
    $d^2\sigma^{\pi}/(dpd\Omega)$, for p--Al interactions at 8,12,12.9~\GeVc.
    Each row refers to a
    different $(p_{\hbox{\small min}} \le p<p_{\hbox{\small max}},
    \theta_{\hbox{\small min}} \le \theta<\theta_{\hbox{\small max}})$ bin,
    where $p$ and $\theta$ are the pion momentum and polar angle, respectively.
    The central value as well as the square-root of the diagonal elements
    of the covariance matrix are given.}
%\begin{tabular}{rrrr|r@{$\pm$}lr{$\pm$}lr{$\pm$}lr{$\pm$}l}
\small{
\begin{tabular}{rrrr|r@{$\pm$}lr@{$\pm$}lr@{$\pm$}l}
\hline
$\theta_{\hbox{\small min}}$ &
$\theta_{\hbox{\small max}}$ &
$p_{\hbox{\small min}}$ &
$p_{\hbox{\small max}}$ &
\multicolumn{6}{c}{$d^2\sigma^{\pi^+}/(dpd\Omega)$}
\\
(rad) & (rad) & (\GeVc) & (\GeVc) &
\multicolumn{6}{c}{(barn/(sr \GeVc ))}
\\
  &  &  &
&\multicolumn{2}{c}{$ \bf{8 \ \GeVc}$}
&\multicolumn{2}{c}{$ \bf{12 \ \GeVc}$}
&\multicolumn{2}{c}{$ \bf{12.9 \ \GeVc}$}
\\
\hline

0.025 &0.050 & 0.50 & 0.75&  0.34 &   0.10&  0.46 &   0.15&  0.28 &   0.08\\ 
      &      & 0.75 & 1.00&  0.30 &   0.07&  0.44 &   0.10&  0.43 &   0.07\\ 
      &      & 1.00 & 1.25&  0.23 &   0.06&  0.31 &   0.08&  0.43 &   0.05\\ 
      &      & 1.25 & 1.50&  0.25 &   0.05&  0.39 &   0.09&  0.36 &   0.05\\ 
      &      & 1.50 & 2.00&  0.31 &   0.05&  0.45 &   0.08&  0.45 &   0.04\\ 
      &      & 2.00 & 2.50&  0.26 &   0.04&  0.50 &   0.09&  0.50 &   0.05\\ 
      &      & 2.50 & 3.00&  0.30 &   0.05&  0.49 &   0.08&  0.47 &   0.04\\ 
      &      & 3.00 & 3.50&  0.18 &   0.03&  0.42 &   0.07&  0.52 &   0.04\\ 
      &      & 3.50 & 4.00&  0.25 &   0.04&  0.42 &   0.06&  0.42 &   0.04\\ 
      &      & 4.00 & 5.00&  0.12 &   0.03&  0.27 &   0.04&  0.30 &   0.02\\ 
      &      & 5.00 & 6.50& 0.027 &  0.013&  0.12 &   0.02&  0.19 &   0.02\\ 
      &      & 6.50 & 8.00&       &       &  0.06 &   0.02& 0.108 &  0.013\\ 
0.050 &0.075 & 0.50 & 0.75&  0.31 &   0.07&  0.44 &   0.10&  0.48 &   0.08\\ 
      &      & 0.75 & 1.00&  0.33 &   0.05&  0.33 &   0.06&  0.45 &   0.04\\ 
      &      & 1.00 & 1.25&  0.24 &   0.04&  0.31 &   0.06&  0.31 &   0.03\\ 
      &      & 1.25 & 1.50&  0.33 &   0.05&  0.37 &   0.07&  0.48 &   0.05\\ 
      &      & 1.50 & 2.00&  0.38 &   0.04&  0.44 &   0.06&  0.47 &   0.04\\ 
      &      & 2.00 & 2.50&  0.37 &   0.04&  0.46 &   0.06&  0.49 &   0.03\\ 
      &      & 2.50 & 3.00&  0.27 &   0.03&  0.48 &   0.06&  0.44 &   0.03\\ 
      &      & 3.00 & 3.50&  0.20 &   0.03&  0.33 &   0.05&  0.44 &   0.03\\ 
      &      & 3.50 & 4.00&  0.12 &   0.02&  0.28 &   0.05&  0.34 &   0.02\\ 
      &      & 4.00 & 5.00&  0.11 &   0.02&  0.18 &   0.03&  0.23 &   0.02\\ 
      &      & 5.00 & 6.50& 0.009 &  0.003& 0.067 &  0.012& 0.111 &  0.008\\ 
      &      & 6.50 & 8.00&       &       & 0.035 &  0.008& 0.049 &  0.006\\ 
0.075 &0.100 & 0.50 & 0.75&  0.32 &   0.06&  0.47 &   0.09&  0.55 &   0.06\\ 
      &      & 0.75 & 1.00&  0.38 &   0.05&  0.57 &   0.08&  0.48 &   0.04\\ 
      &      & 1.00 & 1.25&  0.40 &   0.04&  0.52 &   0.07&  0.48 &   0.04\\ 
      &      & 1.25 & 1.50&  0.41 &   0.04&  0.43 &   0.06&  0.55 &   0.04\\ 
      &      & 1.50 & 2.00&  0.34 &   0.03&  0.51 &   0.05&  0.56 &   0.03\\ 
      &      & 2.00 & 2.50&  0.27 &   0.03&  0.48 &   0.05&  0.45 &   0.03\\ 
      &      & 2.50 & 3.00&  0.20 &   0.02&  0.41 &   0.04&  0.38 &   0.02\\ 
      &      & 3.00 & 3.50&  0.16 &   0.02&  0.25 &   0.03&  0.30 &   0.02\\ 
      &      & 3.50 & 4.00&  0.10 &   0.02&  0.19 &   0.02&  0.23 &   0.02\\ 
      &      & 4.00 & 5.00& 0.056 &  0.008& 0.117 &  0.015& 0.126 &  0.011\\ 
      &      & 5.00 & 6.50& 0.004 &  0.002& 0.034 &  0.007& 0.052 &  0.005\\ 
      &      & 6.50 & 8.00&       &       & 0.010 &  0.003& 0.015 &  0.002\\ 

%%%%%%%%%%%%%%%%%%%%%%%%%%%%%%%%%%%%%%%%%%%%%%%%%%%%%%%%%%%%%%%%%%%%%%%%%%%%%
\hline
\end{tabular}
}
\end{table*}
\begin{table*}[!ht]
  \caption{\label{tab:xsec_results_Alfine2}
    HARP results for the double-differential  $\pi^-$ production
    cross-section in the laboratory system,
    $d^2\sigma^{\pi}/(dpd\Omega)$, for p--Al interactions at 8,12,12.9~\GeVc.
    Each row refers to a
    different $(p_{\hbox{\small min}} \le p<p_{\hbox{\small max}},
    \theta_{\hbox{\small min}} \le \theta<\theta_{\hbox{\small max}})$ bin,
    where $p$ and $\theta$ are the pion momentum and polar angle, respectively.
    The central value as well as the square-root of the diagonal elements
    of the covariance matrix are given.}

%\begin{tabular}{rrrr|r@{$\pm$}lr{$\pm$}lr{$\pm$}l}
\small{
\begin{tabular}{rrrr|r@{$\pm$}lr@{$\pm$}lr@{$\pm$}l}
\hline
$\theta_{\hbox{\small min}}$ &
$\theta_{\hbox{\small max}}$ &
$p_{\hbox{\small min}}$ &
$p_{\hbox{\small max}}$ &
\multicolumn{6}{c}{$d^2\sigma^{\pi^-}/(dpd\Omega)$}
\\
(rad) & (rad) & (\GeVc) & (\GeVc) &
\multicolumn{6}{c}{(barn/(sr \GeVc))}
\\
  &  &  &
&\multicolumn{2}{c}{$ \bf{8 \ \GeVc}$}
&\multicolumn{2}{c}{$ \bf{12 \ \GeVc}$}
&\multicolumn{2}{c}{$ \bf{12.9 \ \GeVc}$}
\\
\hline

0.025 &0.050 & 0.50 & 0.75&  0.11 &   0.05&  0.09 &   0.05&  0.24 &   0.06\\ 
      &      & 0.75 & 1.00&  0.27 &   0.06&  0.24 &   0.07&  0.36 &   0.05\\ 
      &      & 1.00 & 1.25&  0.35 &   0.07&  0.43 &   0.10&  0.37 &   0.05\\ 
      &      & 1.25 & 1.50&  0.20 &   0.05&  0.57 &   0.12&  0.49 &   0.06\\ 
      &      & 1.50 & 2.00&  0.20 &   0.04&  0.28 &   0.07&  0.39 &   0.04\\ 
      &      & 2.00 & 2.50&  0.14 &   0.04&  0.12 &   0.04&  0.27 &   0.04\\ 
      &      & 2.50 & 3.00&  0.10 &   0.02&  0.15 &   0.05&  0.23 &   0.03\\ 
      &      & 3.00 & 3.50&  0.09 &   0.02&  0.20 &   0.05&  0.25 &   0.03\\ 
      &      & 3.50 & 4.00&  0.05 &   0.02&  0.10 &   0.03&  0.15 &   0.03\\ 
      &      & 4.00 & 5.00& 0.015 &  0.007&  0.11 &   0.03& 0.103 &  0.014\\ 
      &      & 5.00 & 6.50&     * &      *& 0.041 &  0.014& 0.065 &  0.009\\ 
      &      & 6.50 & 8.00&       &       & 0.009 &  0.006& 0.021 &  0.006\\ 
0.050 &0.075 & 0.50 & 0.75&  0.19 &   0.05&  0.45 &   0.11&  0.35 &   0.06\\ 
      &      & 0.75 & 1.00&  0.29 &   0.05&  0.39 &   0.07&  0.41 &   0.04\\ 
      &      & 1.00 & 1.25&  0.18 &   0.04&  0.41 &   0.07&  0.42 &   0.04\\ 
      &      & 1.25 & 1.50&  0.15 &   0.03&  0.22 &   0.05&  0.40 &   0.04\\ 
      &      & 1.50 & 2.00&  0.20 &   0.03&  0.30 &   0.05&  0.33 &   0.03\\ 
      &      & 2.00 & 2.50&  0.13 &   0.02&  0.28 &   0.05&  0.31 &   0.03\\ 
      &      & 2.50 & 3.00&  0.07 &   0.02&  0.19 &   0.04&  0.22 &   0.02\\ 
      &      & 3.00 & 3.50& 0.051 &  0.012&  0.16 &   0.03&  0.21 &   0.02\\ 
      &      & 3.50 & 4.00& 0.047 &  0.011&  0.15 &   0.03&  0.14 &   0.02\\ 
      &      & 4.00 & 5.00& 0.015 &  0.007&  0.07 &   0.02& 0.085 &  0.009\\ 
      &      & 5.00 & 6.50& 0.002 &  0.002& 0.037 &  0.009& 0.043 &  0.005\\ 
      &      & 6.50 & 8.00&       &       & 0.006 &  0.004& 0.013 &  0.003\\ 
0.075 &0.100 & 0.50 & 0.75&  0.32 &   0.06&  0.35 &   0.07&  0.34 &   0.05\\ 
      &      & 0.75 & 1.00&  0.29 &   0.04&  0.34 &   0.05&  0.41 &   0.04\\ 
      &      & 1.00 & 1.25&  0.25 &   0.03&  0.39 &   0.06&  0.38 &   0.03\\ 
      &      & 1.25 & 1.50&  0.22 &   0.03&  0.35 &   0.05&  0.40 &   0.03\\ 
      &      & 1.50 & 2.00&  0.18 &   0.02&  0.36 &   0.04&  0.38 &   0.03\\ 
      &      & 2.00 & 2.50&  0.11 &   0.02&  0.26 &   0.04&  0.28 &   0.02\\ 
      &      & 2.50 & 3.00&  0.07 &   0.02&  0.15 &   0.03&  0.18 &   0.02\\ 
      &      & 3.00 & 3.50& 0.032 &  0.007&  0.13 &   0.02& 0.153 &  0.011\\ 
      &      & 3.50 & 4.00& 0.021 &  0.005& 0.072 &  0.014& 0.096 &  0.010\\ 
      &      & 4.00 & 5.00& 0.009 &  0.003& 0.030 &  0.007& 0.057 &  0.006\\ 
      &      & 5.00 & 6.50&     * &      *& 0.008 &  0.003& 0.018 &  0.002\\ 
      &      & 6.50 & 8.00&       &       & 0.001 &  0.001& 0.003 &  0.001\\ 
%%%%%%%%%%%%%%%%%%%%%%%%%%%%%%%%%%%%%%%%%%%%%%%%%%%%%%%%%%%%%%%%%%%%%%%%%%%%%
\hline
\end{tabular}
}
\end{table*}

\begin{table*}[!ht]
  \caption{\label{tab:xsec_results_Cu7} 
    HARP results for the double-differential $\pi^+$  and $\pi^-$ production
    cross-section in the laboratory system,
    $d^2\sigma^{\pi}/(dpd\Omega)$, for $p$--Cu 
    interactions at 8 and 12~\GeVc.
    Each row refers to a
    different $(p_{\hbox{\small min}} \le p<p_{\hbox{\small max}},
    \theta_{\hbox{\small min}} \le \theta<\theta_{\hbox{\small max}})$ bin,
    where $p$ and $\theta$ are the pion momentum and polar angle, respectively.
    A finer angular binning than in the previous set of tables is used.
    The central value as well as the square-root of the diagonal elements
    of the covariance matrix are given.}

%\begin{tabular}{rrrr|r@{$\pm$}lr{$\pm$}lr{$\pm$}lr{$\pm$}l}
\small{
\begin{tabular}{rrrr|r@{$\pm$}lr@{$\pm$}l|r@{$\pm$}lr@{$\pm$}l}
\hline
$\theta_{\hbox{\small min}}$ &
$\theta_{\hbox{\small max}}$ &
$p_{\hbox{\small min}}$ &
$p_{\hbox{\small max}}$ &
\multicolumn{4}{c}{$d^2\sigma^{\pi^+}/(dpd\Omega)$} &
\multicolumn{4}{c}{$d^2\sigma^{\pi^-}/(dpd\Omega)$}
\\
(rad) & (rad) & (\GeVc) & (\GeVc) &
\multicolumn{4}{c}{(barn/(sr \GeVc))} &
\multicolumn{4}{c}{(barn/(sr \GeVc))}
\\
  &  &  &
&\multicolumn{2}{c}{$ \bf{8 \ \GeVc}$}
&\multicolumn{2}{c}{$ \bf{12 \ \GeVc}$}
&\multicolumn{2}{c}{$ \bf{8 \ \GeVc}$}
&\multicolumn{2}{c}{$ \bf{12 \ \GeVc}$}
\\
\hline
0.025 &0.050 & 0.50 & 0.75&  0.58 &   0.18&  0.50 &   0.18&  0.23 &   0.10&  0.28 &   0.13\\ 
      &      & 0.75 & 1.00&  0.40 &   0.10&  0.56 &   0.14&  0.48 &   0.10&  0.43 &   0.12\\ 
      &      & 1.00 & 1.25&  0.31 &   0.08&  0.45 &   0.12&  0.43 &   0.10&  0.69 &   0.15\\ 
      &      & 1.25 & 1.50&  0.47 &   0.10&  0.80 &   0.17&  0.22 &   0.07&  0.89 &   0.18\\ 
      &      & 1.50 & 2.00&  0.73 &   0.10&  0.77 &   0.13&  0.35 &   0.07&  0.64 &   0.13\\ 
      &      & 2.00 & 2.50&  0.34 &   0.07&  0.79 &   0.13&  0.19 &   0.05&  0.28 &   0.09\\ 
      &      & 2.50 & 3.00&  0.44 &   0.08&  1.03 &   0.15&  0.20 &   0.04&  0.38 &   0.09\\ 
      &      & 3.00 & 3.50&  0.18 &   0.05&  0.78 &   0.11&  0.09 &   0.03&  0.24 &   0.07\\ 
      &      & 3.50 & 4.00&  0.13 &   0.06&  0.66 &   0.10&  0.05 &   0.03&  0.17 &   0.05\\ 
      &      & 4.00 & 5.00&  0.04 &   0.03&  0.41 &   0.06& 0.030 &  0.012&  0.20 &   0.04\\ 
      &      & 5.00 & 6.50&  0.02 &   0.02&  0.22 &   0.04&    *  &    *   &  0.08 &   0.02\\ 
      &      & 6.50 & 8.00&       &       &  0.12 &   0.03&       &       & 0.016 &  0.010\\ 
0.050 &0.075 & 0.50 & 0.75&  0.60 &   0.13&  0.89 &   0.19&  0.35 &   0.08&  0.51 &   0.14\\ 
      &      & 0.75 & 1.00&  0.55 &   0.08&  0.70 &   0.12&  0.33 &   0.06&  0.52 &   0.10\\ 
      &      & 1.00 & 1.25&  0.39 &   0.06&  0.59 &   0.11&  0.39 &   0.07&  0.70 &   0.12\\ 
      &      & 1.25 & 1.50&  0.53 &   0.08&  0.74 &   0.13&  0.35 &   0.06&  0.48 &   0.09\\ 
      &      & 1.50 & 2.00&  0.62 &   0.07&  0.69 &   0.09&  0.37 &   0.05&  0.67 &   0.10\\ 
      &      & 2.00 & 2.50&  0.46 &   0.06&  0.64 &   0.09&  0.22 &   0.04&  0.49 &   0.08\\ 
      &      & 2.50 & 3.00&  0.32 &   0.05&  0.64 &   0.08&  0.09 &   0.02&  0.35 &   0.06\\ 
      &      & 3.00 & 3.50&  0.15 &   0.05&  0.53 &   0.07&  0.05 &   0.02&  0.30 &   0.06\\ 
      &      & 3.50 & 4.00&  0.06 &   0.03&  0.41 &   0.07&  0.08 &   0.02&  0.18 &   0.04\\ 
      &      & 4.00 & 5.00&  0.05 &   0.02&  0.29 &   0.04& 0.016 &  0.010&  0.13 &   0.03\\ 
      &      & 5.00 & 6.50& 0.006 &  0.003&  0.11 &   0.02& 0.000 &  0.001& 0.044 &  0.014\\ 
      &      & 6.50 & 8.00&       &       & 0.043 &  0.012&       &       & 0.014 &  0.010\\ 
0.075 &0.100 & 0.50 & 0.75&  0.43 &   0.09&  0.65 &   0.12&  0.52 &   0.09&  0.72 &   0.14\\ 
      &      & 0.75 & 1.00&  0.64 &   0.08&  0.84 &   0.12&  0.53 &   0.07&  0.53 &   0.08\\ 
      &      & 1.00 & 1.25&  0.65 &   0.07&  0.83 &   0.10&  0.36 &   0.05&  0.73 &   0.10\\ 
      &      & 1.25 & 1.50&  0.71 &   0.07&  0.83 &   0.10&  0.41 &   0.06&  0.82 &   0.10\\ 
      &      & 1.50 & 2.00&  0.51 &   0.05&  0.90 &   0.09&  0.28 &   0.04&  0.49 &   0.06\\ 
      &      & 2.00 & 2.50&  0.43 &   0.05&  0.82 &   0.08&  0.14 &   0.02&  0.35 &   0.05\\ 
      &      & 2.50 & 3.00&  0.21 &   0.03&  0.59 &   0.06&  0.08 &   0.02&  0.35 &   0.05\\ 
      &      & 3.00 & 3.50&  0.11 &   0.02&  0.53 &   0.06& 0.033 &  0.009&  0.17 &   0.03\\ 
      &      & 3.50 & 4.00& 0.048 &  0.014&  0.32 &   0.04& 0.032 &  0.009&  0.13 &   0.02\\ 
%%%%%%%%%%%%%%%%%%%%%%%%%%%%%%%%%%%%%%%%%%%%%%%%%%%%%%%%%%%%%%%%%%%%%%%%%%%%%
\hline
\end{tabular}
}
\end{table*}

\begin{table*}[!ht]
  \caption{\label{tab:xsec_results_Sn7} 
    HARP results for the double-differential $\pi^+$  and $\pi^-$ production
    cross-section in the laboratory system,
    $d^2\sigma^{\pi}/(dpd\Omega)$, for $p$--
Sn 
    interactions at 8 and 12~\GeVc.
    Each row refers to a
    different $(p_{\hbox{\small min}} \le p<p_{\hbox{\small max}},
    \theta_{\hbox{\small min}} \le \theta<\theta_{\hbox{\small max}})$ bin,
    where $p$ and $\theta$ are the pion momentum and polar angle, respectively.
    A finer angular binning than in the previous set of tables is used.
    The central value as well as the square-root of the diagonal elements
    of the covariance matrix are given.}

%\begin{tabular}{rrrr|r@{$\pm$}lr{$\pm$}lr{$\pm$}lr{$\pm$}l}
\small{
\begin{tabular}{rrrr|r@{$\pm$}lr@{$\pm$}l|r@{$\pm$}lr@{$\pm$}l}
\hline
$\theta_{\hbox{\small min}}$ &
$\theta_{\hbox{\small max}}$ &
$p_{\hbox{\small min}}$ &
$p_{\hbox{\small max}}$ &
\multicolumn{4}{c}{$d^2\sigma^{\pi^+}/(dpd\Omega)$} &
\multicolumn{4}{c}{$d^2\sigma^{\pi^-}/(dpd\Omega)$}
\\
(rad) & (rad) & (\GeVc) & (\GeVc) &
\multicolumn{4}{c}{(barn/(sr \GeVc))} &
\multicolumn{4}{c}{(barn/(sr \GeVc))}
\\
  &  &  &
&\multicolumn{2}{c}{$ \bf{8 \ \GeVc}$}
&\multicolumn{2}{c}{$ \bf{12 \ \GeVc}$}
&\multicolumn{2}{c}{$ \bf{8 \ \GeVc}$}
&\multicolumn{2}{c}{$ \bf{12 \ \GeVc}$}
\\
\hline
0.025 &0.050 & 0.50 & 0.75&  1.06 &   0.27&  1.06 &   0.25&  0.86 &   0.24&  1.03 &   0.24\\ 
      &      & 0.75 & 1.00&  0.89 &   0.18&  0.93 &   0.17&  0.81 &   0.15&  0.79 &   0.15\\ 
      &      & 1.00 & 1.25&  0.73 &   0.16&  0.72 &   0.14&  0.69 &   0.15&  1.22 &   0.19\\ 
      &      & 1.25 & 1.50&  0.51 &   0.13&  1.10 &   0.18&  0.37 &   0.11&  1.15 &   0.18\\ 
      &      & 1.50 & 2.00&  0.62 &   0.11&  1.00 &   0.13&  0.51 &   0.10&  0.98 &   0.15\\ 
      &      & 2.00 & 2.50&  0.56 &   0.11&  1.00 &   0.13&  0.35 &   0.08&  0.33 &   0.09\\ 
      &      & 2.50 & 3.00&  0.47 &   0.10&  1.15 &   0.15&  0.24 &   0.06&  0.46 &   0.09\\ 
      &      & 3.00 & 3.50&  0.30 &   0.09&  0.87 &   0.11&  0.15 &   0.04&  0.40 &   0.09\\ 
      &      & 3.50 & 4.00&  0.23 &   0.27&  0.73 &   0.10&  0.05 &   0.04&  0.31 &   0.06\\ 
      &      & 4.00 & 5.00&  0.06 &   0.09&  0.64 &   0.07&  0.04 &   0.02&  0.23 &   0.04\\ 
      &      & 5.00 & 6.50&  0.07 &   0.05&  0.27 &   0.05&  0.00 &   0.01&  0.09 &   0.02\\ 
      &      & 6.50 & 8.00&       &       &  0.14 &   0.03&       &       & 0.022 &  0.012\\ 
0.050 &0.075 & 0.50 & 0.75&  0.82 &   0.17&  1.58 &   0.24&  0.51 &   0.12&  0.79 &   0.15\\ 
      &      & 0.75 & 1.00&  0.76 &   0.11&  0.93 &   0.13&  0.64 &   0.11&  0.81 &   0.11\\ 
      &      & 1.00 & 1.25&  0.82 &   0.11&  0.72 &   0.11&  0.49 &   0.10&  1.17 &   0.14\\ 
      &      & 1.25 & 1.50&  0.84 &   0.12&  0.99 &   0.14&  0.38 &   0.08&  0.69 &   0.10\\ 
      &      & 1.50 & 2.00&  0.73 &   0.09&  0.91 &   0.09&  0.49 &   0.07&  0.84 &   0.10\\ 
      &      & 2.00 & 2.50&  0.70 &   0.08&  0.99 &   0.10&  0.34 &   0.05&  0.56 &   0.08\\ 
      &      & 2.50 & 3.00&  0.52 &   0.08&  0.95 &   0.09&  0.16 &   0.04&  0.46 &   0.06\\ 
      &      & 3.00 & 3.50&  0.21 &   0.07&  0.94 &   0.09&  0.13 &   0.03&  0.41 &   0.06\\ 
      &      & 3.50 & 4.00&  0.05 &   0.04&  0.53 &   0.08&  0.06 &   0.02&  0.29 &   0.04\\ 
      &      & 4.00 & 5.00&  0.09 &   0.03&  0.40 &   0.04&  0.03 &   0.02&  0.17 &   0.03\\ 
      &      & 5.00 & 6.50& 0.011 &  0.005&  0.18 &   0.03& 0.006 &  0.005& 0.044 &  0.012\\ 
      &      & 6.50 & 8.00&       &       &  0.08 &   0.02&       &       & 0.014 &  0.006\\ 
0.075 &0.100 & 0.50 & 0.75&  0.72 &   0.13&  1.05 &   0.15&  0.49 &   0.10&  0.94 &   0.15\\ 
      &      & 0.75 & 1.00&  0.89 &   0.11&  1.21 &   0.13&  0.75 &   0.09&  1.06 &   0.12\\ 
      &      & 1.00 & 1.25&  0.72 &   0.09&  1.25 &   0.12&  0.53 &   0.08&  1.15 &   0.12\\ 
      &      & 1.25 & 1.50&  0.95 &   0.11&  1.19 &   0.12&  0.56 &   0.08&  1.03 &   0.10\\ 
      &      & 1.50 & 2.00&  0.75 &   0.07&  1.21 &   0.09&  0.36 &   0.05&  0.80 &   0.07\\ 
      &      & 2.00 & 2.50&  0.66 &   0.07&  1.04 &   0.08&  0.17 &   0.03&  0.46 &   0.05\\ 
      &      & 2.50 & 3.00&  0.28 &   0.06&  0.84 &   0.07&  0.11 &   0.02&  0.46 &   0.05\\ 
      &      & 3.00 & 3.50&  0.09 &   0.02&  0.54 &   0.05&  0.06 &   0.02&  0.22 &   0.03\\ 
      &      & 3.50 & 4.00&  0.06 &   0.02&  0.38 &   0.04& 0.028 &  0.009&  0.14 &   0.02\\ 
%%%%%%%%%%%%%%%%%%%%%%%%%%%%%%%%%%%%%%%%%%%%%%%%%%%%%%%%%%%%%%%%%%%%%%%%%%%%%
\hline
\end{tabular}
}
\end{table*}

\begin{table*}[!ht]
  \caption{\label{tab:xsec_results_Ta7} 
    HARP results for the double-differential $\pi^+$  and $\pi^-$ production
    cross-section in the laboratory system,
    $d^2\sigma^{\pi}/(dpd\Omega)$, for $p$--Ta 
    interactions at 8 and 12~\GeVc.
    Each row refers to a
    different $(p_{\hbox{\small min}} \le p<p_{\hbox{\small max}},
    \theta_{\hbox{\small min}} \le \theta<\theta_{\hbox{\small max}})$ bin,
    where $p$ and $\theta$ are the pion momentum and polar angle, respectively.
    A finer angular binning than in the previous set of tables is used.
    The central value as well as the square-root of the diagonal elements
    of the covariance matrix are given.}

%\begin{tabular}{rrrr|r@{$\pm$}lr{$\pm$}lr{$\pm$}lr{$\pm$}l}
\small{
\begin{tabular}{rrrr|r@{$\pm$}lr@{$\pm$}l|r@{$\pm$}lr@{$\pm$}l}
\hline
$\theta_{\hbox{\small min}}$ &
$\theta_{\hbox{\small max}}$ &
$p_{\hbox{\small min}}$ &
$p_{\hbox{\small max}}$ &
\multicolumn{4}{c}{$d^2\sigma^{\pi^+}/(dpd\Omega)$} &
\multicolumn{4}{c}{$d^2\sigma^{\pi^-}/(dpd\Omega)$}
\\
(rad) & (rad) & (\GeVc) & (\GeVc) &
\multicolumn{4}{c}{(barn/(sr \GeVc))} &
\multicolumn{4}{c}{(barn/(sr \GeVc))}
\\
  &  &  &
&\multicolumn{2}{c}{$ \bf{8 \ \GeVc}$}
&\multicolumn{2}{c}{$ \bf{12 \ \GeVc}$}
&\multicolumn{2}{c}{$ \bf{8 \ \GeVc}$}
&\multicolumn{2}{c}{$ \bf{12 \ \GeVc}$}
\\
\hline
0.025 &0.050 & 0.50 & 0.75&  1.39 &   0.40&  1.74 &   0.48&  0.69 &   0.27&  0.94 &   0.30\\ 
      &      & 0.75 & 1.00&  0.71 &   0.20&  1.44 &   0.32&  0.65 &   0.17&  1.19 &   0.28\\ 
      &      & 1.00 & 1.25&  0.57 &   0.18&  1.07 &   0.26&  0.52 &   0.16&  1.98 &   0.37\\ 
      &      & 1.25 & 1.50&  0.60 &   0.17&  1.28 &   0.28&  0.44 &   0.15&  1.85 &   0.36\\ 
      &      & 1.50 & 2.00&  0.92 &   0.17&  1.27 &   0.23&  0.61 &   0.13&  0.73 &   0.19\\ 
      &      & 2.00 & 2.50&  0.60 &   0.14&  1.18 &   0.22&  0.28 &   0.09&  0.53 &   0.17\\ 
      &      & 2.50 & 3.00&  0.63 &   0.15&  1.47 &   0.24&  0.22 &   0.07&  0.47 &   0.14\\ 
      &      & 3.00 & 3.50&  0.25 &   0.13&  1.09 &   0.19&  0.17 &   0.06&  0.48 &   0.14\\ 
      &      & 3.50 & 4.00&  0.02 &   0.05&  1.04 &   0.18&  0.05 &   0.05&  0.35 &   0.10\\ 
      &      & 4.00 & 5.00&  0.01 &   0.24&  0.71 &   0.12&  0.03 &   0.04&  0.27 &   0.07\\ 
      &      & 5.00 & 6.50&  0.02 &   0.07&  0.32 &   0.07&  0.00 &   0.03&  0.11 &   0.04\\ 
      &      & 6.50 & 8.00&       &       &  0.17 &   0.05&       &       &  0.02 &   0.02\\ 
0.050 &0.075 & 0.50 & 0.75&  0.82 &   0.22&  1.89 &   0.37&  0.61 &   0.17&  0.85 &   0.22\\ 
      &      & 0.75 & 1.00&  0.99 &   0.17&  1.26 &   0.23&  0.77 &   0.15&  1.61 &   0.26\\ 
      &      & 1.00 & 1.25&  0.76 &   0.13&  0.88 &   0.17&  0.79 &   0.16&  1.87 &   0.26\\ 
      &      & 1.25 & 1.50&  0.90 &   0.15&  1.14 &   0.23&  0.51 &   0.12&  0.99 &   0.18\\ 
      &      & 1.50 & 2.00&  0.84 &   0.12&  1.14 &   0.16&  0.48 &   0.09&  0.95 &   0.15\\ 
      &      & 2.00 & 2.50&  0.71 &   0.11&  1.05 &   0.15&  0.35 &   0.07&  0.91 &   0.15\\ 
      &      & 2.50 & 3.00&  0.44 &   0.09&  1.27 &   0.18&  0.13 &   0.05&  0.59 &   0.12\\ 
      &      & 3.00 & 3.50&  0.18 &   0.09&  0.90 &   0.13&  0.05 &   0.02&  0.49 &   0.10\\ 
      &      & 3.50 & 4.00&  0.06 &   0.06&  0.72 &   0.12&  0.13 &   0.04&  0.28 &   0.09\\ 
      &      & 4.00 & 5.00&  0.07 &   0.05&  0.36 &   0.07&  0.01 &   0.02&  0.14 &   0.04\\ 
      &      & 5.00 & 6.50& 0.004 &  0.004&  0.20 &   0.04&    *  &    *   &  0.07 &   0.02\\ 
      &      & 6.50 & 8.00&       &       &  0.07 &   0.02&       &       & 0.021 &  0.013\\ 
0.075 &0.100 & 0.50 & 0.75&  0.55 &   0.14&  2.01 &   0.30&  0.70 &   0.15&  1.49 &   0.26\\ 
      &      & 0.75 & 1.00&  1.12 &   0.16&  1.41 &   0.20&  0.64 &   0.10&  0.89 &   0.14\\ 
      &      & 1.00 & 1.25&  0.96 &   0.13&  1.83 &   0.22&  0.72 &   0.12&  1.16 &   0.17\\ 
      &      & 1.25 & 1.50&  1.10 &   0.14&  1.07 &   0.15&  0.41 &   0.08&  1.01 &   0.15\\ 
      &      & 1.50 & 2.00&  0.77 &   0.09&  1.53 &   0.15&  0.38 &   0.06&  0.84 &   0.11\\ 
      &      & 2.00 & 2.50&  0.49 &   0.08&  1.06 &   0.13&  0.21 &   0.04&  0.93 &   0.12\\ 
      &      & 2.50 & 3.00&  0.27 &   0.06&  1.13 &   0.12&  0.11 &   0.03&  0.39 &   0.08\\ 
      &      & 3.00 & 3.50&  0.13 &   0.05&  0.50 &   0.08&  0.04 &   0.02&  0.38 &   0.06\\ 
      &      & 3.50 & 4.00&  0.05 &   0.02&  0.45 &   0.07&  0.06 &   0.02&  0.21 &   0.04\\ 
%%%%%%%%%%%%%%%%%%%%%%%%%%%%%%%%%%%%%%%%%%%%%%%%%%%%%%%%%%%%%%%%%%%%%%%%%%%%%
\hline
\end{tabular}
}
\end{table*}
\begin{table*}[!ht]
  \caption{\label{tab:xsec_results_Pb7} 
    HARP results for the double-differential $\pi^+$  and $\pi^-$ production
    cross-section in the laboratory system,
    $d^2\sigma^{\pi}/(dpd\Omega)$, for $p$--Pb 
    interactions at 8 and 12~\GeVc.
    Each row refers to a
    different $(p_{\hbox{\small min}} \le p<p_{\hbox{\small max}},
    \theta_{\hbox{\small min}} \le \theta<\theta_{\hbox{\small max}})$ bin,
    where $p$ and $\theta$ are the pion momentum and polar angle, respectively.
    A finer angular binning than in the previous set of tables is used.
    The central value as well as the square-root of the diagonal elements
    of the covariance matrix are given.}

%\begin{tabular}{rrrr|r@{$\pm$}lr{$\pm$}lr{$\pm$}lr{$\pm$}l}
\small{
\begin{tabular}{rrrr|r@{$\pm$}lr@{$\pm$}l|r@{$\pm$}lr@{$\pm$}l}
\hline
$\theta_{\hbox{\small min}}$ &
$\theta_{\hbox{\small max}}$ &
$p_{\hbox{\small min}}$ &
$p_{\hbox{\small max}}$ &
\multicolumn{4}{c}{$d^2\sigma^{\pi^+}/(dpd\Omega)$} &
\multicolumn{4}{c}{$d^2\sigma^{\pi^-}/(dpd\Omega)$}
\\
(rad) & (rad) & (\GeVc) & (\GeVc) &
\multicolumn{4}{c}{(barn/(sr \GeVc))} &
\multicolumn{4}{c}{(barn/(sr \GeVc))}
\\
  &  &  &
&\multicolumn{2}{c}{$ \bf{8 \ \GeVc}$}
&\multicolumn{2}{c}{$ \bf{12 \ \GeVc}$}
&\multicolumn{2}{c}{$ \bf{8 \ \GeVc}$}
&\multicolumn{2}{c}{$ \bf{12 \ \GeVc}$}
\\
\hline
0.025 &0.050 & 0.50 & 0.75&  1.12 &   0.36&  1.66 &   0.56&  0.23 &   0.15&  1.47 &   0.51\\ 
      &      & 0.75 & 1.00&  1.17 &   0.28&  2.34 &   0.57&  1.23 &   0.27&  2.20 &   0.54\\ 
      &      & 1.00 & 1.25&  0.38 &   0.14&  1.31 &   0.39&  0.81 &   0.21&  1.00 &   0.32\\ 
      &      & 1.25 & 1.50&  0.62 &   0.19&  1.23 &   0.37&  0.35 &   0.13&  2.00 &   0.54\\ 
      &      & 1.50 & 2.00&  1.13 &   0.20&  1.59 &   0.36&  0.54 &   0.13&  1.18 &   0.33\\ 
      &      & 2.00 & 2.50&  0.54 &   0.15&  1.17 &   0.30&  0.34 &   0.12&  0.55 &   0.23\\ 
      &      & 2.50 & 3.00&  0.50 &   0.14&  1.39 &   0.34&  0.16 &   0.06&  0.64 &   0.23\\ 
      &      & 3.00 & 3.50&  0.33 &   0.13&  1.69 &   0.35&  0.16 &   0.07&  0.61 &   0.24\\ 
      &      & 3.50 & 4.00&  0.08 &   0.09&  0.73 &   0.20&  0.03 &   0.04&  0.69 &   0.22\\ 
      &      & 4.00 & 5.00&  0.05 &   0.19&  0.66 &   0.15&  0.07 &   0.03&  0.44 &   0.14\\ 
      &      & 5.00 & 6.50&   *   &   *   &  0.41 &   0.11&  0.00 &   0.01&  0.16 &   0.22\\ 
      &      & 6.50 & 8.00&       &       &  0.14 &   0.06&       &       &  0.02 &   0.15\\ 
0.050 &0.075 & 0.50 & 0.75&  0.88 &   0.23&  2.65 &   0.60&  0.47 &   0.15&  1.08 &   0.33\\ 
      &      & 0.75 & 1.00&  1.14 &   0.18&  1.61 &   0.34&  0.73 &   0.15&  1.25 &   0.28\\ 
      &      & 1.00 & 1.25&  0.90 &   0.15&  1.06 &   0.26&  0.58 &   0.14&  1.43 &   0.31\\ 
      &      & 1.25 & 1.50&  0.88 &   0.16&  1.33 &   0.32&  0.56 &   0.13&  1.19 &   0.29\\ 
      &      & 1.50 & 2.00&  0.81 &   0.12&  1.26 &   0.23&  0.60 &   0.10&  0.86 &   0.19\\ 
      &      & 2.00 & 2.50&  0.67 &   0.11&  1.30 &   0.23&  0.33 &   0.07&  0.58 &   0.16\\ 
      &      & 2.50 & 3.00&  0.60 &   0.11&  1.11 &   0.22&  0.18 &   0.06&  0.50 &   0.16\\ 
      &      & 3.00 & 3.50&  0.28 &   0.11&  0.87 &   0.18&  0.07 &   0.03&  0.44 &   0.13\\ 
      &      & 3.50 & 4.00&  0.03 &   0.13&  0.61 &   0.14&  0.10 &   0.04&  0.25 &   0.20\\ 
      &      & 4.00 & 5.00&  0.12 &   0.06&  0.52 &   0.11&  0.02 &   0.02&  0.19 &   0.16\\ 
      &      & 5.00 & 6.50&  0.01 &   0.03&  0.20 &   0.06&    *  &    *  &  0.07 &   0.09\\ 
      &      & 6.50 & 8.00&       &       &  0.03 &   0.02&       &       &  0.02 &   0.10\\ 
0.075 &0.100 & 0.50 & 0.75&  0.72 &   0.17&  1.50 &   0.32&  0.93 &   0.18&  1.53 &   0.35\\ 
      &      & 0.75 & 1.00&  1.21 &   0.16&  1.57 &   0.29&  0.70 &   0.11&  1.44 &   0.26\\ 
      &      & 1.00 & 1.25&  0.75 &   0.11&  1.51 &   0.27&  0.41 &   0.09&  1.60 &   0.28\\ 
      &      & 1.25 & 1.50&  0.92 &   0.13&  0.93 &   0.19&  0.64 &   0.11&  1.21 &   0.22\\ 
      &      & 1.50 & 2.00&  0.81 &   0.10&  1.62 &   0.24&  0.37 &   0.07&  0.96 &   0.16\\ 
      &      & 2.00 & 2.50&  0.66 &   0.10&  1.48 &   0.20&  0.23 &   0.05&  0.91 &   0.16\\ 
      &      & 2.50 & 3.00&  0.28 &   0.07&  0.96 &   0.14&  0.12 &   0.04&  0.38 &   0.09\\ 
      &      & 3.00 & 3.50&  0.06 &   0.03&  0.80 &   0.13& 0.037 &  0.015&  0.50 &   0.19\\ 
      &      & 3.50 & 4.00& 0.027 &  0.015&  0.52 &   0.10& 0.028 &  0.011&  0.31 &   0.16\\ 
%%%%%%%%%%%%%%%%%%%%%%%%%%%%%%%%%%%%%%%%%%%%%%%%%%%%%%%%%%%%%%%%%%%%%%%%%%%%%
\hline
\end{tabular}
}
\end{table*}

\end{appendix}
\end{document}